\documentclass[11pt,a4paper]{article}
\usepackage{jheppub,xcolor}
\usepackage[utf8]{inputenc}
\graphicspath{{./figs/}}

\usepackage{amssymb}
\usepackage{dcolumn}	
\usepackage{bm}			
\usepackage{bbm} 		
\usepackage{multirow}
\usepackage{slashed}
\usepackage{blkarray}
\usepackage{booktabs}
\usepackage{makecell}
\usepackage{longtable}
\usepackage{placeins}
\usepackage{graphicx}
\usepackage{float}
\usepackage{multirow}
\usepackage{verbatim}
\usepackage{hyperref}
\usepackage{lscape} 
\usepackage{tablefootnote} 
\usepackage[font=small,labelfont=bf]{caption}
\usepackage[compat=1.1.0]{tikz-feynman}

\definecolor{Red}{rgb}{1.,0.,0.}
\definecolor{aquamarine}{rgb}{0.6,1.,0.87}

\newcommand{\lag}{\mathcal{L}}
\newcommand{\mcM}{\mathcal{M}}
\newcommand{\mcO}{\mathcal{O}}

\newcommand{\phSpa}{\ensuremath{\phantom{\frac{C^Q}{1}}}\hspace{-1.75em}}
\newcommand{\sss}{\scriptscriptstyle}
\newcommand{\pdp}{\ensuremath{\varphi^\dagger\varphi}}

\newcommand{\Ohqii}{\mcO_{\varphi q,ii}^{(1)}}
\newcommand{\Ohqtii}{\mcO_{\varphi q,ii}^{(3)}}

\newcommand{\OhqTopMinus}{\mcO_{\varphi Q}^{(-)}}
\newcommand{\OhqTopt}{\mcO_{\varphi Q}^{(3)}}
\newcommand{\Ohu}{\mcO_{\varphi u}}
\newcommand{\Oht}{\mcO_{\varphi t}}
\newcommand{\Oth}{\mcO_{t \varphi}}
\newcommand{\Otg}{\mcO_{t G}}
\newcommand{\Ohd}{\mcO_{\varphi d}}
\newcommand{\Odh}{\mcO_{d \varphi}}
\newcommand{\Opg}{\mcO_{\varphi G}}
\newcommand{\op}{\mathcal{O}}
\newcommand{\Opp}[2]{\OO_{\sss #1}^{\sss #2}}
\newcommand{\OO}{\ensuremath{\mathcal{O}}}

\newcommand{\chq}{c_{\varphi q}^{(1)}}
\newcommand{\chqt}{c_{\varphi q}^{(3)}}
\newcommand{\chqTop}{c_{\varphi Q}^{(1)}}
\newcommand{\chqTopt}{c_{\varphi Q}^{(3)}}
\newcommand{\chqTopMinus}{c_{\varphi Q}^{(-)}}
\newcommand{\chu}{c_{\varphi u}}
\newcommand{\cht}{c_{\varphi t}}
\newcommand{\chd}{c_{\varphi d}}
\newcommand{\cth}{c_{t \varphi }}
\newcommand{\ctg}{c_{t G}}
\newcommand{\cpg}{c_{\varphi G}}
\newcommand{\ctW}{c_{t W}}
\newcommand{\ctB}{c_{t B}}
\newcommand{\ctZ}{c_{t Z}}
\newcommand{\cpB}{c_{\varphi B}}
\newcommand{\cpW}{c_{\varphi W}}
\newcommand{\chqtii}{\ensuremath{c_{\varphi q,ii}^{(3)}}}
\newcommand{\chqii}{\ensuremath{c_{\varphi q,ii}^{(1)}}}
\newcommand{\chqminusii}{\ensuremath{c_{\varphi q,ii}^{(-)}}}

\newcommand{\ptmin}{\ensuremath{p_{T,\text{min}}}}

\newcommand{\ptz}{\ensuremath{p_{T}^{Z}}}
\newcommand{\pth}{\ensuremath{p_{T}^{H}}}
\newcommand{\TeV}{\,\mathrm{TeV}}
\newcommand{\GeV}{\,\mathrm{GeV}}

\newcommand{\code}[1]{\mbox{\texttt{#1}}}
\newcommand\logsmtsq{\text{log}^2\big(\frac{s}{m_t^2} \big)}
\newcommand\logsmt{\text{log}\big(\frac{s}{m_t^2} \big)}

\newcommand\logsmzsq{\text{log}^2\big(\frac{s}{m_Z^2} \big)}

\newcommand\sw{s_\text{w}}
\newcommand\cw{c_\text{w}}
\newcommand\ct{c\theta}

\def\lra#1{\overset{\text{\scriptsize$\leftrightarrow$}}{#1}}

\title{Diboson production in the SMEFT from gluon fusion}

\author[a]{Alejo~N.~Rossia,}
\author[a]{Marion~O. A.~Thomas}
\author[a]{and Eleni~Vryonidou}
\affiliation[a]{Dept. of Physics and Astronomy, University of Manchester, Manchester M13 9PL, UK}

\emailAdd{alejo.rossia@manchester.ac.uk}
\emailAdd{marion.thomas@manchester.ac.uk}
\emailAdd{eleni.vryonidou@manchester.ac.uk}

\date{\today}
\abstract{Precision measurements of diboson production at the LHC is an important probe of the limits of the Standard Model. The gluon-fusion channel of this process offers a connection between the Higgs and top sectors. We study in a systematic way gluon-induced diboson production in the Standard Model Effective Field Theory. We compute the amplitudes of double Higgs, double $Z/W$ and associated $ZH$ production at one loop and with up to one insertion of a dimension-6 operator. We study their high-energy limit and identify to which operators each channel could be most sensitive. To illustrate the relevance of these processes, we perform a phenomenological study of associated $ZH$ production. 
We show that for some top operators the gluon-induced channel can offer competitive sensitivity to constraints obtained from top quark production processes. }


\begin{document}

\maketitle
\newpage
\section{Introduction}

The main goal of collider phenomenology at this time is to assimilate the lessons from the early runs of the Large Hadron Collider (LHC) and set the stage to fully exploit the data to be collected during Run 3 and the High-Luminosity LHC (HL-LHC) era.
In light of the absence of new resonances, the search for Beyond the Standard Model (BSM) Physics at the LHC is moving towards the indirect precision measurement paradigm.

The first manifestation of a new particle at energies beyond our direct reach would likely be a rise of the high-energy tails in differential distributions with respect to the Standard Model (SM) prediction. Hence, the precise measurement of such distributions represents an excellent opportunity to test our current best description of Nature and understand how much space is left for New Physics (NP). Not only this could yield powerful insights into the physics that awaits us at higher energies, but also will constitute a key piece in the legacy of the (HL-)LHC program.

Assuming that all BSM particles have a mass larger than the energies we can probe directly, calls for the use of an Effective Field Theory (EFT) in order to parametrise their effects. The Standard Model Effective Field Theory (SMEFT) offers a powerful and systematic method to parametrise deviations from the Standard Model and interpret (HL-)LHC data. It is built from the SM, respecting its field content, gauge symmetries and Electroweak Symmetry Breaking (EWSB) pattern, and adding higher-dimensional operators. As with any other EFT, it is only valid up to a cutoff energy scale $\Lambda$.

Formally, the SMEFT Lagrangian is an infinite series in the energy dimension of the operators,
\begin{equation}
    \lag_{\text{SMEFT}} = \lag_{\text{SM}} +\sum_{d=5}^{\infty}\sum_{k} \frac{c_k}{\Lambda^{d-4}}\mcO_{k}^{(d)}
\end{equation}
where each operator $\mcO_{k}^{(d)}$ has energy dimension $(d)$, $c_k$ is a dimensionless
coupling that we call Wilson coefficient (WC), and the series converges as long as the typical energy of the process obeys $E/\Lambda \ll 1$. Normalising the WCs in this way makes explicit how higher-order operator contributions are further suppressed by the cutoff scale.
The leading deviations with respect to the SM of relevance for collider physics are generated by operators of dimension 6, on which we will focus. 

Not every process is well-suited to constrain SMEFT WCs via precision measurements. Those in which amplitudes with one dimension-6 operator insertion interfere with the SM and generate growth with energy with respect to the SM cross-section are particularly promising. If the process has also low background levels, the high-energy tails of differential distributions could give stringent constraints on the value of the WCs. Drell-Yan~\cite{deBlas:2013qqa,Farina:2016rws,Alioli:2017nzr,Fuentes-Martin:2020lea, Torre:2020aiz, Alioli:2020kez} and diboson production~\cite{Biekotter:2014gup, Butter:2016cvz, Green:2016trm, Baglio:2017bfe, Panico:2017frx, Azatov:2017kzw, Franceschini:2017xkh, Bellazzini:2018paj, Liu:2018pkg, Henning:2018kys,  Biekotter:2018rhp, Banerjee:2018bio,  Grojean:2018dqj,  Baglio:2018bkm, Azatov:2019xxn, Brehmer:2019gmn, Banerjee:2019pks, Banerjee:2019twi, Chiu:2019ksm, Baglio:2019uty, Bishara:2020vix, Chen:2020mev, Baglio:2020oqu, Ethier:2021ydt, Bishara:2020pfx, Bishara:2022vsc} have become the two quintessential processes in this program thanks to combining both features with an interesting breadth of interactions to probe.

Diboson, i.e. $VV(')$, $VH$, and $HH$ production\footnote{Double Higgs production is formally part of the diboson category although its study tends to be separated due to its relevance for Higgs self-interactions. Here, we will not make such distinction.}, allows the study of several EW couplings strictly determined by the SM gauge symmetry.  

At the LHC, $VV(')$ and $VH$ production are largely dominated by the quark-initiated channel whilst $HH$ production is induced by top quark loops. The tree-level quark initiated contributions to $VV(')$ and $VH$ are typically studied to provide information on the light quark couplings to the gauge bosons, the gauge boson self-interaction and the couplings of the Higgs to the gauge bosons. 

The gap between $VV(')$ and $VH$ production and the top sector is bridged by the gluon-fusion channel of diboson production. Though limited to neutral final states and suppressed due to its loop-induced nature, gluon fusion can still give a relevant contribution to $WW$, $ZZ$ and $ZH$ production at the LHC in the high-energy regime~\cite{Glover:1988rg,Glover:1988fe,Binoth:2005ua,Djouadi:2005gi,Englert:2013vua}. In all cases, the cross-section is dominated by top-quark-loop diagrams, sometimes interacting directly with a virtual Higgs boson.
Furthermore, the different energy behaviour of the quark- and gluon-initiated channels offers the opportunity to amplify the contribution of the latter by probing particular kinematic regions. 
 
This link to the top sector is of particular interest since the top quark is another candidate to show signs of BSM physics at colliders. Already in the SM, the top sector is closely connected to the Higgs/EW one via the near-unity top Yukawa coupling, 
and the top EW gauge couplings. Furthermore, several classes of BSM models 
tend to relate closely the BSM fate of the top quark, the Higgs boson and the EW gauge bosons. As a reflection of this, the effort to perform combined fits of the EW, Higgs and top sectors has increased remarkably in recent years~\cite{Ethier:2021bye,Ellis:2020unq}.

In light of this connection to top physics, several groups have studied gluon-initiated diboson production before and stressed its potential to constrain top couplings~\cite{Azatov:2015oxa,Azatov:2016xik,Englert:2016hvy,BessidskaiaBylund:2016jvp,Cao:2020npb}. To fully explore the potential of this class of  processes we aim to systematically analyse the high-energy behaviour of the gluon-initiated diboson production amplitudes generated by dimension-6 SMEFT operators. Such a study will enable us to identify which operators are more promising, especially in the context of high-energy measurements at the LHC, HL-LHC and future colliders. Analyses with the same spirit for quark-initiated diboson and $t\bar t$ production are available in the literature~\cite{Franceschini:2017xkh,Maltoni:2019aot}, as well as studies within the on-shell scattering amplitudes formalism~\cite{Shadmi:2018xan,Liu:2023jbq}.

To then demonstrate in a realistic manner how the loop induced component of the diboson process plays a special role, we focus on $ZH$ production.
Previous studies of this process focused on one of the initial states, either $gg$ or $q\bar q$, and set the other to its SM value to play the role of a background~\cite{Englert:2016hvy,Bishara:2022vsc,Bishara:2020pfx,Banerjee:2019twi}. 
This methodology can be justified by the fact that quark- and gluon-initiated $ZH$ production probe disjoint sets of SMEFT dimension-6 operators. When those sets are linked by a flavour hypothesis such as Flavour Universality, the sensitivity arises mainly from the quark-initiated process due to a higher cross-section and steeper energy growth. 

However, the progress towards global fits of experimental data and more general flavour assumptions calls for a joint study of the different initial-state channels.  In this work, we perform for the first time such joint phenomenological analysis of both the quark and gluon-induced production channels.

 This paper is organised as follows. In Section~\ref{sec:HeliAmp}, we compute analytically the one-loop amplitudes of $gg\to HH$, $gg\to ZH$, and $gg\to ZZ/WW$ in the SMEFT with up to one insertion of dimension-6 operators and study their energy behaviour. Then, in Section~\ref{sec:Pheno_ggZh}, we extend an analysis of $q\bar q\to ZH$ at (HL-)LHC going beyond Flavour Universality and properly accounting for the contribution of the gluon-initiated channel. This allows us to explore the prospects of $ZH$ in probing particular dimension-6 operators. 
 In this way, we perform a truly comprehensive study of the (HL-)LHC reach of $pp\to ZH$ in the SMEFT framework.

\section{Energy-growing helicity amplitudes}
\label{sec:HeliAmp}
\subsection{Conventions and Methodology}
\label{sec:HeliAmp_Conventions}

\renewcommand{\arraystretch}{1.8}
\begin{table}[H]
    \centering
    \begin{tabular}{ | m{2em}  m{2em} m{4.6cm} | m{2em}  m{2em} m{4.6cm} | } 
    \hline
    \multicolumn{6}{|c|}{Purely bosonic} \\
    \hline
    $\mathcal{O}_i$ & $c_i$ & Definition & $\mathcal{O}_i$ & $c_i$ & Definition  \\
    \hline

    $\Opg$&$\cpg$ & $\left(\pdp-\tfrac{v^2}{2}\right)G^{\mu\nu}_{\sss A}\,G_{\mu\nu}^{\sss A}$&
    $\op_{\varphi B}$& $\cpB$ &
     $\left(\pdp-\tfrac{v^2}{2}\right)B^{\mu\nu}\,B_{\mu\nu}$\\

    $\op_{\varphi W}$&
     $\cpW$ &
     $\left(\pdp-\tfrac{v^2}{2}\right)W^{\mu\nu}_{\sss I}\,W_{\mu\nu}^{\sss I}$ &
      $\op_{\varphi}$ &$c_{\varphi}$ & $\left(\pdp-\tfrac{v^2}{2}\right)^3$\\
      
     $\op_{d \varphi}$ &$c_{d \varphi}$& $\partial_\mu (\varphi^\dagger \varphi)\partial^\mu (\varphi^\dagger \varphi)$&&&\\
    \hline
    \hline
     \multicolumn{6}{|c|}{3rd generation quarks} \\
    \hline
    $\mathcal{O}_i$ & $c_i$ & Definition &$\mathcal{O}_i$ & $c_i$ & Definition  \\
    \hline
     $\op_{t\varphi}$ & $\cth$ & $\left(\pdp-\tfrac{v^2}{2}\right) \bar{Q}\,t\,\tilde{\varphi} + \text{h.c.}$&
     $\op_{\varphi t}$ & $\cht$ &
    $i\big(\varphi^\dagger\,\lra{D}_\mu\,\,\varphi\big)
    \big(\bar{t}\,\gamma^\mu\,t\big)$\\

    $\op_{tG}$&$\ctg$ & $g{\sss S}\,\big(\bar{Q}\sigma^{\mu\nu}\,T_{\sss A}\,t\big)\,\tilde{\varphi}\,G^A_{\mu\nu} + \text{h.c.}$&
    $\Opp{\varphi Q}{\sss(1)}$& $\chqTop$ &
    $i\big(\varphi^\dagger\lra{D}_\mu\,\varphi\big)
    \big(\bar{Q}\,\gamma^\mu\,Q\big)$\\

    $\op_{tW}$&$\ctW$ & $\big(\bar{Q}\sigma^{\mu\nu}\,\tau_{\sss I}\,t\big)\,
    \tilde{\varphi}\,W^I_{\mu\nu}
    + \text{h.c.}$&
    $\Opp{\varphi Q}{\sss(3)}$& $\chqTopt$ &
    $i\big(\varphi^\dagger\lra{D}_\mu\,\tau_{\sss I}\varphi\big)
    \big(\bar{Q}\,\gamma^\mu\,\tau^{\sss I}Q\big)$\\

    $\op_{tB}$&$\ctB$&
    $\big(\bar{Q}\sigma^{\mu\nu}\,t\big)
    \,\tilde{\varphi}\,B_{\mu\nu}
    + \text{h.c.}$&
    \ensuremath{\mathcal{O}_{\varphi Q}^{(-)}} & $\chqTopMinus$ & $\chqTop-\chqTopt$\\
    
    $\mathcal{O}_{tZ}$& $\ctZ$ & $-\sin\theta_W\, \ctB + \cos\theta_W\, \ctW$&&&\\

     \hline
     \hline
    \multicolumn{6}{|c|}{1st, 2nd generation quarks} \\
    \hline
    $\mathcal{O}_i$ & $c_i$ & Definition &$\mathcal{O}_i$ & $c_i$ & Definition  \\
    \hline
    $\Opp{\varphi q_i}{\sss(1)}$& $\chqii$ &
    $\sum\limits_{\sss j=1,2} i\big(\varphi^\dagger\lra{D}_\mu\,\varphi\big)
    \big(\bar{q}_j\,\gamma^\mu\,q_j\big)$&
     $\op_{\varphi u}$& $\chu$ &
    $\sum\limits_{\sss j=1,2} i\big(\varphi^\dagger\,\lra{D}_\mu\,\,\varphi\big)
    \big(\bar{u}_j\,\gamma^\mu\,u_j\big)$\\

    $\Opp{\varphi q_i}{\sss(3)}$& $\chqtii$ &
    $\sum\limits_{\sss j=1,2} i\big(\varphi^\dagger\lra{D}_\mu\,\tau_{\sss I}\varphi\big)
    \big(\bar{q}_j\,\gamma^\mu\,\tau^{\sss I}q_j\big)$&
    $\op_{\varphi d}$& $\chd$ &
    $\sum\limits_{\sss j=1,2,3} i\big(\varphi^\dagger\,\lra{D}_\mu\,\,\varphi\big)
    \big(\bar{d}_j\,\gamma^\mu\,d_j\big)$\\
    
    \ensuremath{\mathcal{O}_{\varphi q,i}^{(-)}} & $\chqminusii$ & $\chqii-\chqtii$&&&\\
    \hline     
    \end{tabular}
    \caption{ Dimension-$6$ operators $\mathcal{O}_i$ and their associated Wilson Coefficients $c_i$ entering in $gg \rightarrow HH, ZH, ZZ, WW$. 
    }
    \label{operators}
\end{table}

We use the Warsaw basis of dimension-$6$ operators along with a U$(2)_q\times$U$(3)_d\times$U$(2)_u\times(\text{U}(1)_\ell\times\text{U}(1)_e)^3$ flavour assumption and the operator definitions from the \code{SMEFTatNLO} model \cite{1008.4884,Degrande:2020evl}. The operators relevant to gluon-initiated diboson production are presented in Table~\ref{operators}. We note here that we do not discuss operators which enter only through the modification of the input parameters as these will not lead to energy growth. 
The left-handed quark doublets are denoted by $q_i$ for the first two generations and by $Q$ for the third generation, while $t$ denotes a right-handed top-quark field and $u_i, d_i$ the right-handed $u, c$ and $d, s, b$ fields respectively. All the quarks are considered to be massless except for the top. The Higgs doublet is given by $\varphi$ with vacuum expectation value $v/\sqrt{2}$. We define
\begin{equation}
    \varphi^\dagger\lra{D}_\mu\,\varphi \equiv \varphi^\dagger D_\mu \varphi - (D_\mu \varphi)^\dagger \varphi\ , \qquad \varphi^\dagger\lra{D}_\mu\,\tau_I \,\varphi \equiv \varphi^\dagger\, \tau_I D_\mu \varphi - (D_\mu \varphi)^\dagger \tau_I \varphi,
\end{equation}
where the covariant derivative is given by:
\begin{equation}
    D_\mu \varphi = \Big(\partial_\mu -i \frac{g}{2} \tau_I W_\mu^I -i \frac{g'}{2}B_\mu \Big) \varphi
\end{equation}
$W_\mu,\, B_\mu$ are the gauge bosons fields, $g$ and $ g'$ are the $SU(2)_L$ and $U(1)_Y$ couplings respectively and $\tau_{I}$ are the Pauli sigma matrices. Furthermore $G^A_{\mu\nu}$, $W^{\mu\nu}$ and $B^{\mu\nu}$ stand for the $SU(3)_C$, $SU(2)_L$ and $U(1)_Y$ field strength tensors. $T^A = \frac{1}{2} \lambda^A$ are the $SU(3)$ generators where $\lambda^A$ are the Gell-Mann matrices. The strong coupling constant is denoted by $g_s$, $\theta_W$ represents the weak mixing angle and $\sigma^{\mu\nu} = \frac{i}{2} [\gamma^\mu,\gamma^\nu]$.

Not all operators considered here are generated with WCs of the same order in common BSM scenarios. The current operators, $\mcO_{\varphi t}$, $\mcO_{\varphi Q}^{(1),(3)}$, $\mcO_{\varphi q,ii}^{(1),(3)}$, $\mcO_{\varphi u}$, and $\mcO_{\varphi d}$; the dimension-6 Yukawa operator, $\mcO_{t\varphi}$; and the Higgs operators $\mcO_{d\varphi}$ and $\mcO_{\varphi}$ are generated at tree level by several possible UV completions~\cite{Giudice:2007fh,Contino:2013kra,deBlas:2017xtg}. Assuming the SILH power counting~\cite{Giudice:2007fh,Contino:2013kra}, one expects $c\sim g^2$ for the current operators, $c\sim y_{t}^{SM}$ for the Yukawa operator and $c_\varphi\sim\lambda$, where $\lambda$ is the SM quartic Higgs potential coefficient.
The remaining bosonic and dipole operators can not be generated at tree level via renormalisable interactions\footnote{There is a rare exception: a heavy vector boson that is a weak doublet with hypercharge $1/2$, $\mathcal{L}_1$, can generate $\mcO_{\varphi W}$ and $\mcO_{\varphi B}$ via mixing with the Higgs boson~\cite{deBlas:2017xtg}.} and hence their WCs are expected to be suppressed by $g_{*}^2/16\pi^2$, where $g_{*}$ is the typical coupling strength between SM and UV particles. Some exotic models could partially avoid this suppression for $\mcO_{\varphi W}$ and $\mcO_{\varphi B}$~\cite{Liu:2016idz}. A further suppression of the Yukawa and dipole operators in scenarios endowed with Minimal Flavour Violation (MFV) is avoided due to $y_{t}^{\text{SM}}\simeq 1$~\cite{DAmbrosio:2002vsn}.
However, we consider all operators regardless of their size and our results are independent of these power counting considerations.

In the rest of this section, we present the leading high-energy behaviour of the $g g \rightarrow HH, ZH, ZZ, WW$ amplitudes in the SM and with the insertion of one dimension-6 SMEFT operator. We analyse separately each possible helicity configuration of the initial and final state particles. The analytical expression for the full amplitude of each process was obtained in Wolfram Mathematica $12.3$ using the FeynCalc~\cite{Mertig:1990an,1601.01167,2001.04407}, FeynHelpers~\cite{1611.06793}, Package-X~\cite{1503.01469} and FeynArts~\cite{hep-ph/0012260} packages.

We only consider leading order diagrams, which are one-loop QCD diagrams in all cases except for the insertion of $\Opg$, which introduces a contact term between the Higgs boson and the gluons and hence generates tree-level diagrams for $gg \rightarrow HH, ZZ, WW$. All the loops considered here are finite except when $\op_{tG}$ enters in $HH, ZZ$ and $WW$ production. The UV divergence can be reabsorbed in the renormalisation of $\Opg$, as discussed  in~\cite{Degrande:2012gr,Jenkins:2013zja,Jenkins:2013wua,Alonso:2013hga,Maltoni:2016yxb,Grazzini:2016paz}. In the $\overline{MS}$ scheme, $\cpg$ is renormalised as:
\begin{equation}
    \cpg^0 = \cpg(\mu_{EFT}) + \delta\cpg
\end{equation}
where $\mu_{EFT}$ denotes the renormalisation scale of the EFT and the counterterm is given by:
\begin{equation}
    \delta\cpg = \ctg \frac{g_s^2\, m_t}{4 \sqrt{2} \pi^2 \, v}\Gamma(1+\epsilon)\frac{1}{\epsilon}\Bigg(\frac{4 \pi \mu^2}{\mu_{EFT}^2}\Bigg)^\epsilon
\end{equation}
In this work we consider $\mu = \mu_{EFT}$.
\\

The helicity amplitudes have been obtained using a modified version of the Mathematica package used in ~\cite{Maltoni:2019aot}
where the polarisation vectors are defined in the following way:
\begin{align}
    \varepsilon^{\pm}_{\mu} (p) &= \frac{1}{\sqrt{2}}(0,\text{cos}\theta,\pm i,-\text{sin}\theta)\\
    \varepsilon^0_{\mu} (p) &= \frac{1}{M}(|\textbf{p}|,E\ \text{sin}\theta,0,E\ \text{cos}\theta)
\end{align}
In the above, $\varepsilon^{\pm}_{\mu}$ denotes the transverse polarisation vectors and $\varepsilon^{0}_{\mu}$ stands for the longitudinal one, where $p$ is the 4-momentum of a particle, \textbf{p}, $M$ and $E$ are its 3-momentum, mass and energy respectively, and $\theta$ is the angle of the momentum with respect to the beam axis. Finally, as we are interested in the high-energy limit we performed a series expansion in energy to find the leading contribution to the helicity amplitudes in the high $\sqrt{s}$ limit. In this expansion we also consider the mass of the final state bosons to be significantly smaller than the centre of mass energy. \\

We validated our results by comparing the numerical values obtained from our analytical predictions with the numerical predictions given by the \code{SMEFTatNLO} model in \code{Madgraph5\_aMC@NLO} v$3.4$. The analytical expressions for the full helicity amplitudes match with Madgraph. We note that for each amplitude we provide only the leading term(s) in the energy expansion. The level of agreement between this leading term and the full amplitude depends on the process and operator, with operators that rescale the SM and s-channel contributions tending to the high-energy limit fast. For these amplitudes the leading term provides an estimate to the total amplitude to about 10\% at around 5 TeV. Convergence is slower for the top dipole operators and when the leading term is a constant. Often subleading terms can be numerically significant even at energies around 10 TeV depending on the kinematic configuration. In particular this is the case for $\Otg$ in $ZH, ZZ$ and $WW$ production and for $\Opg$ in $gg \rightarrow ZH$. For conciseness, we report only the leading term(s) as this gives a reasonably reliable expectation for the tails of the distributions at the HL-LHC and FCC-hh. 

When the amplitude behaves at high energies like $s^{\alpha} \log^2(\frac{s}{m_t^2})$, with $\alpha$ a half-integer, we found that the leading term in the high-energy regime can be computed by applying the method of regions~\cite{Beneke:1997zp, Becher:2014oda}. The squared logarithm arises from phase-space regions in which ones of the loop lines becomes soft and collinear, extracted by expanding the loop traces in the corresponding region. This configuration leads to a scaleless integral  
which can be evaluated by neglecting the top masses in the denominator and computing the appropriate residues. The integral is regulated in the IR and UV regions by $m_t$ and $\sqrt{s}$ respectively and yields $\log^2(\frac{s}{m_t^2})-2\pi\,i \log(\frac{s}{m_t^2})$.

This technique applies to both triangle and box diagrams, with either an SM-like Lorentz structure or one insertion of a dipole operator. For loops with planar topology, this procedure is analogous to the s-channel unitarity cut employed in Ref.~\cite{Glover:1988fe,Glover:1988rg}, but it can be extended to non-planar topologies allowing to reproduce the results in a wider variety of helicity configurations. Non-planar diagrams are key in reproducing the right behaviour when one dipole operator is inserted. We note here that the simple s-channel unitarity cut argument is not enough to link the behaviour of the loop-induced $gg\to VV$ amplitude to the one of the underlying $t\bar{t}\to VV$, when there are contributions from the non-planar topologies as is the case in several of the amplitudes that we compute. We have cross-checked several of our results using this method. For clarity we show the detailed computation for an example in Appendix~\ref{app:LogSquared}.

\subsection{$gg \rightarrow HH$}
\label{sec:HeliAmp_ggHH}
\begin{figure}[h!]
    \centering
\begin{centering}
\scalebox{0.75}{
\begin{tikzpicture}
\begin{feynman}
\vertex (i1) {$g$};
\vertex[below=1cm of i1] (i2) {$g$};
\vertex[right=0.866cm of i1, empty dot] (l1) {};
\vertex[below=1cm of l1, empty dot] (l3) {};
\vertex[below right=0.5cm and 0.866cm of l1, empty dot] (l2) {};
\vertex[right=0.866cm of l2, empty dot] (v2) {};
\vertex[right=2.6cm of l1] (f1) {$H$};
\vertex[right=2.6cm of l3] (f2) {$H$};
\diagram* {{[edges={fermion, arrow size=1pt}]
(l1) -- (l2) -- (l3) -- (l1),},
(i1) -- [gluon] (l1),
(i2) -- [gluon] (l3),
(l2) -- [scalar] (v2),
(v2) -- [scalar] (f1),
(v2) -- [scalar] (f2) };
\end{feynman}
\end{tikzpicture}}
\hfill
\hfill
\scalebox{0.75}{
\begin{tikzpicture}
\begin{feynman}
\vertex (i1) {$g$};
\vertex[below=1cm of i1] (i2) {$g$};
\vertex[right=1cm of i1, empty dot] (l1) {};
\vertex[right=1cm of l1, empty dot] (l2) {};
\vertex[below=1cm of l2, empty dot] (l3) {};
\vertex[below=1cm of l1, empty dot] (l4) {};
\vertex[right=1cm of l3] (f1) {$H$};
\vertex[right=1cm of l2] (f2) {$H$};
\diagram* {{[edges={fermion, arrow size=1pt}]
(l1) -- (l2) -- (l3) -- (l4) -- (l1),},
(i1) -- [gluon] (l1),
(i2) -- [gluon] (l4),
(l3) -- [scalar] (f1),
(l2) -- [scalar] (f2) };
\end{feynman}
\end{tikzpicture}}
\hfill
\scalebox{0.75}{
\begin{tikzpicture}
\begin{feynman}
\vertex (i1) {$g$};
\vertex[below=1cm of i1] (i2) {$g$};
\vertex[below right=0.5cm and 0.866cm of i1, dot] (v1) {};
\vertex[right = 0.866cm of v1] (v2);
\vertex[right=2.6cm of i1] (f1) {$H$};
\vertex[right=2.6cm of i2] (f2) {$H$};
\diagram* {
(i1) -- [gluon] (v1),
(i2) -- [gluon] (v1),
(v1) -- [scalar] (v2),
(v2) -- [scalar] (f1),
(v2) -- [scalar] (f2) };
\end{feynman}
\end{tikzpicture}}
\hfill
\hfill
\scalebox{0.75}{
\begin{tikzpicture}
\begin{feynman}
\vertex (i1) {$g$};
\vertex[below=1cm of i1] (i2) {$g$};
\vertex[below right=0.5cm and 0.866cm of i1, dot] (v1) {};
\vertex[right=1.732cm of i1] (f1) {$H$};
\vertex[right=1.732cm of i2] (f2) {$H$};
\diagram* {
(i1) -- [gluon] (v1),
(i2) -- [gluon] (v1),
(v1) -- [scalar] (f1),
(v1) -- [scalar] (f2) };
\end{feynman}
\end{tikzpicture}}
\hfill
\end{centering}

\begin{centering}
\hfill
\scalebox{0.75}{
\begin{tikzpicture}
\begin{feynman}
\vertex (i1) {$g$};
\vertex[below=1cm of i1] (i2) {$g$};
\vertex[right=0.866cm of i1] (l1);
\vertex[below=1cm of l1] (l3);
\vertex[below right=0.5cm and 0.866cm of l1, dot] (l2) {};
\vertex[right=1.732cm of l1] (f1) {$H$};
\vertex[right=1.732cm of l3] (f2) {$H$};
\diagram* {{[edges={fermion, arrow size=1pt}]
(l1) -- (l2) -- (l3) -- (l1),},
(i1) -- [gluon] (l1),
(i2) -- [gluon] (l3),
(l2) -- [scalar] (f1),
(l2) -- [scalar] (f2) };
\end{feynman}
\end{tikzpicture}}
\hfill
\scalebox{0.75}{
\begin{tikzpicture}
\begin{feynman}
\vertex (i1) {$g$};
\vertex[below=1cm of i1] (i2) {$g$};
\vertex[right=1cm of i1] (l1);
\vertex[right=1cm of l1] (l2);
\vertex[below right = 1cm and 0.5cm of l1, dot] (l3) {};
\vertex[right=1cm of l2] (f2) {$H$};
\vertex[below=1cm of f2] (f1) {$H$};
\diagram* {{[edges={fermion, arrow size=1pt}]
(l1) -- (l2) -- (l3) -- (l1),},
(i1) -- [gluon] (l1),
(i2) -- [gluon] (l3),
(l3) -- [scalar] (f1),
(l2) -- [scalar] (f2) };
\end{feynman}
\end{tikzpicture}}
\hfill
\scalebox{0.75}{
\begin{tikzpicture}
\begin{feynman}
\vertex (i1) {$g$};
\vertex[below=1cm of i1] (i2) {$g$};
\vertex[right=1cm of i1] (l1);
\vertex[right=1cm of i1, dot] (ph1) {};
\vertex[right=1cm of i2] (l2);
\vertex[below right=0.5 cm and 0.866cm of l1] (v1);
\vertex[right=2.6cm of i1] (f1) {$H$};
\vertex[right=2.6cm of i2] (f2) {$H$};
\diagram* {{[edges={fermion, arrow size=1pt}]
(l1) -- [out=0, in=0] (l2) -- [out=180, in=180] (l1),},
(i1) -- [gluon] (l1),
(i2) -- [gluon] (l2),
(l1) -- [scalar] (v1),
(v1) -- [scalar] (f1),
(v1) -- [scalar] (f2) };
\end{feynman}
\end{tikzpicture}}
\hfill
\end{centering}

    \caption{Diagram topologies that enter in the computation of $gg\to HH$ in SMEFT at one-loop. The empty dots represent couplings that could be either SM-like or modified by dimension-6 operators. The filled dots represent vertices generated only by dimension-6 operators. Only one insertion of dimension-6 operators is allowed per diagram. }
    \label{fig:HHDiags}
\end{figure}
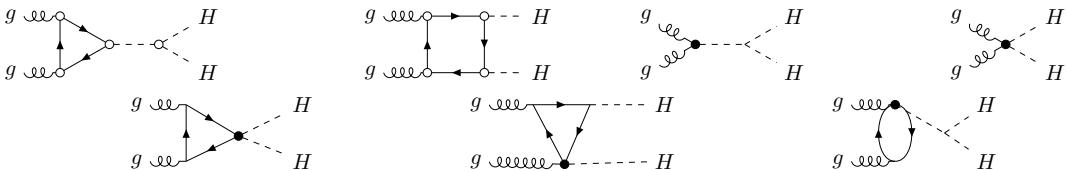

The first process we consider is double Higgs production. This process is of particular interest as it constitutes the simplest process which can probe the Higgs self coupling. From our perspective it is the simplest of the processes we study as it involves two scalars in the final state. 
For the CP even operators we consider, flipping all $\pm \leftrightarrow \mp$ in a helicity combination does not modify the helicity amplitude. This, combined with Bose symmetry
gives two independent helicity configurations for $g g \rightarrow H H$. These two configurations correspond to s-wave and d-wave contributions, with the boxes contributing to both s- and d-wave and triangles only to s-wave amplitudes. Relevant diagram topologies with possible insertions of SMEFT operators are shown in Fig.~\ref{fig:HHDiags}.

The leading behaviour in the high energy limit for the different helicity amplitudes is presented in Table~\ref{HHTable1} and the results are consistent with \cite{Azatov:2015oxa}. For the SM, we give the schematic energy behaviour of all independent helicity configurations.  For the SMEFT operators we give the full analytical expression of the leading term up to a phase. As we are interested in studying amplitudes which grow in the high energy limit, in this section only the growing SMEFT helicity amplitudes are shown. The results for the constant and decaying amplitudes of $gg \rightarrow HH, ZH, ZZ$ can be found in Appendix~\ref{app:helicity_amp}. We note here that  $\mathcal{O}_{\varphi}$, which enters in the triple Higgs vertex, does not lead to any growth with energy as it simply rescales the SM triangle amplitude.\\

We start by commenting on the Yukawa operator, $\op_{t \varphi}$, which leads to a growth in energy when the two incoming gluons have the same helicity and tends to a constant amplitude otherwise. This is because $\op_{t \varphi}$ rescales the $t\bar{t}H$ vertex and introduces a new $t\bar{t}HH$ vertex. This fact combined with the presence of both triangles and boxes which scale differently with the modification of the Yukawa coupling leads to a non-trivial impact of this operator on $HH$. This is in contrast with single Higgs production where the Yukawa operator acts as a simple rescaling of the SM prediction. It is instructive to note that for the SM, the box diagrams tend to a constant at high energy while the triangle ones decrease $\propto$ $\frac{1}{s} \logsmtsq $ for same helicities and do not contribute for opposite helicities \cite{Glover:1987nx,Plehn:1996wb}.  The $t\bar{t}HH$ interaction leads to a new triangle diagram without a Higgs propagator shown in the bottom left of Fig.~\ref{fig:HHDiags}. The absence of the s-channel propagator and of the corresponding  $1/s$ suppression leads to a logarithmic growth. \\

\begin{table}[h]
    \centering
    \resizebox{\textwidth}{!}{
    \begin{tabular}{|c|c|c|c|c|c|c|}
    \hline
         $\small{\lambda_{g_1}, \lambda_{g_2}, \lambda_{H_1}, \lambda_{H_2}}$& SM & $\Oth$ & $\Otg$& $\mathcal{O}_{d \varphi}$&$\Opg$ \\
         \hline
         $+, +, 0, 0$& $s^0$&$\frac{3\,m_t\, v \, g_s^2}{32 \pi^2}\Big[\logsmt-i\pi\Big]^2$&
         
         $s\,\frac{ m_t\, g_s^2}{4 \pi^2 v} \Big[\!\log\! \Big(\frac{s}{\mu_{EFT}^2}\frac{\sqrt{1-\ct^2}}{2}\Big) -2 \Big]$& 
         $\frac{m_t^2 \, g_s^2}{8 \sqrt{2} \, \pi^2} \Big[\logsmt-i\pi\Big]^2$&$s\,\sqrt{2}$ \\
         
         $+, -, 0, 0$&$s^0$&$-$&$s \frac{m_t\, g_s^2}{8\pi^2\,v}$&$-$&$\diagup$\\
         \hline
    \end{tabular}}
    \caption{High energy behaviour of the $gg \rightarrow HH$ helicity amplitudes in the SM and modified by SMEFT operators. The $``-"$ and $``\diagup"$ denote when a helicity amplitude is not growing or is equal to $0$ respectively. $\lambda_{g_1}, \lambda_{g_2}, \lambda_{H_1}, \lambda_{H_2}$ represent the polarisation of the two incoming gluons and the two outgoing Higgs bosons and $\ct$ stands for the cosine of the collision angle in the centre of mass frame. We also keep implicit the overall colour factor $\delta^{ab}$, where $a, b$ are the colours of the incoming gluons, as well as the overall dependence on the WCs and $\Lambda^2$. }
    \label{HHTable1}
\end{table}

The purely bosonic operator $\op_{d \varphi}$ shifts the kinetic term for the Higgs field and the canonical form is restored through the following Higgs field redefinition \cite{Degrande:2020evl,2012.11343}:
\begin{equation}
    H \rightarrow H \,\bigg ( 1 + c_{d \varphi} \frac{v^2}{\Lambda^2}\bigg )
    \label{eq-fieldred}
\end{equation}
As a result of the field redefinition, $\op_{d \varphi}$ shifts all Higgs couplings to fermion and gauge bosons in the same way. The $HHH$ interaction  receives both a rescaling and an additional momentum dependent  correction. This additional correction has two powers of momentum which, in the $(++0\,0)$ helicity configuration, cancel out the $1/s$ dependence of the triangle diagrams, leading to a logarithmic growth of the amplitude. It should be noted that an alternative Higgs field redefinition can be performed \cite{0910.4182} in which the Higgs self coupling does not have any momentum dependence. However in this different redefinition, $\op_{d \varphi}$ introduces a $t\bar{t}HH$ vertex as shown in the bottom-left of Fig.~\ref{fig:HHDiags}. Since this diagram does not have the $1/s$ dependence of the Higgs propagator, it can grow logarithmically as discussed above, reproducing the energy growth that we observe for  $\op_{d \varphi}$ with the field redefinition from Eq.~\eqref{eq-fieldred}.

For both $\op_{t \varphi}$ and $\op_{d \varphi}$, it is worth noting that the energy growth observed in the $gg\to HH$ process is related to the energy growth in the underlying $t\bar{t}\to HH$ process. The sub-process amplitudes for the opposite top helicities grow linearly with energy due to the presence of the contact $t\bar{t}HH$ interaction, see for example Ref. \cite{Maltoni:2019aot} for the crossed $tH\to tH$ process. In this case the leading contributions come from triangle diagrams, by definition planar, and the result is consistent with the s-channel unitarity expectation, confirmed also by our calculation of the corresponding loops in the soft and collinear limits.

The chromomagnetic operator $\op_{tG}$, which alters the $t\bar{t}g$ vertex and introduces a $t\bar{t}gH$ vertex, induces a growth with energy for both helicity configurations. The quadratic growth in the center of mass energy is due to the box diagrams with a modified $t\bar{t}g$ vertex and the triangle diagrams with a $t\bar{t}gH$ vertex. The modified vertex, in both cases adds a power of momentum and a different Lorentz structure than in the SM. It should be noted that we have kept $\mu_{EFT}$ explicit in our expressions. For the reasonable choice of $\mu_{EFT}=\sqrt{s}$ the logarithmic term vanishes and both helicity amplitudes grow quadratically with the energy. 

A direct coupling between the Higgs boson and the gluons is introduced by $\Opg$ and $g g \rightarrow H H$ becomes a tree-level process with either a $ggH$ or a $ggHH$ vertex. In the $(+ + 0\,0)$ helicity configuration the $3$-point vertices diagram goes to a constant in the high energy limit as the $\frac{1}{s}$ behaviour of the Higgs propagator is cancelled by the two powers of momentum in the $ggH$ vertex. Thus the $4$-point vertex diagram which does not have a propagator grows linearly with $s$. The behaviour of the $\Opg$ amplitudes is identical to that expected from the infinite top mass limit of the SM amplitudes. The different energy behaviour compared to the SM and the large energy growth observed demonstrate once more why such an approximation is not appropriate for the double Higgs process, in contrast with single Higgs production. 

To conclude our discussion of double Higgs production, we note that the quadratic growth found for $\op_{tG}$ and $\Opg$ can be particularly useful to probe their otherwise loop-suppressed effects. Additionally, as the SM amplitudes are constant for both helicity configurations, the interference between the SM and the growing SMEFT amplitudes will also grow in the high-energy limit.

\subsection{$gg \rightarrow ZH$}
\label{sec:HeliAmp_ggZH}
 
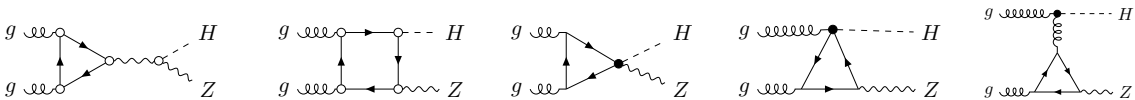
\begin{figure}[h!]
    \centering
\scalebox{0.75}{
\begin{tikzpicture}
\begin{feynman}
\vertex (i1) {$g$};
\vertex[below=1cm of i1] (i2) {$g$};
\vertex[right=0.866cm of i1, empty dot] (l1) {};
\vertex[below=1cm of l1, empty dot] (l3) {};
\vertex[below right=0.5cm and 0.866cm of l1, empty dot] (l2) {};
\vertex[right=0.866cm of l2, empty dot] (v2) {};
\vertex[right=2.6cm of l1] (f1) {$H$};
\vertex[right=2.6cm of l3] (f2) {$Z$};
\diagram* {{[edges={fermion, arrow size=1pt}]
(l1) -- (l2) -- (l3) -- (l1),},
(i1) -- [gluon] (l1),
(i2) -- [gluon] (l3),
(l2) -- [photon] (v2),
(v2) -- [scalar] (f1),
(v2) -- [photon] (f2) };
\end{feynman}
\end{tikzpicture}}
\hfill
\hfill
\scalebox{0.75}{
\begin{tikzpicture}
\begin{feynman}
\vertex (i1) {$g$};
\vertex[below=1cm of i1] (i2) {$g$};
\vertex[right=1cm of i1, empty dot] (l1) {};
\vertex[right=1cm of l1, empty dot] (l2) {};
\vertex[below=1cm of l2, empty dot] (l3) {};
\vertex[below=1cm of l1, empty dot] (l4) {};
\vertex[right=1cm of l3] (f1) {$Z$};
\vertex[right=1cm of l2] (f2) {$H$};
\diagram* {{[edges={fermion, arrow size=1pt}]
(l1) -- (l2) -- (l3) -- (l4) -- (l1),},
(i1) -- [gluon] (l1),
(i2) -- [gluon] (l4),
(l3) -- [photon] (f1),
(l2) -- [scalar] (f2) };
\end{feynman}
\end{tikzpicture}}
\hfill
\scalebox{0.75}{
\begin{tikzpicture}
\begin{feynman}
\vertex (i1) {$g$};
\vertex[below=1cm of i1] (i2) {$g$};
\vertex[right=0.866cm of i1] (l1);
\vertex[below=1cm of l1] (l3);
\vertex[below right=0.5cm and 0.866cm of l1, dot] (l2) {};
\vertex[right=1.732cm of l1] (f1) {$H$};
\vertex[right=1.732cm of l3] (f2) {$Z$};
\diagram* {{[edges={fermion, arrow size=1pt}]
(l1) -- (l2) -- (l3) -- (l1),},
(i1) -- [gluon] (l1),
(i2) -- [gluon] (l3),
(l2) -- [scalar] (f1),
(l2) -- [photon] (f2) };
\end{feynman}
\end{tikzpicture}}
\hfill
\scalebox{0.75}{
\begin{tikzpicture}
\begin{feynman}
\vertex (i1) {$g$};
\vertex[above=1cm of i1] (i2) {$g$};
\vertex[right=1cm of i1] (l1);
\vertex[right=1cm of l1] (l2);
\vertex[above right = 1cm and 0.5cm of l1, dot] (l3) {};
\vertex[right=1cm of l2] (f2) {$Z$};
\vertex[above=1cm of f2] (f1) {$H$};
\diagram* {{[edges={fermion, arrow size=1pt}]
(l1) -- (l2) -- (l3) -- (l1),},
(i1) -- [gluon] (l1),
(i2) -- [gluon] (l3),
(l3) -- [scalar] (f1),
(l2) -- [photon] (f2) };
\end{feynman}
\end{tikzpicture}}
\hfill
\scalebox{0.6}{
\begin{tikzpicture}
\begin{feynman}
\vertex (i1) {$g$};
\vertex[below=1.732cm of i1] (i2) {$g$};
\vertex[right=1.5cm of i1,dot] (v1) {};
\vertex[below=0.866cm of v1] (l1);
\vertex[below right=0.866cm and 0.5cm of l1] (l2);
\vertex[left=1cm of l2] (l3);
\vertex[right=3cm of i2] (f1) {$Z$};
\vertex[above=1.732cm of f1] (f2) {$H$};
\diagram* {{[edges={fermion, arrow size=1pt}]
(l1) -- (l2) -- (l3) -- (l1),},
(i1) -- [gluon] (v1) -- [scalar] (f2),
(v1) -- [gluon] (l1),
(i2) -- [gluon] (l3),
(l2) -- [photon] (f1) };
\end{feynman}
\end{tikzpicture}}
    \caption{Diagram topologies that enter in the computation of $gg\to ZH$ in SMEFT at one-loop. The empty dots represent couplings that could be either SM-like or modified by dimension-6 operators. The filled dots represent vertices generated only by dimension-6 operators. Only one insertion of dimension-6 operators is allowed per diagram.}
    \label{fig:ZHDiags}
\end{figure}

 Using the same symmetries as in the previous subsection, we find that there are $5$ independent helicity configurations for $g g \rightarrow ZH$. The high energy behaviour of the helicity amplitudes is presented for the SM and in the presence of SMEFT operators in Tables~\ref{ZHTable1} and \ref{ZHTable2}. For clarity the tables only include the helicity configurations which lead to a growth for at least one of the operators considered. The amplitudes are given in terms of the vector and axial-vector parts of the SM $t\bar{t}Z$ vertex which can be expressed as:
\begin{equation}
    eZ_\mu \bar{t} \gamma^\mu (c_V +c_A \gamma^5)\,t
\end{equation}
where $c_V = \frac{1}{4\cw\sw} \big(1-\frac{8 \sw^2}{3}\big)$ and $c_A = \frac{-1}{4\cw\sw}$.
The $5$ dimension-$6$ operators of interest in this process are $\op_{t\varphi}$,  $\op_{tG}$, $\Opg$, $\op_{\varphi t}$ and $\Opp{\varphi Q}{\sss(-)}$. Representative diagrams for this process are shown in Fig.~\ref{fig:ZHDiags}. The weak dipole operator $\mathcal{O}_{tZ}$, which enters in the $t\bar{t}Z$ vertex, does not contribute to $gg \rightarrow ZH$ due to charge conjugation invariance \cite{BessidskaiaBylund:2016jvp}. Interestingly $\mathcal{O}_{\varphi W}$, $\mathcal{O}_{\varphi B}$ and $\mathcal{O}_{\varphi WB}$, do not enter either despite modifying the $ZZH$ vertex and introducing a $\gamma ZH$ one. The $ggZ$ loop function, combined with the momentum dependent $ZZH$ interaction contracted with the gauge boson propagator and the external $Z$ polarisation vector lead to a vanishing amplitude. This is in contrast with the tree-level $q\bar{q}\to ZH$ amplitudes in the presence of the Higgs-gauge operators which not only do not vanish but can also lead to energy growing amplitudes for both massless and massive quarks \cite{Franceschini:2017xkh,Bishara:2020vix,Maltoni:2019aot}.

 Finally, we note here that even though massless quarks can potentially enter in the SM $gg \rightarrow ZH$ process, in particular in the triangle loops with a $Z$ propagator, the current-current operators modifying light quark couplings do not affect this process. In fact those operators enter both in the triangle diagram with a $Z$ propagator and in the diagram with a $q\bar{q} ZH$ vertex and those two diagrams cancel each other out as we will also discuss in the following.  \\

\begin{table}[h]
    \centering
    \resizebox{\textwidth}{!}{
    \begin{tabular}{|c|c|c|c|}
    \hline
         $\lambda_{g_1}, \lambda_{g_2}, \lambda_{H}, \lambda_{Z}$& SM & $\Otg$ & $\Opg$\\
         \hline
         $+,+, 0, +$&$\frac{1}{\sqrt{s}} \, \logsmtsq$&

         $\sqrt{s}\, \frac{m_t\,e\,g_s^2\,c_A\, \ct}{2\, \pi^2 \sqrt{1-\ct^2} }\, \logsmt$&
         $-$
         
         \\
         
         $+, +, 0, -$&$\frac{1}{s^{3/2}}\,\logsmtsq $&

         $\sqrt{s}\, \frac{m_t\,e\,g_s^2\,c_A\, \ct}{4\, \pi^2  \sqrt{1-\ct^2} }\, \logsmtsq $&
         $-$
         
         \\
         
         $+ ,+, 0, 0$&$\frac{1}{s}\, \logsmtsq $&

         $\frac{m_t (m_Z^2(1+\ct^2)-m_t^2(5-\ct^2))\,e\,g_s^2\,c_A}{2\sqrt{2}\, \pi^2\,m_Z \,(1-\ct^2) }\logsmtsq $&

         $\frac{m_t^2 \, v\,e\,g_s^2\,c_A }{\pi^2\,m_Z\,(1-\ct^2)}\,\logsmtsq $
           
         \\
         
         $+, -, 0, +$&$\frac{1}{\sqrt{s}}\, \logsmtsq $&

         $\sqrt{s} \,\frac{m_t\,e\,g_s^2\,c_A}{4\, \pi^2 } \, f_1(\ct)$&
         $-$

         \\
         
         $+, -, 0, 0$&$s^0$ &
         $s\, \frac{m_t\,e\,g_s^2\,c_A\, \ct}{2 \sqrt{2}\, \pi^2 m_Z}$
         
         & $\frac{m_t^2 \, v\,e\,g_s^2\,c_A\,\ct }{\pi^2\,m_Z\,(1-\ct^2)}\,\logsmtsq$
         
         \\
         \hline
    \end{tabular}}
    \caption{High energy behaviour of the $gg \rightarrow ZH$ helicity amplitudes in the SM and with modified top-gluon and Higgs-gluon interactions. The cosine and the sine of the weak angle are represented by $\cw$ and $\sw$ respectively. For readability we have defined $f_1(\ct) = \big[1 - 3 \ct^2 +2 \ct^3 + (1-\ct^2)\big(\log(\frac{1+\ct}{2})+ i \pi \big)+(1+\ct)\log(\frac{1+\ct}{2})(\log(\frac{1+\ct}{2})+2i \pi)\big]/\big[\sqrt{1-\ct^2}(1-\ct)\big]$ }
    \label{ZHTable1}
\end{table}

\begin{table}[h]
    \centering
    \resizebox{\textwidth}{!}{
    \begin{tabular}{|c|c|c|c|}
    \hline
         $\lambda_{g_1}, \lambda_{g_2}, \lambda_{H}, \lambda_{Z}$ &  $\Oht$ & $\OhqTopMinus$&$\Oth$ \\
         \hline          
         $+ ,+, 0, 0$&
         $ \frac{m_t^2 \,v \, e \,g_s^2}{32\pi^2\,m_Z\,\cw\,\sw}\Big[\logsmt-i\pi\Big]^2 $
         
         &$ \frac{m_t^2 \,v \, e \,g_s^2}{32\pi^2\,m_Z\,\cw\,\sw}\Big[\logsmt-i\pi\Big]^2$

         &$\frac{m_t \, v^2\,e\,g_s^2}{32\sqrt{2}\pi^2\,m_Z\,\cw\,\sw}\Big[\logsmt-i\pi\Big]^2 $\\
         \hline
    \end{tabular}}
    \caption{High energy behaviour of the $gg \rightarrow ZH$ helicity amplitudes with modified top-Z and top-Higgs interactions.}
    \label{ZHTable2}
\end{table}

The $t \bar{t} Z$ vertex can be modified by $\op_{\varphi t}$ and $\Opp{\varphi Q}{\sss(-)}$, which rescale the SM couplings of the Z boson to the right-handed top singlet and the left-handed third generation doublet respectively. Only the axial vector coupling of the top to the Z enters in this process, and therefore the modification of this coupling is what determines the impact of these operators on the amplitudes. These two operators also introduce a $t \bar{t} Z H$ vertex as well as a $b \bar{b} Z H$ one from $\Opp{\varphi Q}{\sss(-)}$.  The modified triangle diagrams with a Z propagator and with a $t \bar{t} Z H$ vertex cancel each other exactly, invalidating naive expectations from tree-level $t\bar t\to ZH$~\cite{Azatov:2016xik}. More precisely in $q\bar q\to ZH$ the longitudinal part of the $Z$ propagator vanishes for external massless quarks leading to energy-growing amplitudes, while its contraction with the one-loop form factor gives a contribution that cancels the ones of the transverse propagator and the $t\bar t ZH$ vertex. 

Given this cancellation, the behaviour  of $\op_{\varphi t}$ and $\Opp{\varphi Q} {\sss(-)}$  amplitudes can be simply understood from the SM box diagrams with a rescaled $t\bar t Z$ interaction.  Boxes grow logarithmically in the $(++0\,0)$ helicity configuration and decrease in all other cases. Their growth is not observed in the SM due to the logarithmic terms being exactly cancelled by the triangle diagrams. Both operators therefore lead to a logarithmic growth with energy when the two incoming gluons have the same polarisation and the Z boson is longitudinally polarised. 

Another consequence of the cancellation between triangle diagrams with $\op_{\varphi t}$ and $\Opp{\varphi Q}{\sss(-)}$ is that they generate the same behaviour as the Yukawa operator $\op_{t\varphi}$. The latter can only enter in box diagrams with a rescaled $t\bar t H$ interaction. Hence, $gg\to ZH $  is only sensitive to the linear combination $\chqTopMinus-\cht +\frac{\cth}{y_t}$. We will discuss this degeneracy further in our phenomenological analysis of this process in Section \ref{sec:Pheno_ggZh}.

 The chromomagnetic dipole operator $\Otg$ leads to a growth in energy for all the helicity configurations, which is due to the new $t\bar{t}gH$ vertex and the modified Lorentz structure of the $t\bar{t}g$ one. The leading growth happens for $(+-0\,0)$ which grows quadratically. In  $(++0\,0)$ the quadratic growth of the box, triangles with a $Z$ propagator, and $t\bar{t}gH$ vertex diagrams cancel each other out such that the helicity amplitude grows logarithmically. We note here that amplitudes for this operator are typically more complex functions of the scattering angle, due to the different possibilities of this operator entering the Feynman diagrams. 

We conclude our discussion of $gg\to ZH $ by briefly mentioning  $\Opg$. The process remains loop induced as shown in the right-most diagram of Fig. \ref{fig:ZHDiags} and its amplitude grows when the Z is longitudinally polarised. Finally it should be noted that only $\Otg$ and $\Opg$ lead to growing interferences with the SM: this is the case in the $(++0\,0)$ helicity configuration for both operators, and additionally in the $(++0\,+)$ and $(+-0\,+)$ configurations for $\Otg$.

\subsection{$gg \rightarrow ZZ$}
\label{sec:HeliAmp_ggZZ}

\begin{figure}[h!]
    \centering
\scalebox{0.75}{
\begin{tikzpicture}
\begin{feynman}
\vertex (i1) {$g$};
\vertex[below=1cm of i1] (i2) {$g$};
\vertex[right=0.866cm of i1, empty dot] (l1){};
\vertex[below=1cm of l1, empty dot] (l3){};
\vertex[below right=0.5cm and 0.866cm of l1, empty dot] (l2) {};
\vertex[right=0.866cm of l2, empty dot] (v2) {};
\vertex[right=2.6cm of l1] (f1) {$Z$};
\vertex[right=2.6cm of l3] (f2) {$Z$};
\diagram* {{[edges={fermion, arrow size=1pt}]
(l1) -- (l2) -- (l3) -- (l1),},
(i1) -- [gluon] (l1),
(i2) -- [gluon] (l3),
(l2) -- [scalar] (v2),
(v2) -- [photon] (f1),
(v2) -- [photon] (f2) };
\end{feynman}
\end{tikzpicture}}
\hfill
\hfill
\scalebox{0.75}{
\begin{tikzpicture}
\begin{feynman}
\vertex (i1) {$g$};
\vertex[below=1cm of i1] (i2) {$g$};
\vertex[right=1cm of i1, empty dot] (l1){};
\vertex[right=1cm of l1, empty dot] (l2) {};
\vertex[below=1cm of l2, empty dot] (l3) {};
\vertex[below=1cm of l1, empty dot] (l4){};
\vertex[right=1cm of l3] (f1) {$Z$};
\vertex[right=1cm of l2] (f2) {$Z$};
\diagram* {{[edges={fermion, arrow size=1pt}]
(l1) -- (l2) -- (l3) -- (l4) -- (l1),},
(i1) -- [gluon] (l1),
(i2) -- [gluon] (l4),
(l3) -- [photon] (f1),
(l2) -- [photon] (f2) };
\end{feynman}
\end{tikzpicture}}
\hfill
\scalebox{0.75}{
\begin{tikzpicture}
\begin{feynman}
\vertex (i1) {$g$};
\vertex[below=1cm of i1] (i2) {$g$};
\vertex[right=1cm of i1] (l1);
\vertex[right=1cm of i1, dot] (ph1) {};
\vertex[right=1cm of i2] (l2);
\vertex[below right=0.5 cm and 0.866cm of l1] (v1);
\vertex[right=2.6cm of i1] (f1) {$Z$};
\vertex[right=2.6cm of i2] (f2) {$Z$};
\diagram* {{[edges={fermion, arrow size=1pt}]
(l1) -- [out=0, in=0] (l2) -- [out=180, in=180] (l1),},
(i1) -- [gluon] (l1),
(i2) -- [gluon] (l2),
(l1) -- [scalar] (v1),
(v1) -- [photon] (f1),
(v1) -- [photon] (f2) };
\end{feynman}
\end{tikzpicture}}
\hfill
\scalebox{0.75}{
\begin{tikzpicture}
\begin{feynman}
\vertex (i1) {$g$};
\vertex[below=1cm of i1] (i2) {$g$};
\vertex[below right=0.5cm and 0.866cm of i1, dot] (v1) {};
\vertex[right = 0.866cm of v1] (v2);
\vertex[right=2.6cm of i1] (f1) {$Z$};
\vertex[right=2.6cm of i2] (f2) {$Z$};
\diagram* {
(i1) -- [gluon] (v1),
(i2) -- [gluon] (v1),
(v1) -- [scalar] (v2),
(v2) -- [photon] (f1),
(v2) -- [photon] (f2) };
\end{feynman}
\end{tikzpicture}}
    \caption{Diagram topologies that enter in the computation of $gg\to ZZ$ in SMEFT at one-loop. The empty dots represent couplings that could be either SM-like or modified by dimension-6 operators. The filled dots represent vertices generated only by dimension-6 operators. Only one insertion of dimension-6 operators is allowed per diagram.}
    \label{fig:ZZDiags}
\end{figure}
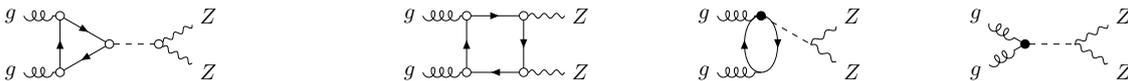

\begin{table}[h]
    \centering
    \resizebox{\textwidth}{!}{
    \begin{tabular}{|c|c|c|c|}
    \hline
         $\lambda_{g_1}, \lambda_{g_2}, \lambda_{Z_1}, \lambda_{Z_2}$ & SM & $\Otg$& $\mathcal{O}_{tZ}$  \\
         \hline
         $+,+, +, +$& $s^0$
        &
        
        $\frac{m_t\,v\,e^2\,g_s^2}{8\pi^2\,\cw^2\,\sw^2}\,
        \log \Big(\frac{\mu_{EFT}^2}{m_t^2}\Big)$
        &$-$
        \\
        
         $+, +, +, -$& $s^0$
        &
 
        $\frac{m_t\,v\,e^2\,g_s^2\,(c_A^2-c_V^2)}{4\pi^2} \,\logsmtsq $
        & 
         $\frac{m_t\,v\,e\,g_s^2\,c_V}{8\pi^2}\Big[\logsmt-i\pi\Big]^2$
        \\
          
         $+, +, +, 0$& $\frac{1}{\sqrt{s}}\, \logsmtsq$
         &
         
         $\sqrt{s}\,\frac{m_t\, v\,e^2\,g_s^2\,c_A^2\,\ct}{ \sqrt{2}\,\pi^2\,m_Z\,\sqrt{1-\ct^2}}\,\logsmt$
         &$-$
         \\
                  
         $+, +, - ,-$& $s^0$
         &
         
         $\frac{m_t\,v\,e^2\,g_s^2\,(3\,c_V^2+c_A^2)}{4\pi^2 }\,\logsmtsq$
         &$\frac{m_t\,v\,e\,g_s^2\,c_V}{4\pi^2} \Big[3\,\logsmtsq -2\,\logsmt\big(\log\!\big(\frac{4}{1-\ct^2} \big)+i \pi\big)\!\Big]$
         \\
         
         $+, +, -, 0$&$\frac{1}{s^{3/2}}\,\logsmtsq$
         &
         
         $\sqrt{s}\,\frac{m_t\,v\,e^2\,g_s^2\,c_A^2\,\ct}{2 \sqrt{2}\,\pi^2\,m_Z\,\sqrt{1-\ct^2}}\, \logsmtsq$
         &$-$
         \\
         
        $+, +, 0, 0$ & $s^0$
        &
        
        $s \,\frac{m_t\, v\,e^2\,g_s^2}{16\pi^2\,m_Z^2\,\cw^2\,\sw^2} \Big[\!\log \Big(\frac{s}{\mu_{EFT}^2} \frac{\sqrt{1-\ct^2}}{2}\Big)\!-2\Big]$ 
        &$-$
         \\
        $+, -, -, +$ 
         &$s^0$&$-$&$-$
         \\
         
         $+, -, -, -$&  $s^0$
         &
         
         $\frac{m_t\, v\,e^2\,g_s^2\,(c_A^2-c_V^2)}{8 \pi^2} \logsmtsq $
         &$\frac{m_t\,v\,e\,g_s^2\,c_V}{4\pi^2} \Big[\logsmtsq-\logsmt \log\!\big(\frac{4}{1-\ct^2}\big)\Big]$
         \\
         
         $+, -, - ,0$&$\frac{1}{\sqrt{s}}\, \logsmtsq$
         &
         
         $ \sqrt{s} \,\frac{m_t\, v\,e^2\,g_s^2\,c_A^2}{2\sqrt{2}\pi^2\,m_Z} \,f_2(\ct)$
         &$-$
       
         \\
         
          $+, -, 0 ,0$& $s^0$
          &
          
          $s \,\frac{m_t\, v\,e^2\,g_s^2\,c_A^2}{2\pi^2\,m_Z^2}$
          &$-$
         
        \\
         \hline

    \end{tabular}}
    \caption{High energy behaviour of the  $gg \rightarrow ZZ$ helicity amplitudes in the SM and in the presence of top dipole operators. $\lambda_{g_1}, \lambda_{g_2}, \lambda_{Z_1}, \lambda_{Z_2}$ represent the polarisations of the two incoming gluons and the two outgoing Z bosons respectively. 
   For readability we define $f_2(\ct) = \big[(1-\ct)^2+(1-\ct^2)(\log(\frac{1+\ct}{2})+i \pi) +(1+\ct)\log(\frac{1+\ct}{2})(\log(\frac{1+\ct}{2}) +2 i \pi )\big]/[\sqrt{1-\ct^2}(1-\ct)]$.}
    \label{ZZTable1}
\end{table}

\begin{table}[h]
    \centering
    \resizebox{\textwidth}{!}{
    \begin{tabular}{|c|c|c|c|}
    \hline
         $\lambda_{g_1}, \lambda_{g_2}, \lambda_{Z_1}, \lambda_{Z_2}$ &  $\Oth$ & $\Oht$ &$\OhqTopMinus$\\
         \hline

        $+, +, 0, 0$& 
        
        $\frac{m_t\, v^3 \, e^2 \, g_s^2}{128 \pi^2\,m_Z^2\,\cw^2\,\sw^2}\Big[\logsmt-i\pi\Big]^2 $&
        
        $\frac{m_t^2\, v^2\, e^2\,g_s^2}{32 \sqrt{2}\pi^2\,m_Z^2\,\cw^2\,\sw^2}\Big[\logsmt-i\pi\Big]^2  $&

        $\frac{m_t^2\,v^2\,e^2\,g_s^2}{32\sqrt{2}\,\pi^2\,m_Z^2\,\cw^2\,\sw^2}\Big[\logsmt-i\pi\Big]^2$
        \\
         \hline
    \end{tabular}}
    \caption{High energy behaviour of the  $gg \rightarrow ZZ$ helicity amplitudes modified by top operators.}
    \label{ZZTable2}
\end{table}

\begin{table}[h]
    \centering
    \begin{tabular}{|c|c|c|c|}
    \hline
        $\lambda_{g_1}, \lambda_{g_2}, \lambda_{Z_1}, \lambda_{Z_2}$  & $\mathcal{O}_{\varphi B}$&$\mathcal{O}_{\varphi W}$&  $\Opg$ \\
         \hline
         $+,+, +, +$  &
         
             $\frac{m_t^2 \, \sw^2 \, g_s^2}{8 \sqrt{2}\, \pi^2} \Big[\logsmt-i\pi\Big]^2 $&
             
             $\frac{m_t^2 \, \cw^2 \, g_s^2}{8 \sqrt{2}\, \pi^2}\Big[\logsmt-i\pi\Big]^2$ & $-$
        \\
         
         $+, +, - ,-$&
         
         $\frac{m_t^2 \, \sw^2 \, g_s^2}{8 \sqrt{2}\, \pi^2}\Big[\logsmt-i\pi\Big]^2$&
         
         $\frac{m_t^2 \, \cw^2 \, g_s^2}{8 \sqrt{2}\, \pi^2}\Big[\logsmt-i\pi\Big]^2$&$-$
         
         \\
        $+, +, 0, 0$&
       $-$&$-$& $s\,\frac{v^2\,e^2}{2\sqrt{2}\,m_Z^2\,\cw^2\,\sw^2}$
        \\
         \hline
    \end{tabular}
    \caption{High energy behaviour of the  $gg \rightarrow ZZ$ helicity amplitudes modified by the purely bosonic operators. }
    \label{ZZTable3}
\end{table}

We now turn our attention to ZZ production. There are $36$ possible helicity combinations for $gg \rightarrow ZZ$, but using the Bose symmetry of the initial state gluons and final state Zs and the fact that all operators considered are CP-even leads to $10$ independent helicity combinations. \footnote{In the SM a permutation of the external $Z$ momenta corresponds to a flip of the helicity $+ \leftrightarrow -$, which reduces the number of independent helicity configurations \cite{1503.08835}. However this argument requires that the two $t\bar{t}Z$ vertices are identical which is not the case in the presence of SMEFT operators such as  $\mathcal{O}_{tZ}$. This argument also holds for $W$ pair production.}.

The operators probed by $gg \rightarrow ZZ$ can be divided into three categories. First, $\op_{tG}$, $\op_{tZ}$, $\op_{t\varphi}$, $\op_{\varphi t}$ and $\Opp{\varphi Q} {\sss(-)}$ enter in the top quark couplings with the Higgs, Z bosons and gluons. Then $\op_{\varphi B}$, $\op_{\varphi W}$ and $\Opg$ modify the bosonic Higgs couplings. Finally, $\op_{\varphi u}$, $\op_{\varphi d}$, $\Opp{\varphi q_i}{\sss(-)}$, $\Opp{\varphi q_i}{\sss(3)}$ and $\Opp{\varphi Q}{\sss(3)}$ all modify the light quark couplings with the Z boson. Example diagrams are shown in Fig.~\ref{fig:ZZDiags}. The high energy behaviour of the SM helicity amplitudes, first discussed in \cite{Kniehl:1990yb,Kniehl:1990iva}, and of the SMEFT growing amplitudes are presented in Tables~\ref{ZZTable1}-\ref{ZZTable3}. 

The operators $\Oht$, $\OhqTopMinus$ and $\Oth$ enter $gg \rightarrow ZZ$ by rescaling the $Z t_R \bar{t}_R$, the $Z t_L \bar{t}_L$ and the $t \bar{t}H$ interactions respectively. In addition $\OhqTopMinus$ also modifies the $Z b_L \bar{b}_L$ vertex. All three operators induce a growth with energy in the $(++0\,0)$ helicity configuration. This is because in the SM, the top boxes and top triangles each either tend to a constant or decrease with energy except in the $(++0\,0)$ configuration where they each grow logarithmically \cite{Glover:1988rg,Glover:1988fe,Azatov:2016xik,Cao:2020npb}. Those growths cancel each other out such that the full SM $(++0\,0)$ amplitude tends to a constant. However $\Oht$, $\OhqTopMinus$ only enter in the box diagrams and $\Oth$ only enters in the triangle ones, thus the logarithmic growths are not cancelled by any other diagrams. In this case our results obtained by expanding the loop calculation in the soft and collinear region confirm that only planar topologies end up contributing in the high-energy region (in particular the region where the top propagator between the initial state gluons becomes soft) and the s-channel unitarity cut argument applies. 

From Table \ref{ZZTable2} we notice that there is a degeneracy in the high-energy behaviour of these three operators. We note though that this degeneracy is only present in the leading term, and subleading terms differ for these operators. In passing we also comment on the corresponding  light quark operator amplitudes  ($\Opp{\varphi Q}{\sss(3)}, \Opp{\varphi q_i}{\sss(3)},\Opp{\varphi u}{},\Opp{\varphi d}{}$) that  tend to at most a constant in the high energy limit. As such, their contributions are suppressed compared to those of the top operators in the high-energy region that we are interested in.

The top chromomagnetic operator, $\Otg$, generates helicity amplitudes with the largest growths in $(++0\,0)$ and $(+-0\,0)$, which rise quadratically with the energy. As expected, the amplitudes which involve a Higgs propagator depend on $\mu_{EFT}$ after the renormalisation of the corresponding UV divergence. Those diagrams enter in the $(++++)$, $(++--)$ and $(++0\,0)$ amplitudes, however in the $(++--)$ configuration the leading term ($\logsmtsq$) comes from the box diagrams and thus does not depend on $\mu_{EFT}$. While most helicity configurations grow with energy, the dominant growths happen when the two $Z$ bosons are longitudinal. 

The weak dipole operator $\mathcal{O}_{tZ}$ modifies the $t\bar{t}Z$ vertex by adding a power of momentum and changing its Lorentz structure compared to the SM boxes. The different gamma matrix structure of the loop leads to a logarithmic growth in the $(+++-)$, $(++--)$ and $(+---)$ helicity configurations. For this operator the amplitude is proportional to the vector coupling of the top to the Z and interestingly the leading growth arises for transverse Z bosons rather than longitudinal ones. This is consistent also with observations made in the tree level $t\bar{t}\to ZZ$ amplitudes, where no energy growth is observed in purely longitudinal boson states \cite{Maltoni:2019aot}. The analytic computation with the method of regions, introduced in Subsection~\ref{sec:HeliAmp_Conventions}, shows that the 
growth for the $(+++-)$ and $(++--)$ helicity configurations is generated exclusively by box diagrams with planar topology, while the $(+---)$ amplitude receives logarithmic contributions only from non-planar boxes. Both planar and non-planar diagrams generate logarithmically-growing $(++00)$ amplitudes, however, each set of diagrams cancels exactly the other one. Furthermore, this computation explains the relative factor of three in front of the log$^2$ between the  $(+++-)$ and $(++--)$ configurations: in the first one, the squared logarithm arises only when the line between the gluons is soft meanwhile, in the second growth is generated when any of the lines between gluons or between one gluon and one $Z$ become soft.

The gauge operators $\op_{\varphi B}$ and $\op_{\varphi W}$ modify the $ZZH$ vertex and enter in the same three helicity configurations as the SM s-channel diagrams. In the SM, the $ZZH$ vertex contracted with the $Z$ bosons polarisation vectors goes to a constant when the Z are transverse and grows $\propto s$ when they are longitudinal. However $\op_{\varphi B}$ and $\op_{\varphi W}$ change the Lorentz structure of the $ZZH$ vertex such that its contraction with the $Z$ polarisation vectors grows $\propto s$ when the Z  are transverse and tends to a constant when they are longitudinal. Adding the $\frac{1}{s}$ and $\logsmtsq$ dependence of the Higgs propagator and the top loop respectively, leads to the observed logarithmic growth of the triangle diagrams in the $(++++)$ and $(++--)$ configurations, similarly to what is observed in the underlying $t\bar{t}\to ZZ$ process. 

A direct coupling between the gluons and the Higgs boson is induced by $\Opg$ and $gg \rightarrow ZZ$ becomes a tree level process which, like the SM triangle diagrams, is only non-zero in the $(++++)$, $(++--)$ and $(++0\,0)$ helicity configurations. The $ggH$ vertex has two powers of momentum which cancel the $1/s$ coming from the Higgs propagator such that the energy behaviour of the diagram is determined by the contraction of the $ZZH$ vertex with the $Z$ polarisation vectors. This contraction tends to a constant when the Z bosons are transverse and to a quadratic growth when the Z bosons are longitudinal, leading to the observed amplitude growth \cite{Azatov:2016xik}.

For this process, almost all the SMEFT growing helicity amplitudes lead to a growing interference with the SM. The only exceptions are for $\Otg$ in the $(++-\,0)$, $(+--+)$ and $(++++)$ configurations where the growth of the SMEFT amplitude is not enough to overcome the suppression of the SM amplitude.

\subsection{$gg \rightarrow WW$}
\label{sec:HeliAmp_ggWW}

\begin{figure}[h!]
    \centering
\scalebox{0.75}{
\begin{tikzpicture}
\begin{feynman}
\vertex (i1) {$g$};
\vertex[below=1cm of i1] (i2) {$g$};
\vertex[right=0.866cm of i1, empty dot] (l1){};
\vertex[below=1cm of l1, empty dot] (l3){};
\vertex[below right=0.5cm and 0.866cm of l1, empty dot] (l2) {};
\vertex[right=0.866cm of l2, empty dot] (v2) {};
\vertex[right=2.6cm of l1] (f1) {$W$};
\vertex[right=2.6cm of l3] (f2) {$W$};
\diagram* {{[edges={fermion, arrow size=1pt}]
(l1) -- (l2) -- (l3) -- (l1),},
(i1) -- [gluon] (l1),
(i2) -- [gluon] (l3),
(l2) -- [scalar] (v2),
(v2) -- [photon] (f1),
(v2) -- [photon] (f2) };
\end{feynman}
\end{tikzpicture}}
\hfill
\hfill
\scalebox{0.75}{
\begin{tikzpicture}
\begin{feynman}
\vertex (i1) {$g$};
\vertex[below=1cm of i1] (i2) {$g$};
\vertex[right=1cm of i1, empty dot] (l1){};
\vertex[right=1cm of l1, empty dot] (l2) {};
\vertex[below=1cm of l2, empty dot] (l3) {};
\vertex[below=1cm of l1, empty dot] (l4){};
\vertex[right=1cm of l3] (f1) {$W$};
\vertex[right=1cm of l2] (f2) {$W$};
\diagram* {{[edges={fermion, arrow size=1pt}]
(l1) -- (l2) -- (l3) -- (l4) -- (l1),},
(i1) -- [gluon] (l1),
(i2) -- [gluon] (l4),
(l3) -- [photon] (f1),
(l2) -- [photon] (f2) };
\end{feynman}
\end{tikzpicture}}
\hfill
\hfill
\scalebox{0.75}{
\begin{tikzpicture}
\begin{feynman}
\vertex (i1) {$g$};
\vertex[below=1cm of i1] (i2) {$g$};
\vertex[right=1cm of i1] (l1);
\vertex[right=1cm of i1, dot] (ph1) {};
\vertex[right=1cm of i2] (l2);
\vertex[below right=0.5 cm and 0.866cm of l1] (v1);
\vertex[right=2.6cm of i1] (f1) {$W$};
\vertex[right=2.6cm of i2] (f2) {$W$};
\diagram* {{[edges={fermion, arrow size=1pt}]
(l1) -- [out=0, in=0] (l2) -- [out=180, in=180] (l1),},
(i1) -- [gluon] (l1),
(i2) -- [gluon] (l2),
(l1) -- [scalar] (v1),
(v1) -- [photon] (f1),
(v1) -- [photon] (f2) };
\end{feynman}
\end{tikzpicture}}
\hfill
\scalebox{0.75}{
\begin{tikzpicture}
\begin{feynman}
\vertex (i1) {$g$};
\vertex[below=1cm of i1] (i2) {$g$};
\vertex[below right=0.5cm and 0.866cm of i1, dot] (v1) {};
\vertex[right = 0.866cm of v1] (v2);
\vertex[right=2.6cm of i1] (f1) {$W$};
\vertex[right=2.6cm of i2] (f2) {$W$};
\diagram* {
(i1) -- [gluon] (v1),
(i2) -- [gluon] (v1),
(v1) -- [scalar] (v2),
(v2) -- [photon] (f1),
(v2) -- [photon] (f2) };
\end{feynman}
\end{tikzpicture}}
    \caption{Diagram topologies that enter in the computation of $gg\to WW$ in SMEFT at one-loop. The empty dots represent couplings that could be either SM-like or modified by dimension-6 operators. The filled dots represent vertices generated only by dimension-6 operators. Only one insertion of dimension-6 operators is allowed per diagram.}
    \label{fig:WWDiags}
\end{figure}
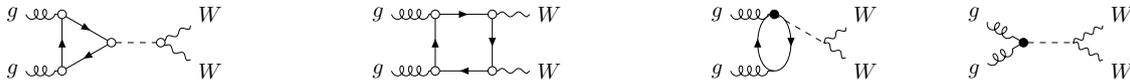

\begin{table}[h]
    \centering
    \resizebox{\textwidth}{!}{
    \begin{tabular}{|c|c|c|c|}
    \hline
         \small{$\lambda_{g_1}, \lambda_{g_2}, \lambda_{W^+}, \lambda_{W^-}$} & SM  & $\Otg$&$\mathcal{O}_{t W}$  \\
         \hline
         $+, +, +,+$&$s^0$ &
        
        $\frac{m_t\, v\,e^2  \,g_s^2}{4 \sqrt{2} \pi^2\, s_\text{w}^2}\,\log \Big(\frac{\mu_{EFT}^2}{m_t^2}\Big)$
        &$-$\\
         
         $+,+, +, -$&$s^0$ &
         $\frac{m_t\, v\, e^2\, g_s^2}{\pi^2 \sw^2}\,g_1(\ct)\,\logsmt$
         & $-$\\  
         
         $+,+, -, +$&$s^0$ &
         $\frac{\, m_t\, v\,e^2  \,g_s^2}{8 \sqrt{2} \pi^2\, s_\text{w}^2}\, \logsmt$
         &$\frac{m_t\, v\,e  \,g_s^2}{8 \sqrt{2} \pi^2\, s_\text{w}} \,\logsmtsq $
         \\
         
          $+,+,-,-$&$s^0$ &
          $\frac{\, m_t\, v\,e^2  \,g_s^2}{8 \sqrt{2} \pi^2\, s_\text{w}^2} \logsmtsq$ 
          &$\frac{3\, m_t\, v\,e  \,g_s^2}{8 \sqrt{2} \pi^2\, s_\text{w}}\Big[3\,\logsmtsq -2\,\logsmt\big(\!\log\!\big(\frac{4}{1-\ct^2} \big)+i \pi\big)\!\Big] $
          \\

         $+, +, +, 0$&$\frac{1}{\sqrt{s}} \, \logsmtsq $ &$
         \sqrt{s}\, \frac{m_t\,v\,e^2\,g_s^2}{64 \pi^2 \,m_W \, \sw^2}\,f_3(\ct)$
         &$-$\\
         
         $+, +, 0, +$&$\frac{1}{\sqrt{s}} \, \logsmtsq $&

         $\sqrt{s}\,\frac{\, m_t\, v\,e^2  \,g_s^2\,c\theta}{8 \pi^2\,m_W\, s_\text{w}^2\, \sqrt{1-c\theta^2 }} \logsmt$ 
         &$-$
         \\
         
         $+,+,-,0$&$\frac{1}{s^{3/2}} \, \logsmtsq $&$-$&$-$\\
         
         $+,+,0,-$&$\frac{1}{s^{3/2}} \, \logsmtsq$&
         $\sqrt{s}\,\frac{\, m_t\, v\,e^2  \,g_s^2\,c\theta}{16 \pi^2\,m_W\, s_\text{w}^2\,\sqrt{1-c\theta^2 }} \,\logsmtsq$
         &$-$\\
         
         $+ ,+, 0, 0$& $s^0$  &
         $s\,\frac{\, m_t\, v\,e^2  \,g_s^2}{8 \sqrt{2} \pi^2\,m_W^2\, s_\text{w}^2}\Big[\!\log \Big(\frac{s}{\mu_{EFT}^2} \frac{\sqrt{1-\ct^2}}{2}\Big)-2\Big] $    
         &$-$\\
         
         $+, -, -, +$&$s^0$&$-$&$-$\\
         
         $+, -, +, -$&$s^0$&$-$&$-$\\
         
         $+, -, - ,-$&$s^0$
         &$\frac{m_t\, v\,e^2  \,g_s^2}{8 \sqrt{2} \pi^2\, s_\text{w}^2}\, \logsmt$
         &$\frac{\, m_t\, v\,e  \,g_s^2}{8 \sqrt{2} \pi^2\, s_\text{w}} \,\logsmtsq $\\
         
         $+, -, -, 0$&$\frac{1}{\sqrt{s}} \, \logsmtsq $&
         $\sqrt{s}\,\frac{m_t\,v\,e^2\,g_s^2}{8 \sqrt{2}\pi^2\,m_W\,\sw^2}\, g_2(\ct)$
         &$-$ \\
         
         $+, -, 0, -$&$\frac{1}{\sqrt{s}} \,\logsmtsq $&
         $\sqrt{s}\,\frac{\, m_t\, v\,e^2  \,g_s^2\, \ct \,\sqrt{1+c\theta}}{16 \pi^2\,m_W\, s_\text{w}^2\,\sqrt{1-c\theta}}$
         &$-$\\
         
         $+, -, 0, 0$&$s^0$ &
         $s\,\frac{\, m_t\, v\,e^2  \,g_s^2\,(1+c\theta)}{16 \sqrt{2} \pi^2 \,m_W^2\, s_\text{w}^2}$
         &$-$\\
         
         \hline
    \end{tabular}}
    \caption{High energy behaviour of the $gg \rightarrow W^+ W^-$ helicity amplitudes in the SM and modified by top dipole operators. For readability we define $f_3(\ct) = \sqrt{1-\ct^2}\Big[\ct\big(\pi^2 +\log^2\big(\frac{1+\ct}{1-\ct}\big)\big)- 4 \log(\frac{1+\ct}{1-\ct})\Big]$}
    \label{WWTable1}
\end{table}

As in $gg \rightarrow ZZ$ there are $36$ possible helicity combinations for $gg \rightarrow W^+W^-$ and the Bose symmetry of the initial state and CP invariance lead to $15$ independent helicity configurations. For this process we only focus on the operators which lead to growing amplitudes. This is the only process considered which probes the $tbW$ interaction and thus is sensitive to the $\Opp{\varphi Q}{\sss(3)}$ and $\mathcal{O}_{tW}$ operators. Example diagrams are shown in Fig.~\ref{fig:WWDiags} and the schematic high energy behaviour of the SM amplitudes and the SMEFT growing helicity amplitudes are presented in Tables~\ref{WWTable1} and ~\ref{WWTable2}. Some of the $\Otg$ results are given up to a function of $\ct$, $g_{1/2}(\ct)$, which we have extracted numerically. 

\begin{table}[h]
    \centering
    \resizebox{\textwidth}{!}{
    \begin{tabular}{|c|c|c|c|c|}
    \hline
         $\lambda_{g_1}, \lambda_{g_2}, \lambda_{W^+}, \lambda_{W^-}$ &  $\OhqTopt$ & $\Oth$ &$\Opg$&$\mathcal{O}_{\varphi W}$\\
         \hline
         $+, +, +, +$&$-$& $-$&$-$&
         $\frac{m_t^2\, g_s^2}{8 \pi^2}\Big[\logsmt-i\pi\Big]^2$
         
         \\
         $+, +, -, -$&$-$& $-$&$-$&
         $\frac{m_t^2\, g_s^2}{8 \pi^2}\Big[\logsmt-i\pi\Big]^2$
         
         \\

         $+ ,+, 0, 0$&$\frac{ m_t^2\, v^2\,e^2  \,g_s^2}{32 \pi^2\,  m_W^2 \, s_\text{w}^2}\Big[\logsmt-i\pi\Big]^2$&

         $\frac{ m_t\, v^3\,e^2  \,g_s^2}{64 \sqrt{2} \pi^2\,  m_W^2 \, s_\text{w}^2}\Big[\logsmt-i\pi\Big]^2 $&

         $s \,\frac{v^2\,e^2}{2\, m_W^2\, \sw^2}$& $-$
         \\
         
         \hline
    \end{tabular}}
    \caption{High energy behaviour of the $gg \rightarrow W^+ W^-$ helicity amplitudes modified by top and purely bosonic operators.}
    \label{WWTable2}
\end{table}

All the operators considered in this subsection lead to similar energy behaviours in $gg \rightarrow WW$ and $gg \rightarrow ZZ$. Similarly to $ZZ$ production, all the SMEFT growing helicity amplitudes lead to a growing interference with the SM except for some $\Otg$ amplitudes. Most $WW$ helicity amplitudes modified by $\Otg$ grow with energy and the leading growths are the quadratic ones of the $(++0\,0)$ and $(+-0\,0)$ configurations, for which both the box diagrams and the diagrams with a $H$ propagator grow $\propto s$. As in $ZZ$ production, the diagrams with a Higgs propagator only enter in the $(++++)$, $(++--)$ and $(++0\,0)$ helicity configurations and are UV divergent. Hence those three amplitudes depend on $\mu_{EFT}$, but the renormalisation scale is not part of the leading energy behaviour for $(++--)$. Next, $\mathcal{O}_{tW}$ affects the $tbW$ vertex by adding a power of momentum and modifying its Lorentz structure compared to the SM boxes and it is therefore similar to $\mathcal{O}_{tZ}$.  Hence $\mathcal{O}_{tW}$ has a similar energy behaviour as $\mathcal{O}_{tZ}$ and it grows logarithmically for the $(++--)$, $(++-+)$ and $(+-++)$ helicity configurations. 

 $\Opp{\varphi Q}{\sss(3)}$ and $\Oth$ rescale the SM $tbW$ and $t\bar{t}H$ and  vertices respectively and both lead to a logarithmic growth when the two $W$ bosons are longitudinally polarised. As in the previous processes this can be understood from the SM diagrams: in the $(++0\,0)$ helicity configuration the box and triangle diagrams each grow logarithmically such that they cancel each other out and the overall $(++0\,0)$ amplitude tends to a constant \cite{Glover:1988fe,Kao:1990tt}. As $\Oth$ only enters in the triangle diagrams, there is no other diagram to cancel the logarithmic growth. The same applies for $\Opp{\varphi Q}{\sss(3)}$ which only enters in the box diagrams.

Finally $\Opg$ and $\mathcal{O}_{\varphi W}$, which induce a $ggH$ vertex and modify the $WWH$ one respectively, only enter when both gluons and both $W$ bosons have the same helicity. The behaviour of the amplitudes is determined by the contraction of the $WWH$ vertex with the $W$ polarisation vectors and the discussion from Subsection~\ref{sec:HeliAmp_ggZZ} applies here as well.

\section{A probe of top couplings at (HL-)LHC }
\label{sec:Pheno_ggZh}
Some of the growths discussed in the previous section can be observed in differential distributions of the relevant diboson processes.  Typically the energy growing amplitudes  lead to harder tails of differential distributions, dominated by the dimension-$6$ squared contributions. The impact of these is a larger deviation from the SM in the higher energy regions compared to the threshold, even when considering energies as low as $\sqrt{s}=400-1000$ GeV  as studied for example for $gg\to ZZ,WW$ in \cite{2203.02418}. In some cases top operators lead to harder distributions also at the interference level, one such example being $\Otg$ which induces a growth for several helicity configurations of all processes considered here. 

The energy growths at the amplitude level are thus particularly interesting as they could, under certain circumstances, offer a handle to improve our sensitivity and probe operators otherwise poorly constrained, as hinted in previous studies~\cite{Englert:2016hvy,BessidskaiaBylund:2016jvp,Azatov:2016xik}. Whilst similar sensitivity studies can be performed for all processes discussed above, we focus on $gg\to ZH$ to explore the potential impact of these growing amplitudes in probing top couplings in a realistic analysis. 

 This section is devoted to revisiting the possibility of using $gg\to ZH$ at HL-LHC to improve the constraints on dimension-6 SMEFT WCs, in particular those related to the top quark. Under the flavour symmetry U$(2)_q\times$U$(3)_d\times$U$(2)_u\times(\text{U}(1)_\ell\times\text{U}(1)_e)^3$, this process is sensitive to the operators $\OhqTopMinus$, $\Oht$, $\Oth$, $\Otg$, and $\Opg$, as shown before. We neglect the effect of the last 2 since they are stringently constrained by current measurements\footnote{We checked that the current constraints are one or two orders of magnitude better than the possible reach from $gg\to ZH$ at HL-LHC.}
For convenience, during the rest of this work, we will employ dimensionful WCs obtained by absorbing the $1/\Lambda^2$ factor into their definition. All the results presented in this section include the cross-section terms that are quadratic on the WCs unless stated otherwise.

To assess the sensitivity of this process to the aforementioned operators, we add the gluon-initiated signal contribution to an analysis originally focused on quark-initiated diboson production~\cite{Bishara:2022vsc}. This addition is meaningful only after relaxing the flavour assumption in said analysis from its original Flavour Universality to the one mentioned in the previous paragraph. Thus, our results are fully compatible with the \code{SMEFiT} fit~\cite{Degrande:2020evl,Ethier:2021bye}. Moreover, this allows us to study the flavour-assumption dependence of the light-quark operator bounds.

We discuss the general analysis strategy and how we simulated the collider events in Subsection~\ref{sec:Pheno_ggZH_analysis}. Then, in Subsection~\ref{sec:NLO_01jet_LO}, we ponder the relevance of higher-order QCD corrections for gluon-initiated $ZH$ production and explain how we accounted for them. The latter is crucial since it determines the interplay between the different initial state channels, which we discuss in Subsection~\ref{sec:qq_gg_flavour}. There, we also explain how the quark-initial state process is affected by the change in flavour assumptions. The projected bounds on the WCs from this analysis at HL-LHC are presented in Subsection~\ref{sec:LHC_bounds}.

\subsection{Analysis strategy}
\label{sec:Pheno_ggZH_analysis}

The final-state configuration of $ZH/WH$ production that is most useful for precision differential measurements is that in which the weak gauge boson decays to charged leptons or neutrinos and the Higgs boson decays to a $b$-quark pair.
Several studies have highlighted this process as a powerful BSM probe, in particular in the boosted Higgs regime~\cite{Banerjee:2018bio, Liu:2018pkg, Banerjee:2019pks, Banerjee:2019twi, Banerjee:2021efl, Bishara:2022vsc}. However, the limited cross-section at high energies allows for a noticeable gain of sensitivity when the boosted and resolved regimes are combined, as shown in Ref.~\cite{Bishara:2022vsc}. We adopt the analysis strategy developed in the latter reference, which exploits the high-energy tails of $p_T$ distributions and reproduces published ATLAS $ZH/WH$ analyses~\cite{ATLAS:2020fcp, ATLAS:2020jwz}. 

We take the quark-initiated signal and background simulations from Ref.~\cite{Bishara:2022vsc}, adapting the former to our flavour assumption as discussed in detail in Subsection~\ref{sec:qq_gg_flavour}. Those simulations were performed at NLO in QCD, except for the $t\bar t$ process, which is the sub-leading background process in the 0-lepton channel and was simulated at LO with one additional hard jet. We add to the signal the contribution of the $gg\to ZH$ process, which we simulate at LO in QCD and for a centre-of-mass energy of 13 TeV to ensure compatibility with the $q\bar q \to ZH$ results. The difference in cross sections between 13 and 14 TeV is negligible in comparison with the impact of the increased luminosity from LHC Run 3 to HL-LHC. This also allows us to obtain results for both LHC Run 3 and HL-LHC by a simple luminosity rescaling. For more details on the simulations, see App.~\ref{app:Sim_details} and Ref.~\cite{Bishara:2022vsc}.

The collider events are classified into two categories, boosted and resolved, according to the presence of a boosted Higgs candidate or two resolved $b$-jets respectively. This classification is done following an adapted version of the scale-invariant tagging procedure~\cite{Gouzevitch:2013qca, Bishara:2016kjn}. Furthermore, the events are split in three channels according to the number of charged leptons in the final state, ranging from 0 to 2. Selection cuts and bins are optimised independently for each of these 6 categories. 

\begin{table}[ht]
	\centering{
		\renewcommand{\arraystretch}{1.25}
  \begin{scriptsize}
		\begin{tabular}{|  c | c | c |}
		\hline
		\multicolumn{2}{|c|}{Categories} & $\ptmin\in$ \\
		\hline
		 & boosted & $\{0, 300, 350,\infty\}$ \\
		\multirow{-1}{*}[1.9ex]{0-lepton} & resolved & $\{0, 160, 200, 250,\infty\}$ \\
		\hline
		& boosted & $\{250, \infty\}$ \\
		\multirow{-1}{*}[1.9ex]{2-lepton} & resolved & $\{175, 200,\infty\}$ \\
		\hline
		\end{tabular}
  \end{scriptsize}}
		\caption{$\ptmin$ bins used in the (HL-)LHC analysis of the 0- and 2-lepton channels.}
		\label{tab:bin_boundaries}
\end{table}

Here, we only consider the 4 categories comprised by the 0- and 2-lepton channels since the 1-lepton channel concerns only $WH$ production. Both 0- and 2-lepton channels use a binning in $\ptmin=\min\lbrace \ptz,\pth \rbrace$, but the bin limits are tailored to each category as seen in Table~\ref{tab:bin_boundaries}. This reflects their different cross-sections and energy distributions. Among the different selection cuts, the most effective to reduce the background is the cut on the invariant mass of the Higgs candidate. Additionally, a veto on untagged jets helps to control the background in the 0-lepton channel, while in the 2-lepton channel a cut on the $p_T$ imbalance of the charged leptons has a similar effect. More details on the selection cuts can be found in App.~\ref{app:Analysis_details} and Ref.~\cite{Bishara:2022vsc}.

\subsection{Higher order QCD corrections}
\label{sec:NLO_01jet_LO}

The full computation at NLO in QCD of $gg\to ZH$ in the SM shows significant corrections in the tails of the $p_T$ distributions with respect to the LO result~\cite{Chen:2022rua}. This could have a significant impact on precision measurements that seize on them, such as the one considered here. This process has not been computed at QCD NLO in SMEFT yet. However, it is known that a sizeable part of the corrections introduced at NLO come from the real emission, in particular in the high-energy tails of $p_T$ distributions~\cite{Chen:2022rua,Hespel:2015zea}. This allows us to partially account for the NLO corrections in the presence of dimension-6 operators by resorting to the simulation of the process with $0$ and $1$ jet merged. Details on the simulations can be found in Appendix~\ref{app:Sim_details_0plus1jet}. 

\begin{figure}[htb!]
\centering
\includegraphics[width=0.5\textwidth]{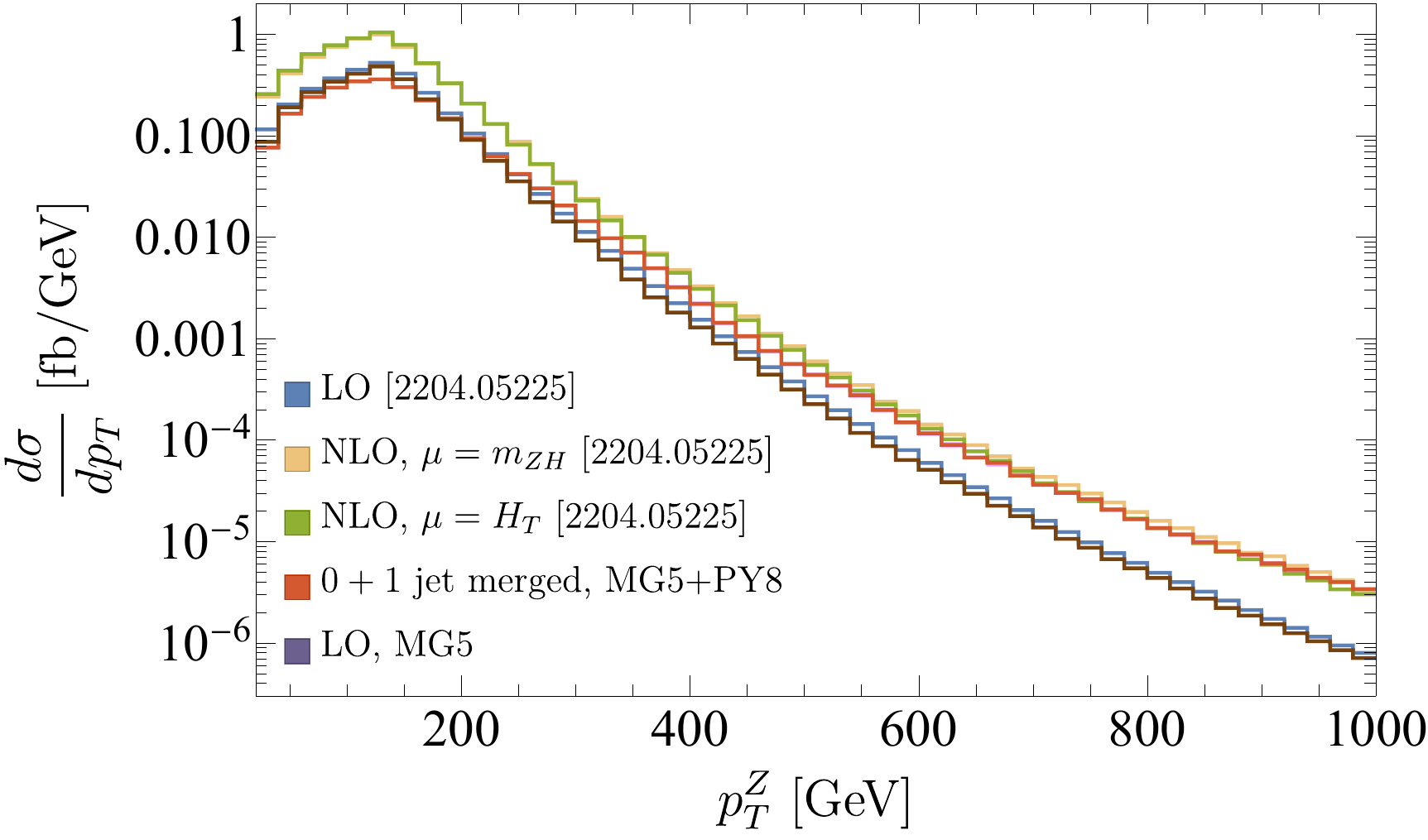}\hfill
\includegraphics[width=0.5\textwidth]{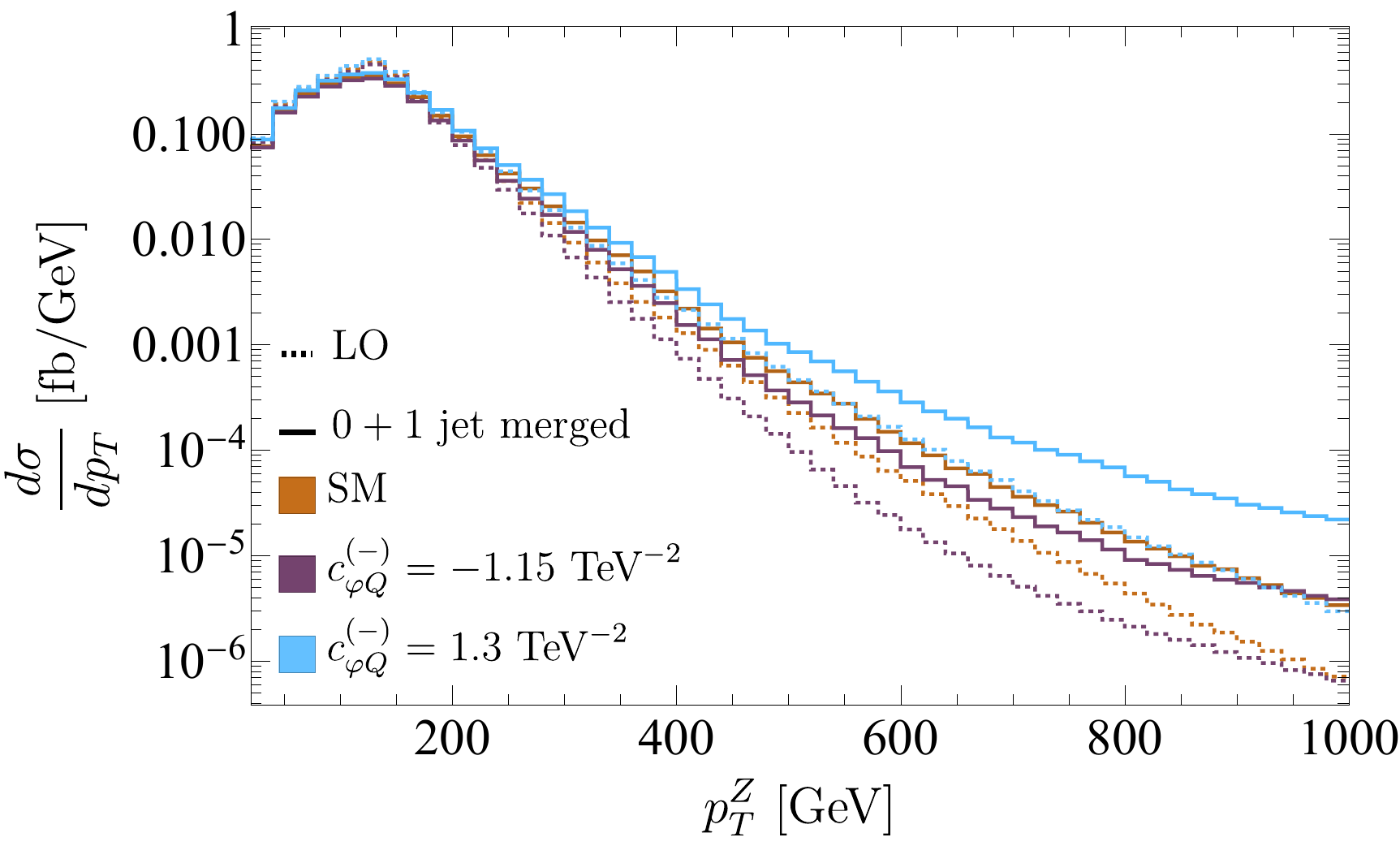}
\caption{Differential cross-section of $gg\to ZH$ with respect to $\ptz$ at different orders. \textbf{Left panel:} SM $p_T^Z$ distribution at LO and NLO from Ref.~\cite{Chen:2022rua}, the latter for 2 different renormalization scales $\mu$. We also show SM distributions at LO and $0+1$ jet merged obtained with \textsc{MadGraph5} (MG5) and \textsc{Pythia8} (PY8). \textbf{Right panel:} $p_T^Z$ distribution at LO and $0+1$ jet merged in the SM and when turning on the operator $\OhqTopMinus$ for 2 different values of its WC, while setting all other WCs to zero. Results obtained with \textsc{MadGraph5} and \textsc{SMEFTatNLO}.} 
\label{fig:plot_NLO_LO_jetMatched_ptZ_SM_cpQ3}
\end{figure}

We show in the left panel of Fig.~\ref{fig:plot_NLO_LO_jetMatched_ptZ_SM_cpQ3} the \ptz~distributions in the SM at LO, NLO and from the $0+1$ jet merged samples. We show the NLO results from Ref.~\cite{Chen:2022rua} for 2 different choices of the renormalization scale $\mu$, $\mu=m_{ZH}$ and $\mu=H_T$. In the case of LO, we include the result from Ref.~\cite{Chen:2022rua} and our parton-level results obtained with \textsc{MadGraph5}. NLO corrections give a harder $p_T$ distribution and that effect is well reproduced by the $0+1$ jet merged samples even at very high energies. However, the latter underestimates the cross-section w.r.t. LO at energies below $\sim200\GeV$.

The result of turning on the dimension-6 operator $\OhqTopMinus$ 
can be seen in the right panel of Fig.~\ref{fig:plot_NLO_LO_jetMatched_ptZ_SM_cpQ3}. The chosen values of $\chqTopMinus$ are similar to the HL-LHC bounds to be presented in Subsection~\ref{sec:LHC_bounds}.
The positive interference between $\OhqTopMinus$ and the SM generates a concavity in the LO $\ptz$ distribution for negative $\chqTopMinus$. As the energy increases, the contribution of the squared EFT amplitude becomes more relevant and the total $\OhqTopMinus$ cross-section exceeds the SM curve for energies at the edge of our plot. All the merged samples produce harder tails than their LO counterparts. Notice that the merged cross section with negative $\chqTopMinus$ also shows the interference effects, since it is below the merged SM curve for $\ptz\in[200-900]\GeV$.

\begin{figure}[htb!]
    \centering
    \includegraphics[width=0.5\textwidth]{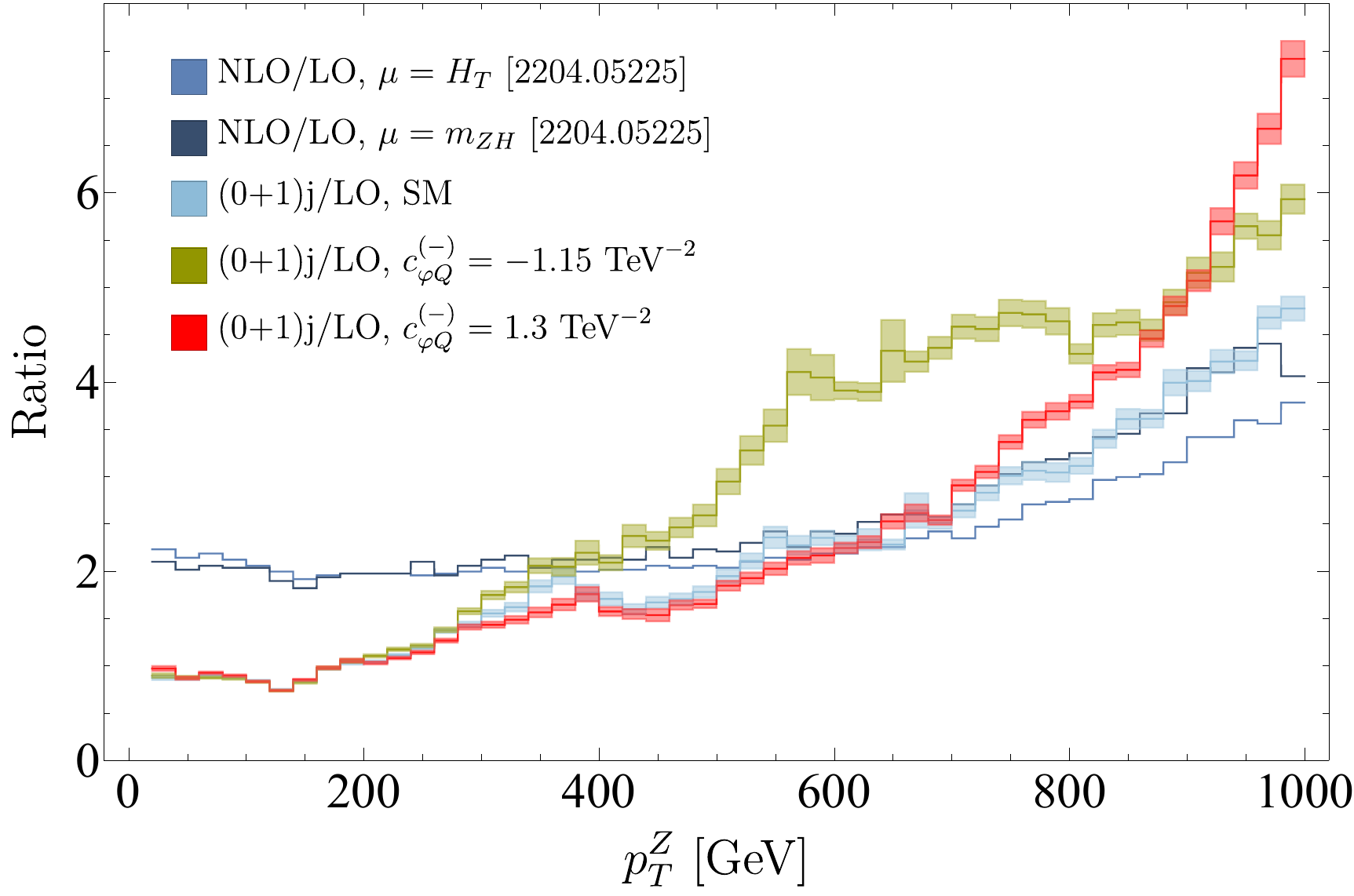}\hfill
    \includegraphics[width=0.5\textwidth]{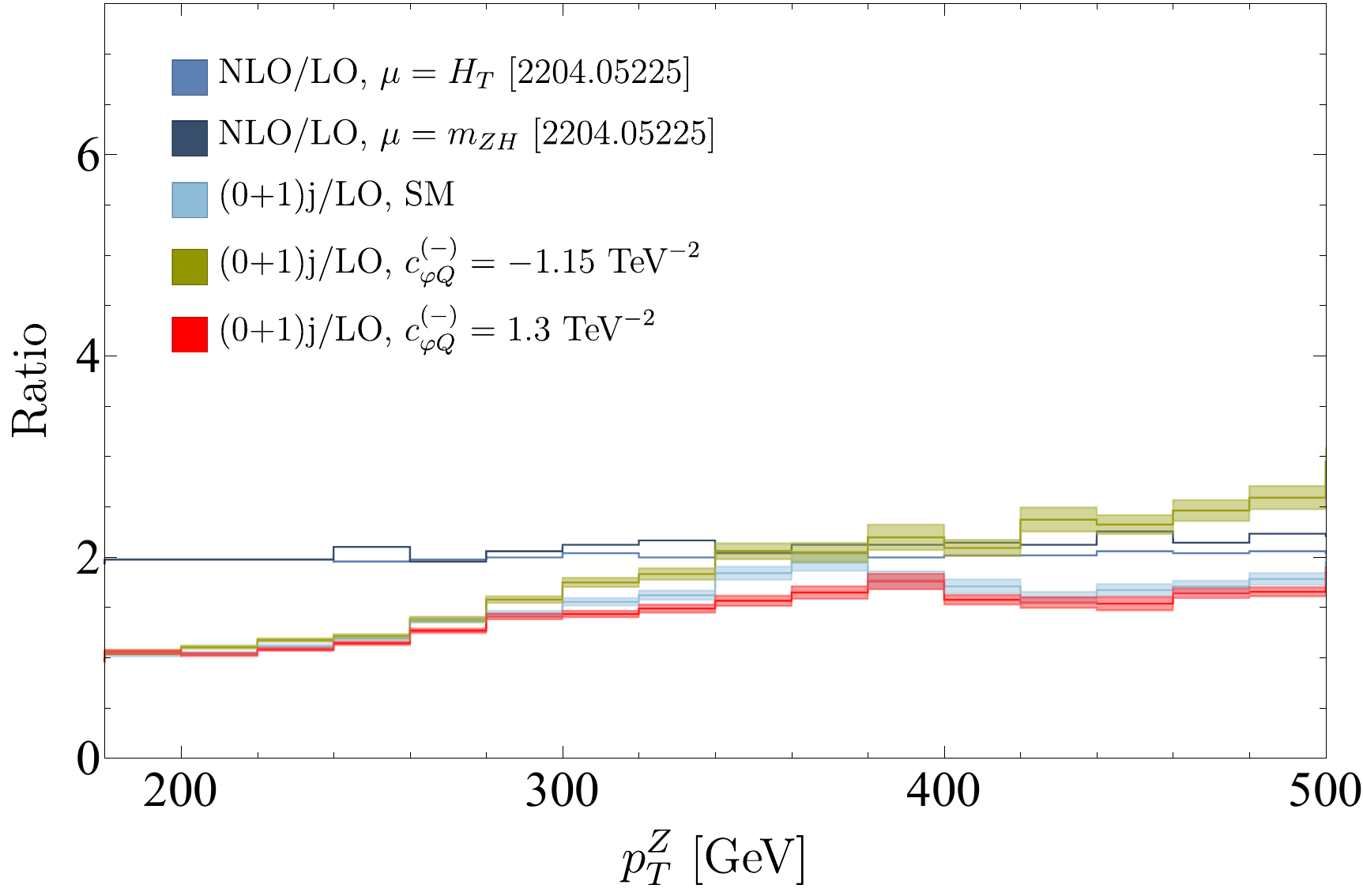}\
    \caption{Ratio of differential cross sections for $gg\to ZH$ for different computation orders. 
The blue curves show the Standard Model results, with the different shades representing different computation orders. The NLO/LO ratios were extracted from Ref.~\cite{Chen:2022rua}. The red and green curves show the result in SMEFT for two representative values of $\chqTopMinus$, while setting all other WCs to zero. These representative values are near the $95\%$ C.L. bound on this operator. The shaded regions indicate the statistical uncertainty. The right panel shows a zoomed-in version of the figure on the left.}
    \label{fig:plot_NLO_LO_jetMatched_2}
\end{figure}

The interference effects are however reduced in the merged samples. This is clearly visible in Fig.~\ref{fig:plot_NLO_LO_jetMatched_2}, where we show the ratio between different simulation orders for the SM and the two $\chqTopMinus$ values used before. The aforementioned interference effect causes the sizeable difference between the SM and negative $\chqTopMinus$ curves in the region $\ptz\in[300-800]\GeV$. At $\ptz\gtrsim 700\GeV$, both values of $\chqTopMinus$ show a higher ratio than the SM curve. This suggests a possible dependence of the NLO/LO k-factor on the WC $\chqTopMinus$, in particular at very high energies.

In Fig.~\ref{fig:plot_NLO_LO_jetMatched_2}, we also include the SM ratio and the NLO/LO k-factors from Ref.~\cite{Chen:2022rua}. From this figure, it is clear that the merged $0+1$ jet samples in the SM underestimate the NLO corrections for $\ptz<500\GeV$ but constitute a very good approximation for higher $\ptz$. The right panel of Fig.~\ref{fig:plot_NLO_LO_jetMatched_2} shows a zoom-in on the mid-energies region that is most relevant for (HL-)LHC analyses. The NLO/LO k-factor is roughly constant in that region, while the rate $0+1$ jet merged over LO is close to it only for $p_T\gtrsim 350$~GeV.

\begin{figure}[htb!]
    \centering
    \includegraphics[width=0.5\textwidth]{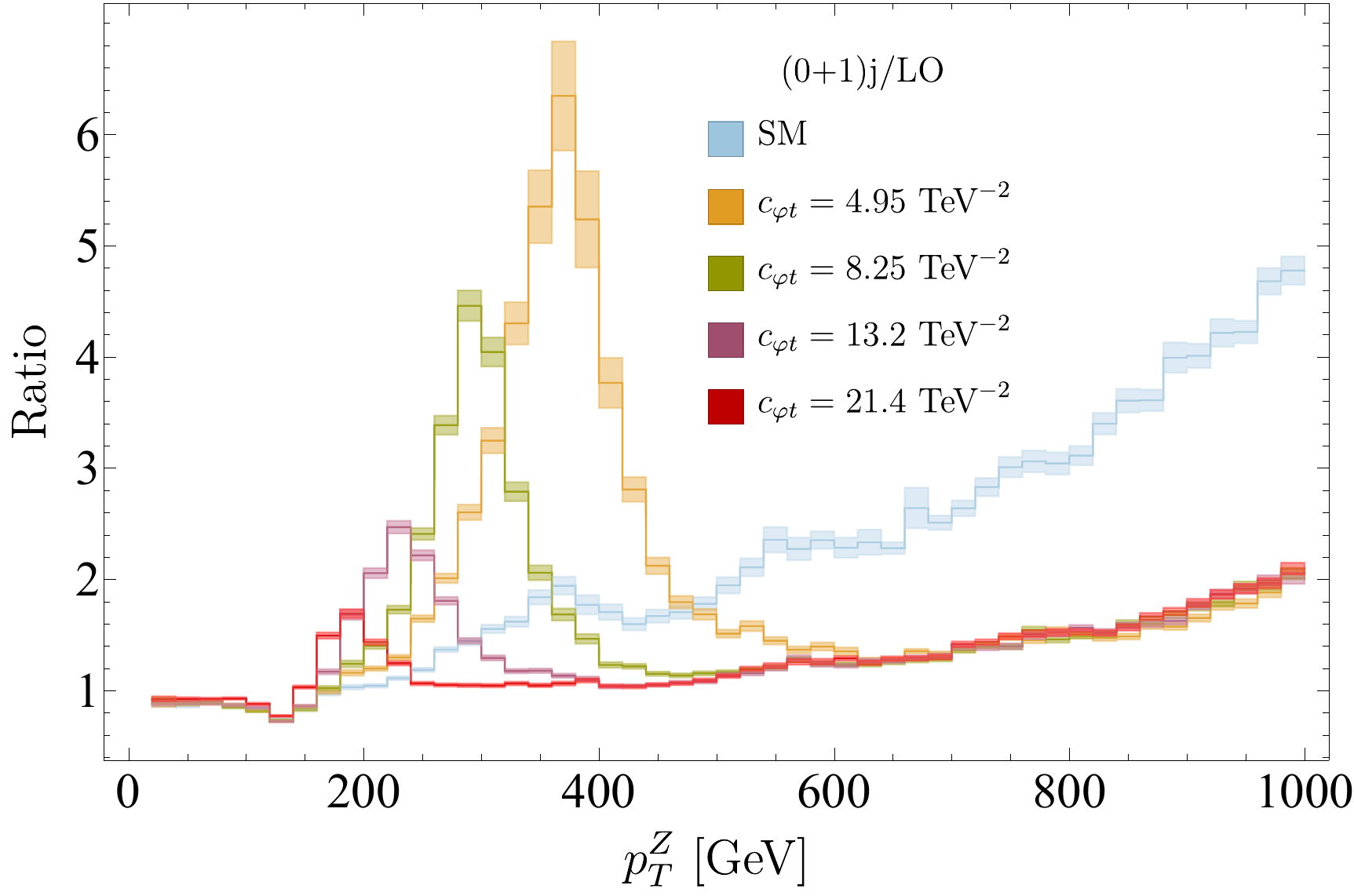}\hfill
    \includegraphics[width=0.5\textwidth]{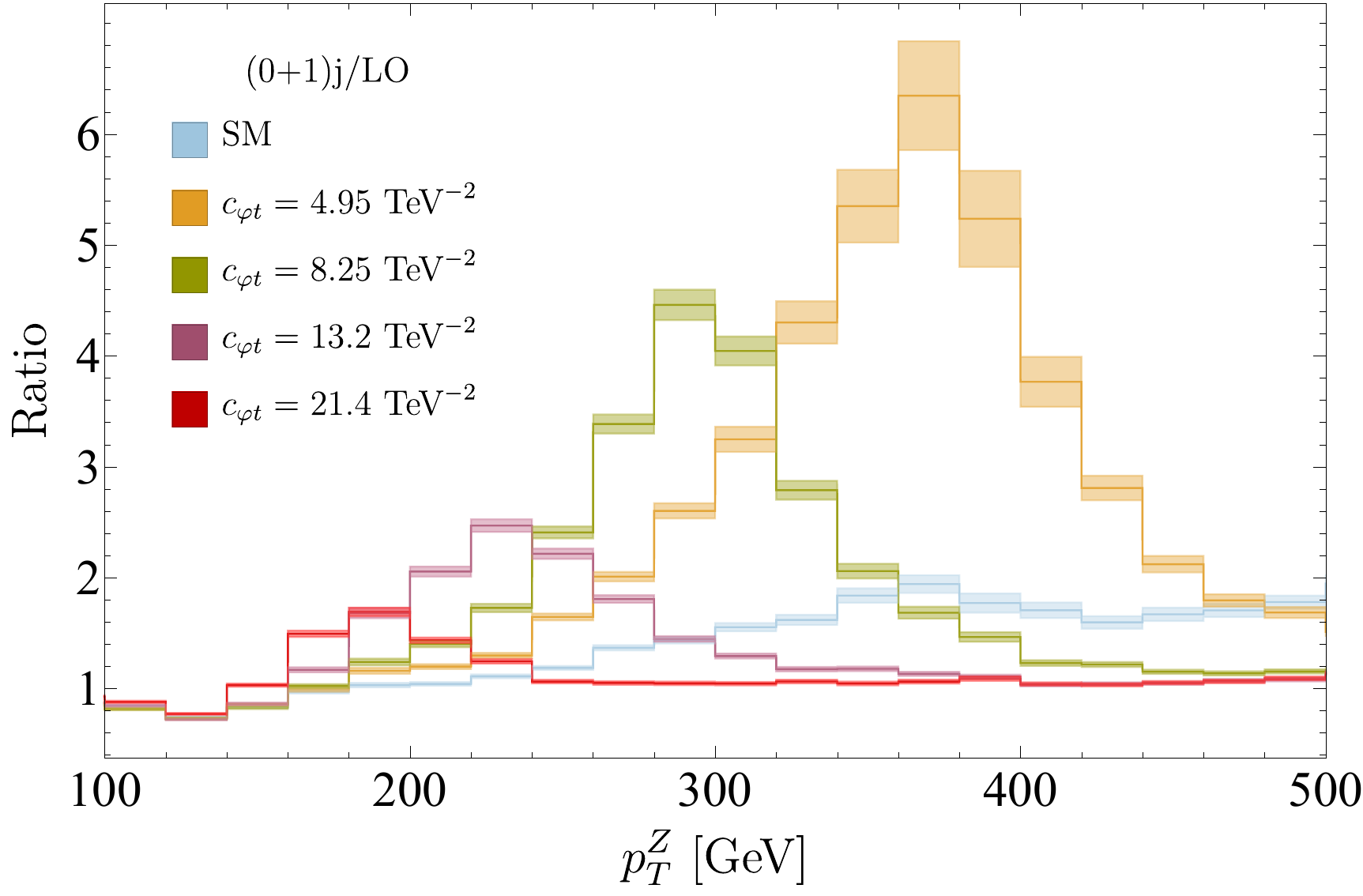}\
    \caption{Ratio of differential cross sections for $gg\to ZH$ between $0+1$ jet merged and LO. The light blue curve shows the Standard Model result. The remaining curves show the result in SMEFT for several positive values of $\cht$ with all other WCs set to zero. The shaded regions indicate the statistical uncertainty. The right panel shows a zoom-in of the plot on the left.}
    \label{fig:plot_NLO_LO_jetMatched_3}
\end{figure}

Higher values of the WC can yield starker differences between the LO and merged samples. Such values would be excluded for $\chqTopMinus$ already from this $ZH$ analysis, but not for $\cht$ or $\cth$. Hence, we show in Fig.~\ref{fig:plot_NLO_LO_jetMatched_3}, the ($0+1$ jet)/LO ratio for several positive values of $\cht$ around the expected HL-LHC bounds, as will be shown in Subsection~\ref{sec:LHC_bounds}. These ratios show a sharp peak at mid-low energies, which decreases in height and moves towards lower energies for increasing values of the WC. 

The peak is caused by the negative interference between $\Oht$ and the SM and the partial deletion of its effect by the hard emission and shower.
At higher energies, the interference becomes less relevant and is overtaken by the $\op (\Lambda^{-4})$ piece of the cross section. The energies at which the interference effect becomes noticeable and at which is surpassed by the quadratic piece move downwards as $\cht$ increases, shifting the peak position accordingly. This non-trivial behaviour of the ratios hints at a possible strong dependence of the NLO/LO k-factors on the value of the WCs, in particular when the interference is relevant. Its detailed study is beyond the scope of this work.

Thus, higher-order corrections due to real emissions might have a sizeable effect on the bounds that can be put on WCs. Whilst different simulations can lead to ratios which can be as large as 10, one has to explore the impact of these corrections in a realistic analysis setup.

\begin{figure}[htb!]
    \centering
    \includegraphics[width=0.49\textwidth]{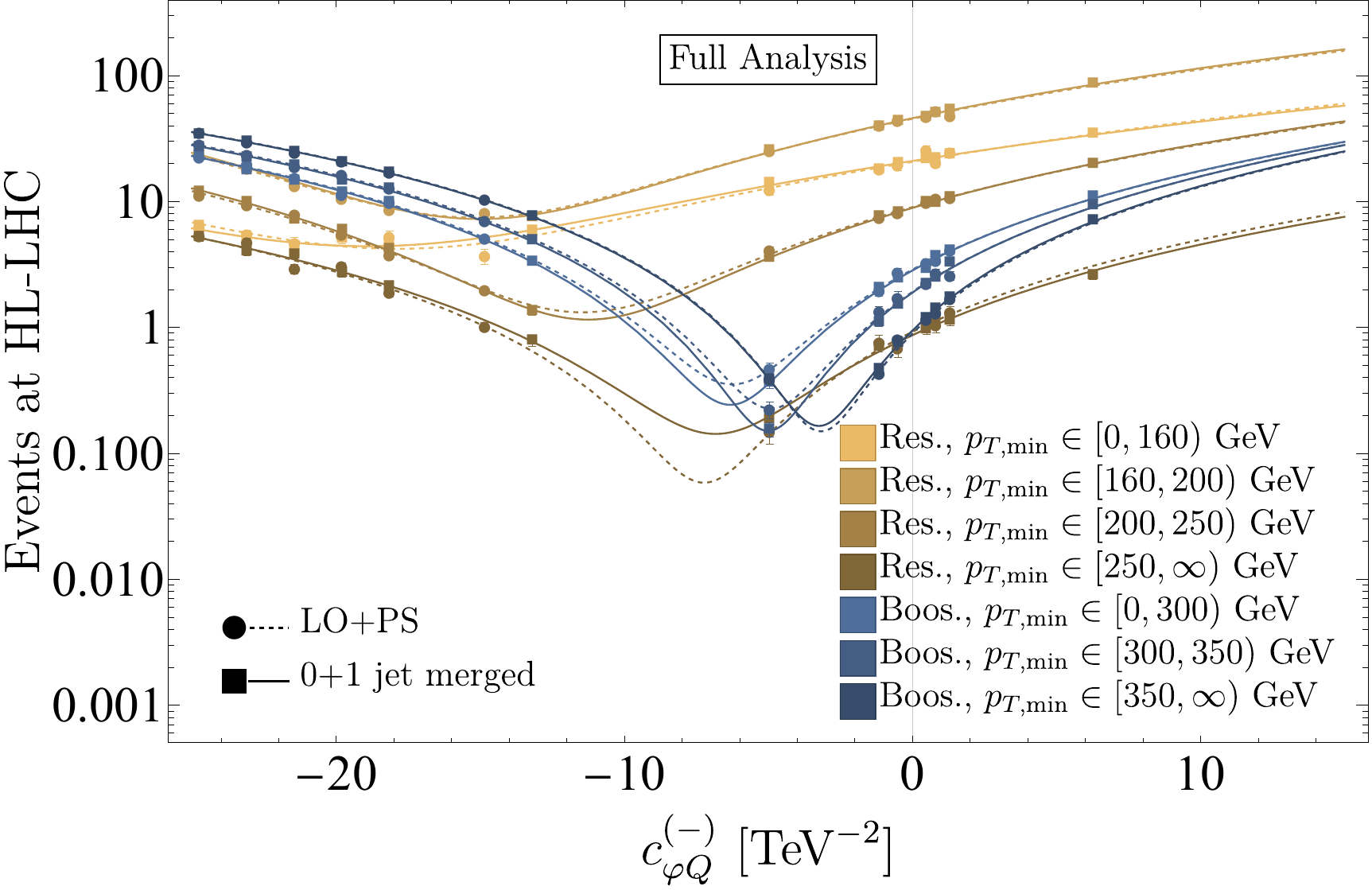}\hfill
    \includegraphics[width=0.49\textwidth]{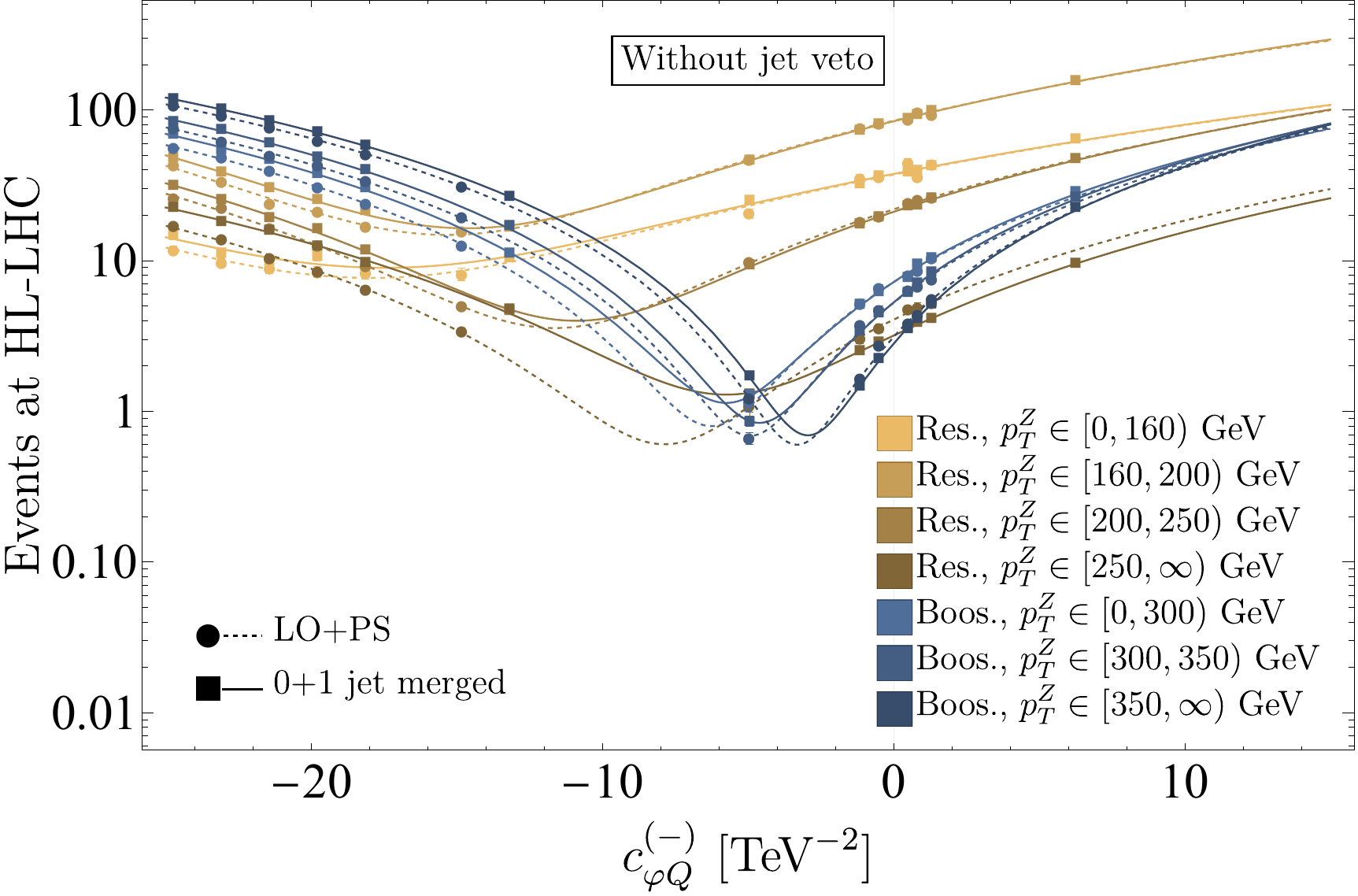}\
    \caption{Number of events at HL-LHC from $gg\to ZH\to \nu\bar\nu b\bar b$ in the different categories and bins as function of $\chqTopMinus$, with all other WCs to zero. The different colours identify the different categories and bins. The squares (circles) represent the results obtained with simulations of 0+1 jet merged (at LO+PS) and the continuous (dashed) line shows the quadratic fit to them. \textbf{Left panel:} Analysis with jet veto and binning in $\ptmin$, as used in the rest of this work. \textbf{Right panel:} Analysis without jet veto and binning on $\ptz$. 
    }
    \label{fig:plot_fits_0lep_LO_01jetmatched}
\end{figure}

In the left panel of Fig.~\ref{fig:plot_fits_0lep_LO_01jetmatched}, we show the number of events expected at HL-LHC for $gg\to ZH\to\nu\bar\nu b\bar b$ as a function of $\chqTopMinus$ obtained with LO and $0+1$ jet merged simulations and the analysis strategy presented in Section~\ref{sec:Pheno_ggZH_analysis}. The dots represent the WC values for which we simulated and analyzed the signal, and the lines are the quadratic functions fitted to them. We show only the $0$-lepton channel since it drives the sensitivity of the analysis and checked that using a different WC would yield qualitatively similar results. The differences between the LO and 0+1 jet results are negligible in most bins, except near the minimum, where the difference is relatively bigger but irrelevant for HL-LHC luminosities since both cases would yield 0 events. 
This is due in part to the presence of a jet-veto in the analysis and the binning in $\ptmin$, as can be deduced from the right panel of Fig.~\ref{fig:plot_fits_0lep_LO_01jetmatched}, in which we show the number of events after removing the jet veto and binning in $\ptz$ instead of $\ptmin$. The difference between LO and $0+1$ jet merged remains small but the gap widens in two situations. First, when $\chqTopMinus$ is negative, due to the reduced impact of the destructive SM interference in the merged samples. Second and most notably, in the highest-energy resolved bin, where the resolved regime favours a low $\pth$ while the bin definition requires a high $\ptz$, giving place to a $p_T$ imbalance that can be compensated by a hard jet.

\begin{figure}[htb!]
    \centering
    \includegraphics[width=0.49\textwidth]{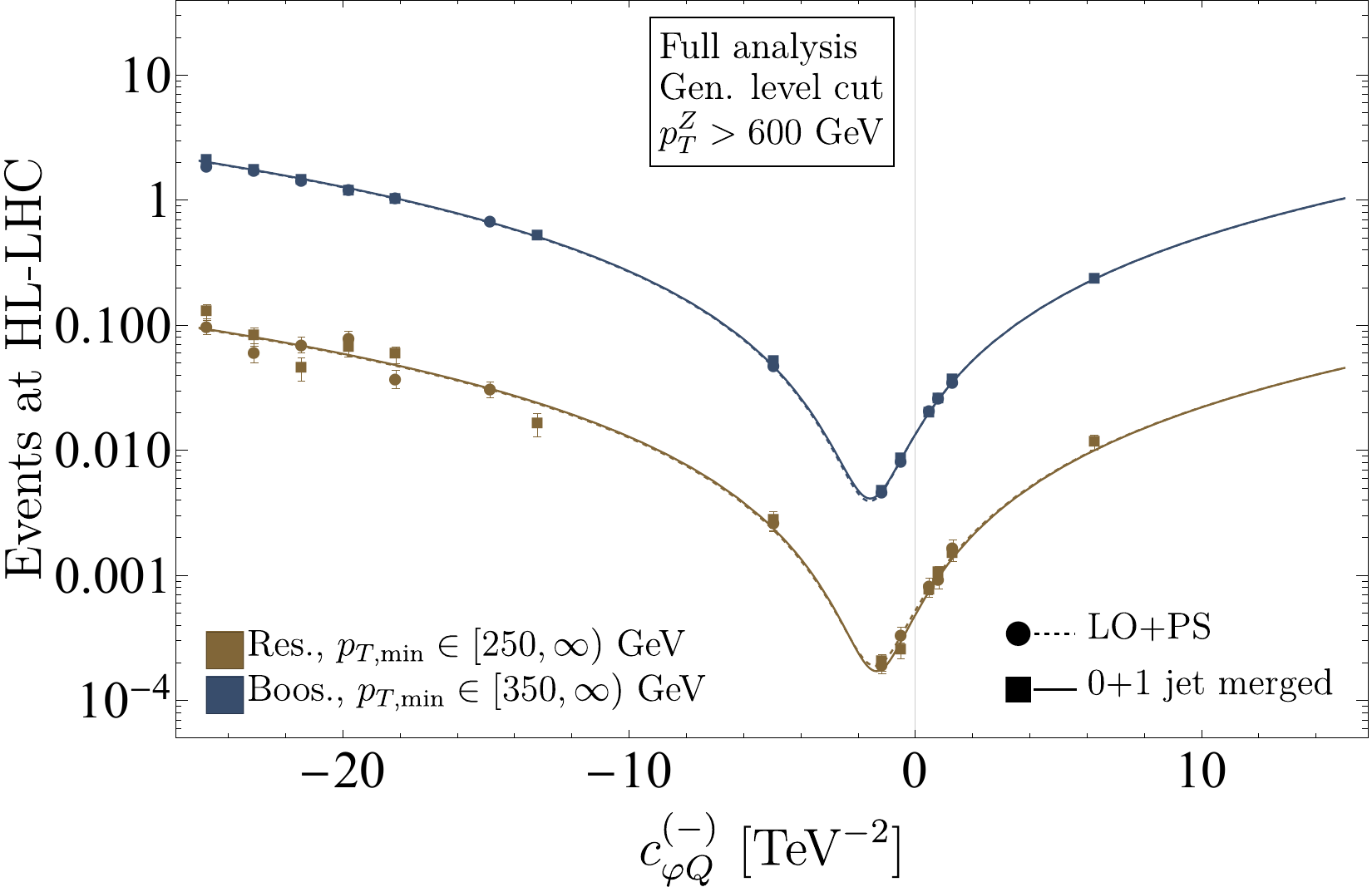}\hfill
    \includegraphics[width=0.49\textwidth]{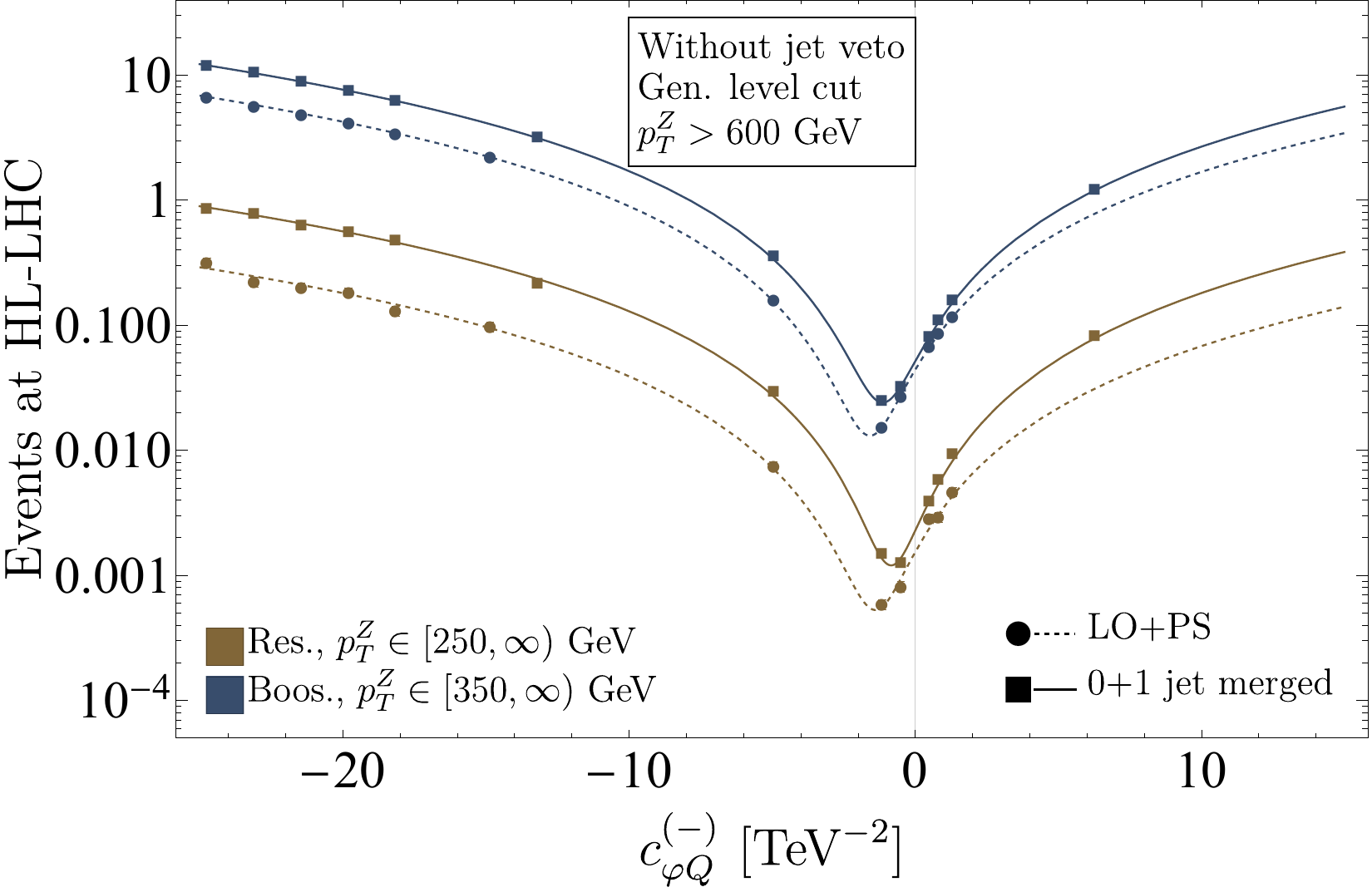}\
    \caption{Number of events at HL-LHC from $gg\to ZH\to \nu\bar\nu b\bar b$ from the high-energy regime as function of $\chqTopMinus$, with all other WCs set to zero. The high-energy condition was enforced by applying  a generation-level cut $\ptz>600$~GeV. The different colours identify the different categories and bins. Only the highest energy bins receive relevant contributions in this regime. The squares (circles) represent the results obtained with simulations of 0+1 jet merged (at LO) and the continuous (dashed) line shows the quadratic fit to them. \textbf{Left panel:} Analysis with jet veto and binning in $\ptmin$, as used in the rest of this work. \textbf{Right panel:} Analysis without jet veto and binning on $p_T^Z$.}
    \label{fig:plot_fits_0lep_LO_01jetmatched_he}
\end{figure}

Another factor that reduces the impact of the real emission is that the (HL-)LHC analysis is mostly sensitive to the $[250,\,500]$~GeV energy region, where the $0+1$ jet merged samples underestimate the NLO corrections, as can be seen in Fig.~\ref{fig:plot_NLO_LO_jetMatched_ptZ_SM_cpQ3}  and~\ref{fig:plot_NLO_LO_jetMatched_2}. As a check  of this phenomenon, we obtained the post-analysis cross-sections but using only events generated with a parton-level cut $\ptz>600$~GeV. The results are presented in Fig.~\ref{fig:plot_fits_0lep_LO_01jetmatched_he}, where the left (right) panel shows the result of using the analysis with (without) jet veto and binning in $\ptmin$ ($\ptz$). Only the highest-energy bins of both the boosted and resolved modes receive significant contributions from the  $\ptz>600$~GeV region. The analysis with a jet veto and the binning in $\ptmin$ shows negligible differences between the simulation orders.

For the case of no jet veto and binning in $\ptz$, shown on the right panel of Fig.~\ref{fig:plot_fits_0lep_LO_01jetmatched_he}, both simulation orders differ by between $\sim50\%$ and $\sim300\%$, depending on the value of $\chqTopMinus$ and whether it is the resolved or boosted mode. The biggest differences are found in the resolved mode for the same reasons as before and near the minimum of the cross-section, where the interference is more relevant. Furthermore, we checked that for this analysis version and in this energy regime the degeneracy among $\chqTopMinus$ and $\cht$
present at LO is broken by the additional jet as the $ZHj$ amplitudes do not simply depend on the axial vector coupling of the top. However, such effect becomes negligible in the full analysis with a jet veto, binning in $\ptmin$ and including lower energies. 

Thus, the number of events at HL-LHC obtained with our analysis strategy is insensitive to the use of $0+1$ jet merged samples. Instead, a simple rescaling of the LO cross-section by a constant SM k-factor offers a better approximation of the impact of NLO corrections for bounds at (HL-)LHC.
Due to the sensitivity of our (HL-)LHC analysis to the $p_{T}\in[250,\,500]$~GeV region, we average the k-factor computed in Ref.~\cite{Chen:2022rua} over those energies. This procedure yields a k-factor of $2.0-2.1$ ($3.0-3.4$) when averaging over $\ptz$ ($\pth$) depending on the renormalization scale. We pick the conservative value of $2.0$ for our analysis and briefly comment on the results obtained with other choices.
From the previous discussion, we expect this simple rescaling to be inaccurate at higher energies, and hence the simulation of $0+1$ jet merged samples could be preferable for future colliders such as FCC-hh. 

\subsection{Flavour assumptions and interplay with the $q\bar{q}$ channel.}
\label{sec:qq_gg_flavour}

\begin{figure}[htb!]
    \centering
    \includegraphics[width=0.5\textwidth]{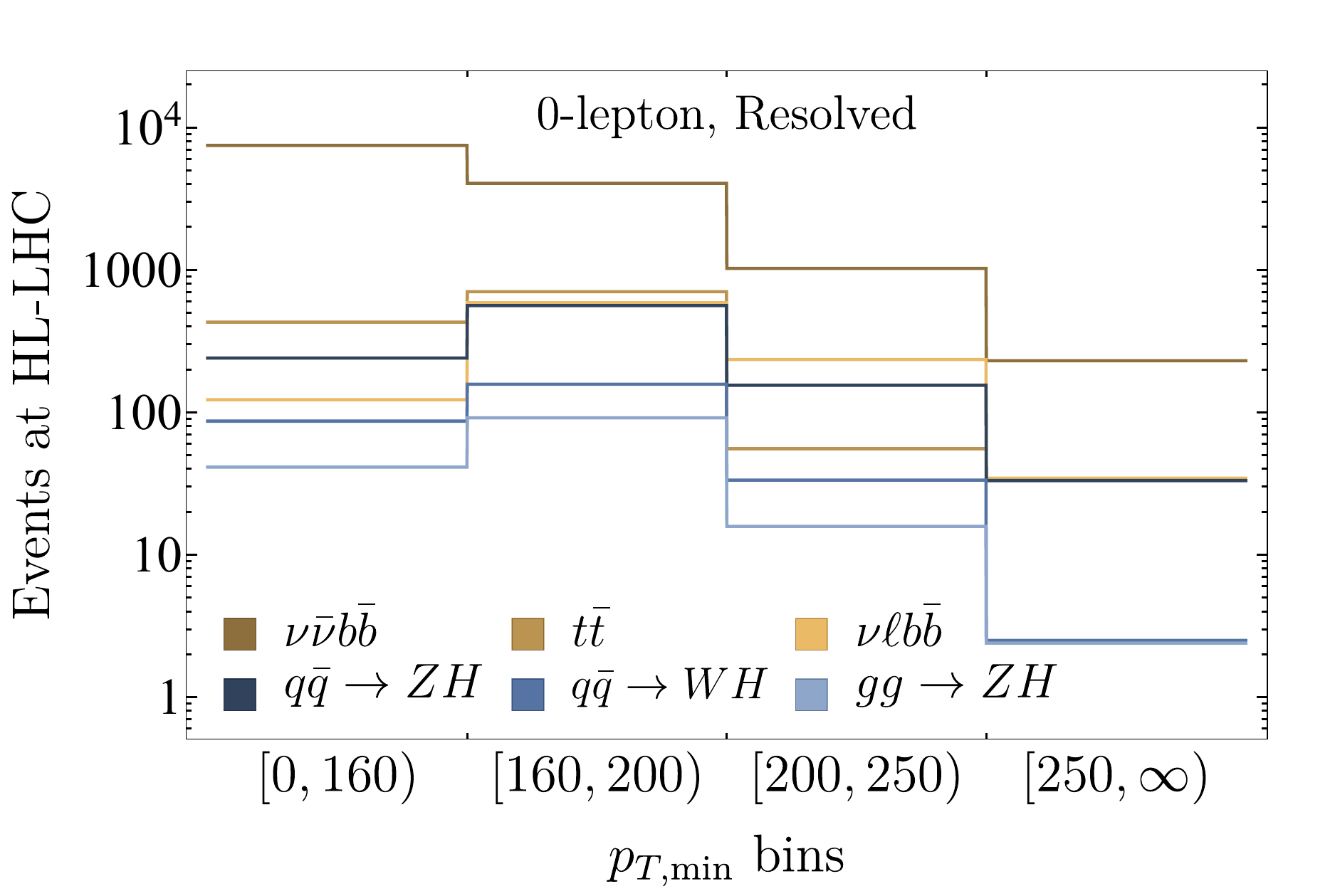}\hfill
    \includegraphics[width=0.5\textwidth]{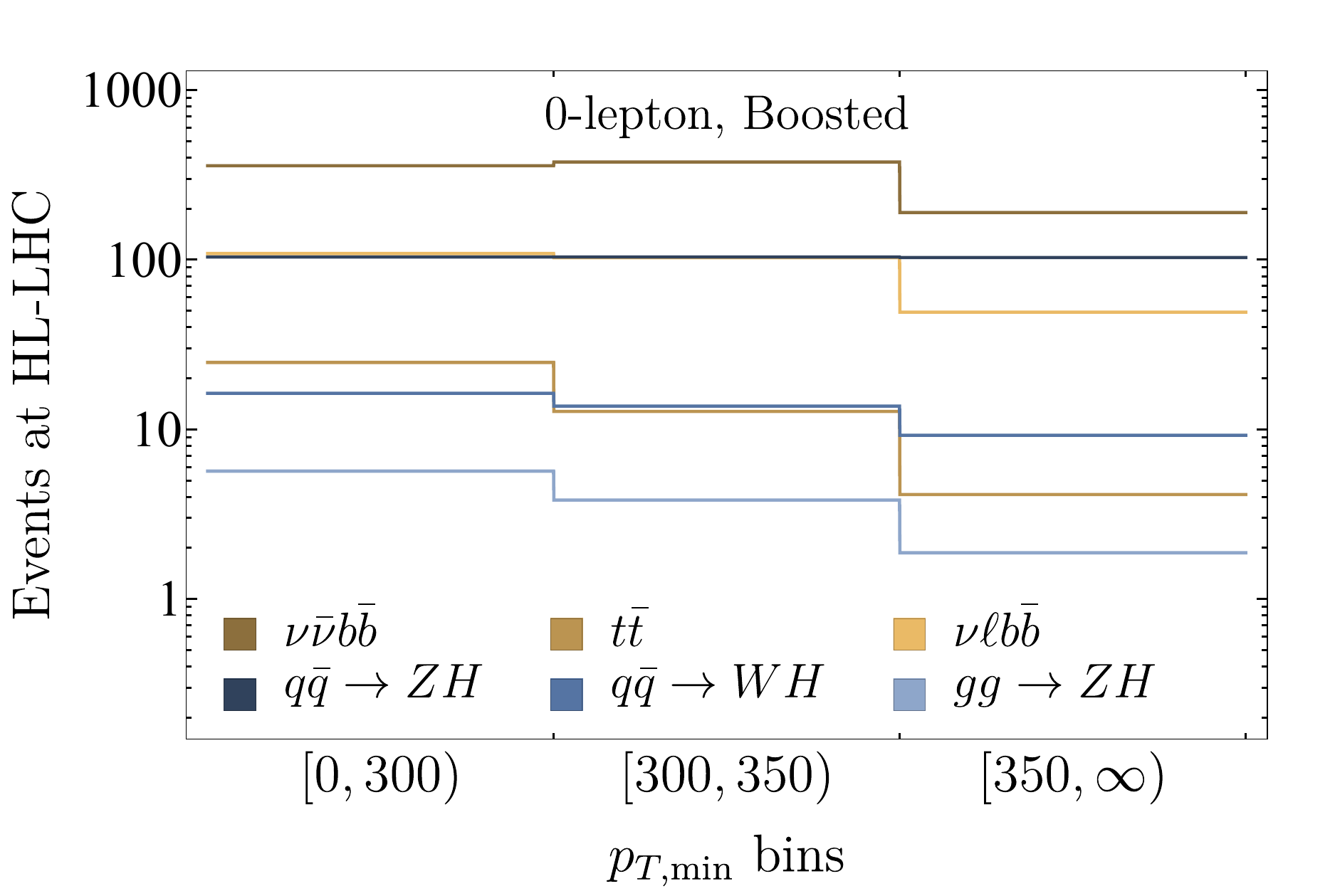}\vfill
    \includegraphics[width=0.5\textwidth]{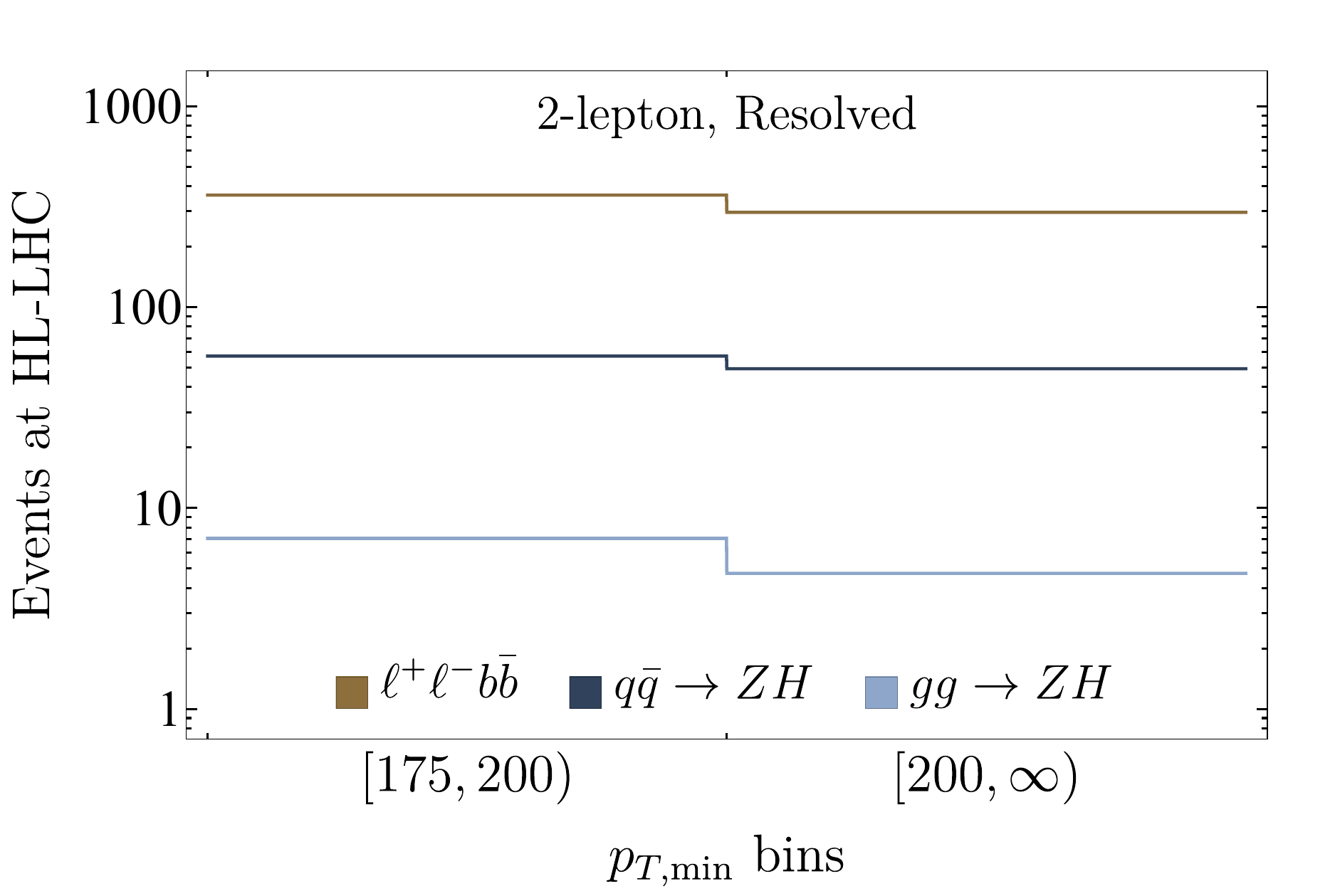}\hfill
    \includegraphics[width=0.5\textwidth]{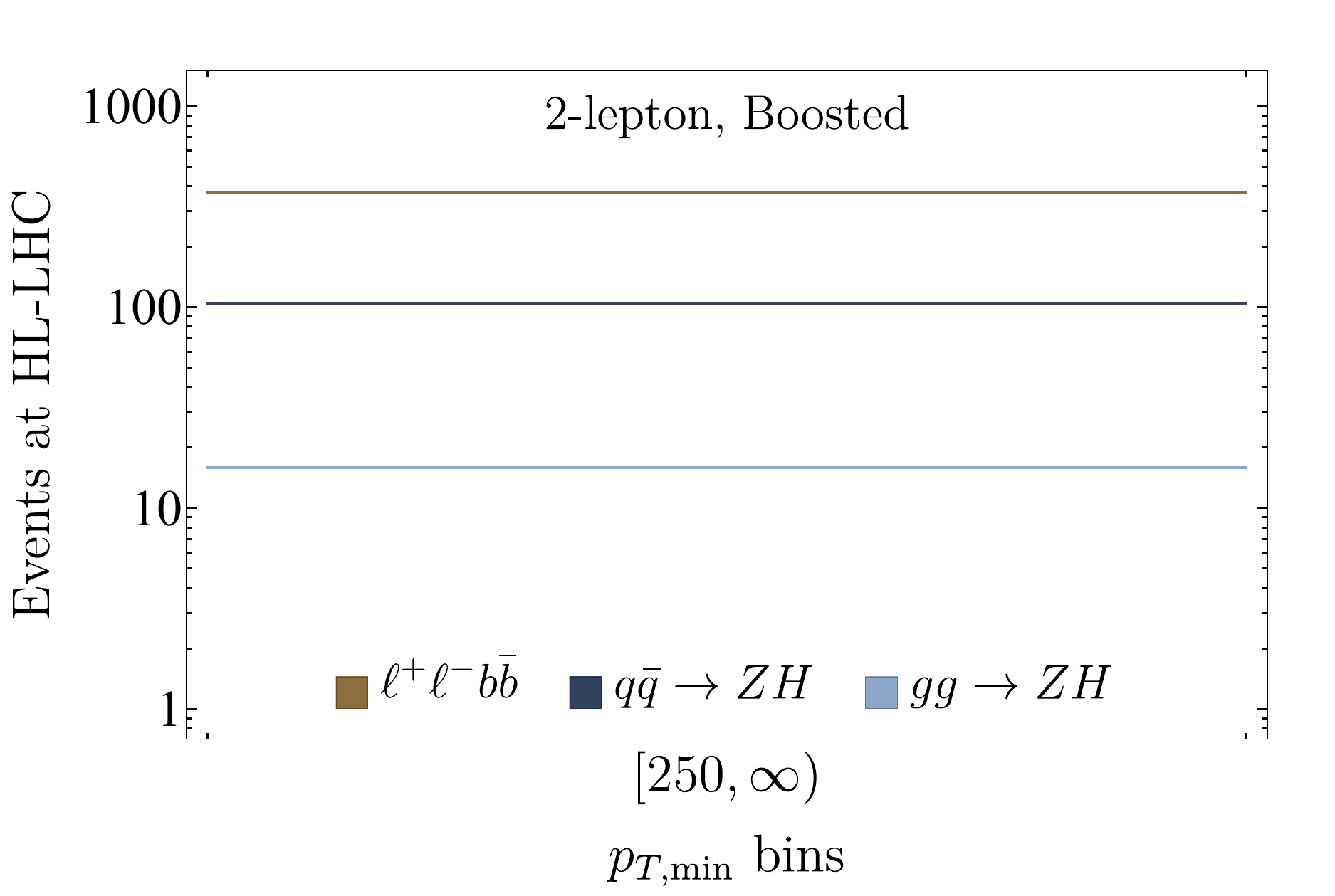}\
    \caption{Number of SM events at HL-LHC for each of the processes contributing to the signal and background in our analysis of $pp\to ZH$. The processes in shades of brown and orange are background processes, while the ones in blue are considered part of the signal. The x-axis represents the analysis bins used in each channel and category. \textbf{Top left:} 0-lepton channel, resolved category. \textbf{Top right:} 0-lepton channel, boosted category. \textbf{Bottom left:} 2-lepton channel, resolved category. \textbf{Bottom right:} 2-lepton channel, boosted category.}
    \label{fig:App_xs_bins_sig_bkgd_SM_0+2lep}
\end{figure}

The gluon-initiated channel is subdominant in $ZH$ production at hadron colliders where it formally enters at NNLO. This can be seen explicitly in Fig.~\ref{fig:App_xs_bins_sig_bkgd_SM_0+2lep}, where we show the number of events at HL-LHC per bin in each channel and category generated by each of the signal and background processes in the SM. In all cases, the gluon-initiated contribution to the signal is one order of magnitude smaller than the one from the quark-initiated channel. Additionally, the background overwhelms the signal in most bins. Therefore, the contribution from gluon-initiated $ZH$ production would not modify ostensibly the flavour-universal results in Ref.~\cite{Bishara:2022vsc}.

Which dimension-6 operators are constrained and how much each production channel contributes to the total sensitivity is a flavour-symmetry dependent statement.
Here, we assume the U$(2)_q\times$U$(3)_d\times$U$(2)_u\times(\text{U}(1)_\ell\times\text{U}(1)_e)^3$ flavour symmetry, 
which allows $gg\to ZH$ to be sensitive to a linear combination of the dimension-6 WCs $\chqTopMinus-\cht+\frac{\cth}{y_{t}}$, as seen in Subsection~\ref{sec:HeliAmp_ggZH}.
In contrast, $qq\to ZH$ is capable to constrain four WCs of the first- and second-generation operators: $\Ohqii$, $\Ohqtii$, $\Ohu$ and $\Ohd$ as well as two of the third-generation ones: $\OhqTopt$, and $\OhqTopMinus$. While additional operators modify the inclusive $qq\to ZH$ cross section, they do not generate growth with energy and consequently, our sensitivity to them is negligible~\cite{Rossia:2021fsi,Bishara:2022vsc}. Since only the $b\bar b$ initial state is sensitive to the heavy-quark operators, $qq \to ZH$ can only probe the combination $2 \chqTopt + \chqTopMinus$. When combined with the gluon-initiated channel, this leaves two unconstrained directions in the space of the heavy-quark operators. There are no flat directions in the space of the light-quark operators and the discussion on the sensitivity to each of their Flavour-Universal versions in Ref.~\cite{Bishara:2022vsc,Bishara:2020pfx} applies straightforwardly.

The abandonment of Flavour Universality implies that the cross-section of the $q\bar{q}$ channel as a function of the WCs could map onto the new flavour scenario in a non-trivial way.
Parton-level analysis shows that the contribution of the $b\bar{b}$ initial state at (HL-)LHC is $\lesssim 1\%$ in the energy regime that matters for sensitivity to NP, $p_{T}\in[200,600]\GeV$, and increases slightly at lower energies. Applying the full analysis method explained in Subsection~\ref{sec:Pheno_ggZH_analysis}, one finds that the contribution of the $b\bar{b}$ initial state to the (HL-)LHC bounds on $\chqt$, $\chq$, and $\chd$ is $\lesssim 0.7\%$. We checked that translating the cross-section of $q\bar q$ in the different bins as a function of the WCs from the Flavour-Universal case to the U$(2)_q\times$U$(3)_d\times$U$(2)_u$ scenario by simply replacing $c_{\varphi q}^{(1),(3)}\to c_{\varphi q,ii}^{(1),(3)}$ means incurring in an error $\lesssim 1\%$, which is of the same order than the uncertainty with which those cross-sections were obtained. In the case of the 0-lepton channel, the error is even smaller since its dependence on $\chqt$ receives a contribution from $WH$ production, for which the replacement $\chqt\to\chqtii$ is exact.

The interference and BSM squared pieces for the heavy-quark WCs arising from $b\bar b\to ZH$ is the difference between the corresponding terms for light-quark and flavour-universal WCs, since,
\begin{equation}
    \sigma\supset \sigma_{\text{int}}^{(1)}\, \chqtii + \sigma_{\text{int}}^{(2)}\, \chqTopt \xrightarrow[\chqTopt=\chqtii]{} (\sigma_{\text{int}}^{(1)} + \sigma_{\text{int}}^{(2)})\, \chqt,
\end{equation}
where we focus on the interference for $\chqt$, $\chqtii$, and $\chqTopt$ for brevity. The minimal differences between the flavour scenarios found previously could lead to conclude that $b\bar b\to ZH$ has negligible sensitivity to $\chqTopt$ and $\chqTopMinus$. However, its dependence on the latter one turns out to be of the same order and slightly bigger than the one of $gg \to ZH$. Whilst being suppressed by the $b$-quark PDFs (and thus being much smaller than the contribution  of the corresponding light quark operators) this contribution comes at tree level and competes with the loop-induced channel. This can be checked explicitly by looking at the tables in Appendix~\ref{app:EvtNumbersVH}, where we give the number of events at HL-LHC in each bin and category as a function of the WCs, and using that each channel contributes to a particular linear combination of WCs.

Hence, we expect the sensitivity to the light-quark operators at (HL-)LHC to be indistinguishable from their counterparts in the Flavour-Universal case, while relaxing the flavour symmetry offers sensitivity to two linear combinations of $\chqTopMinus$, $\chqTopt$, $\cht$, and $\cth$. One of those linear combinations, $\chqTopMinus-\cht+\frac{\cth}{y_{t}}$, is probed by the gluon-initiated process and only $\chqTopMinus$ is probed by both production channels. In the context of a global fit, the two flat directions that remain in the heavy-quark operator space can be constrained by processes such as $tj$ production, which is only sensitive to $\chqTopt$, $t\bar t H$, which probes mainly $\cth$~\cite{Maltoni:2016yxb}, as well as $tZj$ and $t\bar{t}Z$ which probe all of $\chqTopMinus,\,\,\chqTopt,$ and $\cht$ \cite{1601.01167,Degrande:2018fog}. 

Future hadron colliders with higher centre-of-mass energies could change the previous picture.
At $100\,\TeV$, the $b\bar{b}$ channel contributes $\sim 5\%$ of the partonic cross section of $q\bar{q}\to ZH$ in the region $p_{T}^{H}\in[400,\,600]\GeV$~\cite{Bishara:2020pfx}. The uncertainties on the fits and experimental measurements would be of the same order or less, so the bounds on the light-quark operators could be distinguishable from the ones on the flavour-universal ones. Additionally, NLO QCD corrections to $gg\to ZH$ could become more relevant and help to lift flat directions. We leave a detailed study of this case for future work.

\subsection{Constraints at (HL-)LHC}
\label{sec:LHC_bounds}

Here, we present the projected $95\%$ C.L. bounds on dimension-6 WCs at the HL-LHC from the process $pp\to ZH$ for the flavour assumption U$(2)_q\times$U$(3)_d\times$U$(2)_u\times(\text{U}(1)_\ell\times\text{U}(1)_e)^3$. We assume SM-like measurements, uncorrelated observables and three different systematic uncertainties: $1\%$, $5\%$, and $10\%$. Our HL-LHC projections for 1-operator fits are summarised in Table~\ref{tab:bounds_summary_NFU_0plus2lep_HLLHC_kF2} (see App.~\ref{app:Bounds_LHCRun3} for the projections corresponding to LHC Run 3 luminosity). 
At the HL-LHC, $5\%$ syst. uncertainties start overshadowing the statistical ones. This value is the most realistic assumption for HL-LHC according to current measurements
~\cite{ATLAS:2021wqh}. 
The bounds at LHC Run 3 are still statistically limited, hence HL-LHC can improve them by $30-50\%$.

A comparison of the 1-op. bounds on $\chqtii$, $\chqii$, $\chu$ and $\chd$ against the bound on their flavour-universal counterparts from Ref.~\cite{Bishara:2022vsc} reveals that the new flavour assumption and the addition of $gg\to ZH$ has had no relevant effect on $\chqii$, $\chu$ and $\chd$, as expected. The worsening of the bounds on $\chqtii$ is due to the exclusion from this work of the 1-lepton channel, which probes only this coefficient.

Heavy-quark operators are less stringently constrained than their light counterparts. Those that affect the $b\bar b\to ZH$ process, $\chqTopt$ and $\chqTopMinus$, have bounds between one and two orders of magnitude worse than the ones for light-quark operators. $\cht$ and $\cth$, probed only via gluon fusion, are constrained even more loosely and their bounds are up to two orders of magnitude worse than the ones for $\chqTopt$. As discussed before, in many BSM scenarios, these four WCs are expected to be of similar order~\cite{Giudice:2007fh,Contino:2013kra} and therefore the most stringent bounds on the UV model would arise from the $b\bar b$ initiated channel.

\begin{table}[H]
	\begin{centering}
		\begin{tabular}{|c|c|}
			\hline
			WC  & $95\%$ C.L. Bound \tabularnewline
			\hline
			$c_{\varphi q,ii}^{(3)}$ &
					\begin{tabular}{ll}
				$[-1.4,\,1.3]\times10^{-2}$&$1\%$ syst.\\[-0.65em]
				$[-2.1,\,1.8]\times10^{-2}$&$5\%$ syst.\\[-0.65em]
				$[-3.8,\,2.7]\times10^{-2}$&$10\%$ syst.
			\end{tabular}
			\tabularnewline
			\hline
			$c_{\varphi q,ii}^{(1)}$ &
				\begin{tabular}{ll}
				$[-4.6,\,6.4]\times10^{-2}$ & $1\%$ syst.\\[-0.65em]
				$[-5.5,\,7.4]\times10^{-2}$ & $5\%$ syst.\\[-0.65em]
				$[-7.1,\,9.0]\times10^{-2}$ & $10\%$ syst.
			\end{tabular}
			\tabularnewline
			
			\hline
			$c_{\varphi u}$ &
			\begin{tabular}{ll}
				$[-11.3,\,4.1]\times10^{-2}$ & $1\%$ syst.\\[-0.65em]
				$[-12.6,\,5.2]\times10^{-2}$ & $5\%$ syst.\\[-0.65em]
				$[-14.6,\,7.1]\times10^{-2}$ & $10\%$ syst.
			\end{tabular}
			\tabularnewline
			\hline
			$c_{\varphi d}$ &
			\begin{tabular}{ll}
				$[-6.8,\,10.7]\times10^{-2}$&$1\%$ syst.\\[-0.65em]
				$[-8.3,\,12.2]\times10^{-2}$&$5\%$ syst.\\[-0.65em]
				$[-1.08,\,1.47]\times10^{-1}$&$10\%$ syst.
			\end{tabular} 
			\tabularnewline
                \hline
		\end{tabular}
        \begin{tabular}{|c|c|}
			\hline
			WC & $95\%$ C.L. Bound \tabularnewline
			\hline
			$c_{\varphi Q}^{(3)}$ &
			\begin{tabular}{ll}
				$[-6.2,\,4.7]\times10^{-1}$ & $1\%$ syst.\\[-0.65em]
				$[-7.2,\,5.7]\times10^{-1}$ & $5\%$ syst.\\[-0.65em]
				$[-8.9,\,7.4]\times10^{-1}$ & $10\%$ syst.
			\end{tabular}
			\tabularnewline
			\hline
			$c_{\varphi Q}^{(-)}$ &
			\begin{tabular}{ll}
				$[-12.8,\,8.8]\times10^{-1}$ & $1\%$ syst.\\[-0.65em]
				$[-1.5,\,1.1]\phantom{\times10^{-1}}$ & $5\%$ syst.\\[-0.65em]
				$[-1.8,\,1.4]\phantom{\times10^{-1}}$ & $10\%$ syst.
			\end{tabular}	
			\tabularnewline
			\hline
			$c_{\varphi t}$ &
			\begin{tabular}{ll}
				$[-5.8,\,\phantom{1}17.1]\phantom{\times10^{-1}}$ & $1\%$ syst.\\[-0.65em]
				$[-8.1,\,\phantom{1}19.6]\phantom{\times10^{-1}}$ & $5\%$ syst.\\[-0.65em]
				$[ -11.4,\,23.0]\phantom{\times10^{-1}}$ & $10\%$ syst.
			\end{tabular}
			\tabularnewline
			
			\hline
			$c_{t \varphi }$ &
			\begin{tabular}{ll}
				$[-16.9,\,\phantom{1} 5.7]$ & $1\%$ syst.\\[-0.65em]
				$[-19.4,\,\phantom{1} 8.0]$ & $5\%$ syst.\\[-0.65em]
				$[-22.7,\,11.3]\phantom{\times10^{-1}}$ & $10\%$ syst.
			\end{tabular} \\
			\hline
		\end{tabular}
	  \end{centering}
	\caption{Projected bounds at $95\%$ C.L.~from one-dimensional fits on the dimension-6 SMEFT WCs probed by $pp\to ZH$ at HL-LHC with integrated luminosity of $3\,\mathrm{ab}^{-1}$. The WCs are in units of TeV$^{-2}$. The gluon-initiated channels were rescaled with a constant k-factor of $2.0$ to account for NLO QCD corrections. All the results include the quadratic EFT corrections.
		{\bf Left: } Light-quark WCs. {\bf Right column:} Heavy-quark WCs. }
	\label{tab:bounds_summary_NFU_0plus2lep_HLLHC_kF2}
\end{table}

Among the heavy-quark operators, $\chqTopt$ is the best constrained one, despite modifying only the $b\bar b$ channel. The same process is expected to give a worse bound on $\chqTopMinus$ by a factor of  $2$, but this should be partially compensated by the contribution of the gluon-fusion process. In practice, the latter contribution is small and its main effect is a shift of the bounds toward negative values. The dominance of the quark-initiated channel can also be seen in the great difference between the bounds on $\chqTopMinus$ and $\cht$ given that the latter is probed only via top loops. 

Since the gluon-initiated channel is only sensitive to $\chqTopMinus-\cht+\frac{\cth}{y_t}$, the bounds on $\chqTopMinus$ and $\cht$ from $gg\to ZH$ would be the same up to a sign. Moreover, the bounds on $\cth$ should be marginally better and opposite with respect to the ones on $\cht$, as confirmed by our results.
The sensitivity to $\chqTopt$ and $\chqTopMinus$ is led by the piece of the cross-section that is quadratic on the WCs, while for $\cth$ and $\cht$, the cross-section is in an intermediate regime between interference- and quadratic-domination, as evidenced by the high asymmetry of the bounds around $0$, and despite the lack of energy growth in the interference with the SM. Attempts to recover the growth in the interference would be hampered by low statistics at HL-LHC but could be fruitful at future colliders.

For the results presented here, we have rescaled the $gg\to ZH$ cross section by a constant k-factor of $2$ to account for NLO QCD corrections, see Subsection~\ref{sec:NLO_01jet_LO} for details. The use of a higher k-factor, such as $3.37$ which corresponds to the $p_{T}^h$ distributions in Ref.~\cite{Chen:2022rua}, would tighten the bounds on $\cht$ and $\cth$ by $\sim 30\%$.  All the other bounds would be modified by $\lesssim 1\%$. Additionally, the sensitivity to $\chqTopMinus$ ensures that we are in a regime in which the corrections from real emissions are negligible, as can be seen by looking at the left panel in Fig.~\ref{fig:plot_fits_0lep_LO_01jetmatched} around the values corresponding to our bounds.

The importance of $ZH$ production to probe heavy-quark operators is put in perspective when comparing our projections against current bounds. 
Global fits of LHC data, including data with luminosity up to $139$~fb$^{-1}$, are able to set bounds $|\chqTopt|\lesssim 0.6$~TeV$^{-2}$, $|\chqTopMinus|\lesssim 2.9$~TeV$^{-2}$~\cite{Ethier:2021bye}
. Our HL-LHC projections for those WCs are similar and slightly better respectively, and are still competitive against the HL-LHC projections $|\chqTopt|\lesssim 0.6$~TeV$^{-2}$, $|\chqTopMinus|\lesssim 1.3$~TeV$^{-2}$~\cite{deBlas:2022ofj}. In the case of $\cht$, LHC data constrain it to $\cht\in[-13.3,4.0]$~TeV$^{-2}$~\cite{Ethier:2021bye}, which is better than the reach of $gg\to ZH$ at HL-LHC with $5\%$ systematic uncertainty for positive values but worse for negative values, while the HL-LHC projection of the global fit indicates $|\cht|\lesssim 5$~TeV$^{-2}$~\cite{deBlas:2022ofj}. This process is less competitive to probe $\cth$ since the current bound is $[-2.3,2.8]$~TeV$^{-2}$~\cite{Ethier:2021bye} and is expected to improve by a factor $\sim2$ at HL-LHC~\cite{deBlas:2022ofj}. Overall, $ZH$ production could have a significant impact on future global fits, in particular for the WC $\chqTopMinus$, to which it is sensitive thanks to the contributions of both the quark- and gluon-initiated process. Additionally, our bounds on $\cht$ and $\cth$ are $\sim 6$ and $\sim 2$ looser than the ones derived in Ref.~\cite{Englert:2016hvy}. 

\begin{figure}[htb!]
    \centering
    \includegraphics[width=0.48\textwidth]{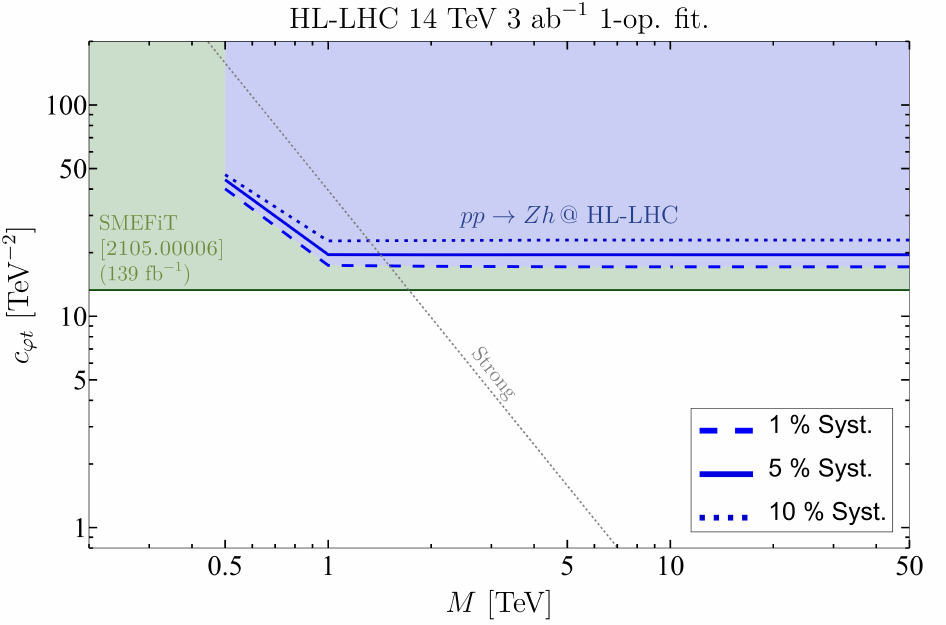}\hfill
    \includegraphics[width=0.48\textwidth]{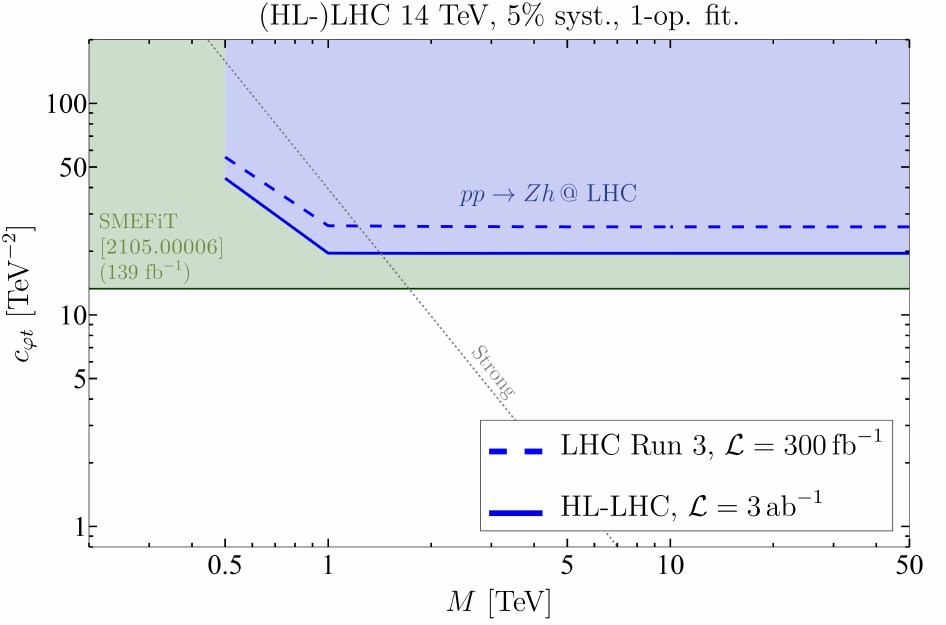}\vfill
    \includegraphics[width=0.48\textwidth]{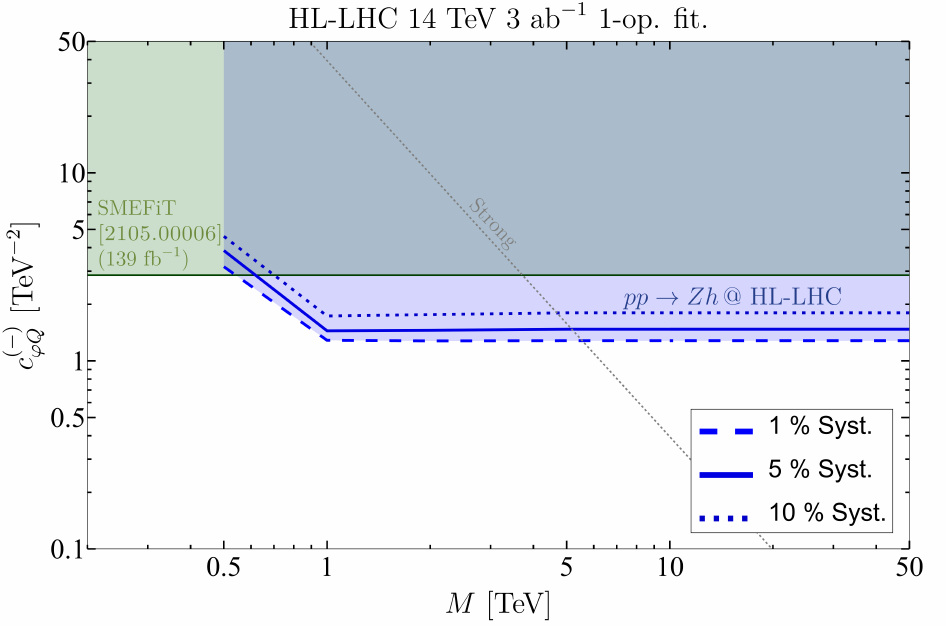}\hfill
    \includegraphics[width=0.48\textwidth]{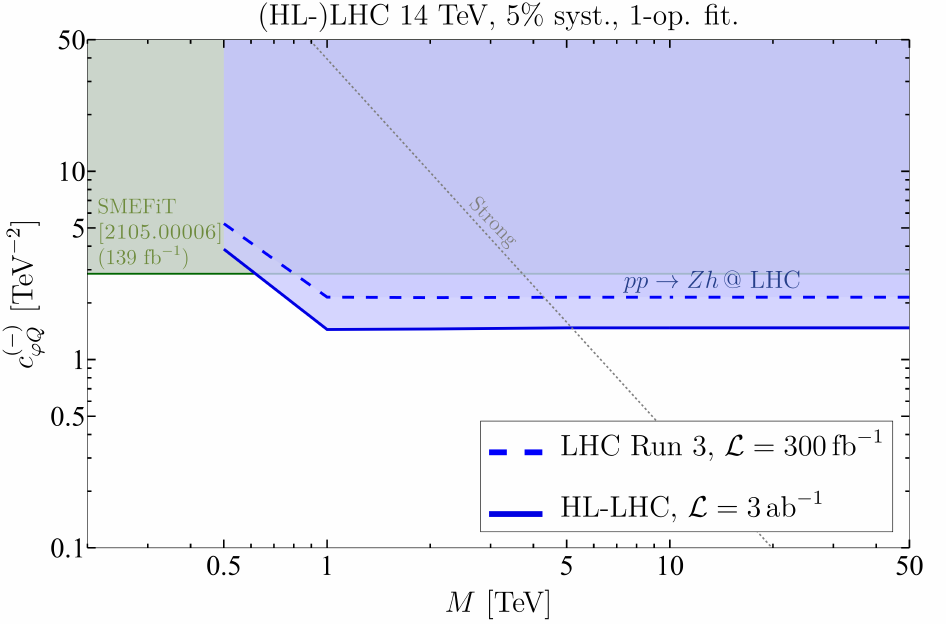}
    \caption{Projected $95\%$ C.L. bounds on $\cht$ (upper row) and $\chqTopMinus$ (lower row) at LHC and HL-LHC from a one-operator fit as a function of the maximal-invariant-mass cut $M$.  We also show the bound from the global fit to LHC data performed by the \code{SMEFiT} collaboration~\cite{Ethier:2021bye} with a dark green line. The dashed grey line shows the expected value of the WC in strongly-coupled NP scenarios, $c\sim 4\pi^2/M^2$. \textbf{Left column:} Projection for HL-LHC with different levels of systematic uncertainty. \textbf{Right column:} Projection for LHC and HL-LHC with a fixed $5\%$ syst. uncertainty and different integrated luminosities, $\lag$.}
    \label{fig:plot_bound_cpQ1vsM}
\end{figure}

The validity of the EFT description can be assessed by studying the dependence of the bounds on the maximal invariant mass of the $ZH$ system, $M$. We show it in Fig.~\ref{fig:plot_bound_cpQ1vsM} for the bounds on $\cht$ (upper panel) and $\chqTopMinus$ (lower panel). For bounds that are asymmetric around $0$, we plot the maximum absolute value. The maximal invariant mass $M$ acts as a proxy for the cutoff of the EFT, hence the bounds in tables~\ref{tab:bounds_summary_NFU_0plus2lep_HLLHC_kF2} and \ref{tab:bounds_summary_NFU_0plus2lep_LHCR3_kF2} are valid as long as the cutoff of the EFT is $\gtrsim\,1\TeV$, while they degrade significantly for lower cutoffs. In this figure, we also plot the current \code{SMEFiT} bound for reference. 

The left column of Fig.~\ref{fig:plot_bound_cpQ1vsM} shows the bounds at HL-LHC for different systematic uncertainties. The sizeable variation of the bound is a reflection of the relatively high level of backgrounds. The right column of the same figure shows the bounds for a fixed $5\%$ syst. uncertainty but with two different integrated luminosities, $\lag=300\,\text{fb}^{-1}$ and $\lag=3\,\text{ab}^{-1}$, which correspond to LHC Run 3 and HL-LHC respectively. As mentioned before, the bounds improve by $30-50\%$ from LHC Run 3 to HL-LHC due to the former being statistically limited. Further increases in luminosity yield smaller gains in sensitivity.

\section{Conclusions}
\label{sec:Conclusions}

In this paper, we have studied the gluon-initiated diboson production process in the SMEFT framework, limiting ourselves to dimension-6 operators and assuming a U$(2)_q\times$U$(3)_d\times$U$(2)_u$
$\times(\text{U}(1)_\ell\times\text{U}(1)_e)^3$ flavour symmetry. We identified all the operators that can modify the process in each of the 4 final states under consideration: $HH$, $ZH$, $ZZ$, and $WW$.
We computed the helicity amplitudes for $gg\to HH,\,ZH,\,ZZ,\, WW$ in the SMEFT with up to one insertion of the relevant dimension-6 operators analytically and studied their high-energy behaviour.

The presence of dimension-6 operators can produce a quadratic growth with energy even when the process remains loop-induced, for example $\Otg$ in $HH$ production. A similar growth is induced by $\Opg$, but in a different way since this operator generates a contact interaction between gluons and Higgses. Two additional operators, $\Oth$ and $\Odh$, modify SM vertices, even changing the Lorentz structure, and induce a milder logarithmic growth.

In the case of $ZH$ production, we find five operators which generate energy growth. Only one of them, $\Otg$, induces amplitudes that are linear in either $\sqrt{s}$ or $s$. The $\sqrt{s}$ behaviour arises for three different helicity configurations with a transversely polarised $Z$ boson, while the fastest growth appears when the $Z$ is longitudinally polarised and the incoming gluons have opposite helicities. The other $4$ operators, $\Opg$, $\Oht$, $\OhqTopMinus$, and $\Oth$, induce a logarithmic growth for a longitudinal $Z$. The first of them does it regardless of the gluon helicities, while the rest require same-helicity gluons. An exact cancellation of the triangle diagrams renders $ZH$ sensitive to only the linear combination $\chqTopMinus-\cht +\frac{\cth}{y_t}$. 

Double $Z$ production is the process with the largest number of growth-inducing dimension-6 operators. $\Otg$ generates amplitudes that behave $\propto s$ at high energies whenever both $Z$ bosons are longitudinal. If only one is longitudinally polarised, this same operator induces a linear growth with energy, and if both $Z$ bosons have transverse polarisations, their amplitudes grow logarithmically.  The only other operator capable of producing amplitudes that grow quadratically with energy is $\Opg$, but it enters only for the $(+,+,0,0)$  helicity configuration. Three operators, $\mcO_{\varphi B}$, $\mcO_{\varphi W}$, and $\mcO_{tZ}$, generate a logarithmic growth for amplitudes with only transversely polarised $Z$ bosons and the last two further require same-helicity gluons. The configuration with same-helicity gluons and longitudinally polarised $Z$ bosons also allows the operators $\Oth$, $\Oht$ and $\OhqTopMinus$ to generate logarithmically-growing amplitudes.

Only $\mcO_{\varphi W}$, $\mcO_{t W}$, $\Oth$, $\OhqTopt$ and $\Otg$ generate energy-growing amplitudes in $WW$ production. The first four of them generate logarithmic growths, with $\Oth$ and $\OhqTopt$ doing it only for longitudinally polarised $W$ bosons with same-helicity gluons, whilst $\mcO_{\varphi W}$ and $\mcO_{t W}$ require transversely polarised $W$ bosons. The amplitudes with $\Otg$ behave $\sim s$ in the high-energy limit when both $W$s are longitudinal,  while one longitudinal and one transverse $W$ makes them $\propto \sqrt{s}$ and the behaviour is logarithmic when all bosons are transverse.

Motivated by the analytical results and the absence of combined phenomenological analysis of the quark and gluon-initiated $ZH$ production, we perform a complete analysis of $ZH$ production at (HL-)LHC. We assessed how the flavour assumptions must be relaxed to make the addition of the gluon-fusion process meaningful and its effects on the reach of the quark-initiated channel. We also discussed in detail the possible effects of NLO QCD corrections to $gg\to ZH$ and approximately included them in our analysis.

Our results show that the flavour-universal bounds on 2-fermion operators constrained by $VH$ production can be safely translated to bounds on light-quark operators thanks to the small contribution of $b\bar b\to ZH$. Nevertheless, the $b\bar b$ initial-state contribution is large enough to provide competitive bounds on $\chqTopt$ and dominate the sensitivity to $\chqTopMinus$. The latter WC is the only one probed by both quark and gluon initial states, albeit the inclusion of the gluon channel does not improve the bound sizeably. Gluon-fusion $ZH$ production offers sensitivity to $\cht$ and $\cth$, otherwise inaccessible in diboson processes, with the former being competitive against current bounds. 

The interplay of the two initial states is not enough to constrain all directions in the space spanned by the heavy-quark operators, since $b\bar b \to ZH$ is sensitive to $2\chqTopt+\chqTopMinus$ and $gg\to ZH$ probes $\chqTopMinus-\cht +\frac{\cth}{y_t}$. The remaining flat directions could be lifted with the full inclusion of NLO QCD corrections to $gg\to ZH$, although with limited sensitivity. 

Several research directions remain open and deserve further exploration. 
The use of a differential distribution to distinguish better between quark- and gluon-initiated processes is an interesting possibility to explore. Going beyond $ZH$, the combination of various loop-induced channels, several of which exhibit energy-growing amplitudes for the same operators is an obvious next step. One can envision that a combination of the $ZH$, $ZZ$ and $WW$ processes could improve constraints on several of the coefficients discussed here and break the degeneracies which plague the $ZH$ channel. Tailored analyses would then be designed for the other processes. Eventually, this class of processes will be included fully in global EFT fits, where a large number of processes is considered to extract maximal information on the coefficients.

Finally, the study of dimension-8 contributions to diboson processes is a lightly trodden path that might be of special relevance to gluon-initiated processes. We have seen that only one dimension-6 operator can promote these processes from loop induced to tree-level and hence offer a steeper growth with energy. Several dimension-8 operators can generate $gg\to VV('),\,VH,\,HH$ at tree-level with a distinctive energy growth, as found recently via on-shell amplitude techniques~\cite{Liu:2023jbq}. We leave a detailed study of those operators for future work.

\section*{Acknowledgments}
We are particularly grateful to F. Bishara, P. Englert, C. Grojean and G. Panico for allowing us to reuse the simulations and analysis techniques from Ref.~\cite{Bishara:2022vsc} and to J. Gaunt for discussions on the method of regions and comments on the relevant parts of the manuscript. We thank H. El Faham, C. Severi, E. Salvioni, T. McKelvey, and C. Englert for useful discussions and C. Grojean and G. Panico for their comments on an early version of this manuscript.
A. R. and E. V. work is supported by the European Research Council (ERC) under the European Union’s Horizon 2020 research and innovation programme (Grant agreement No. 949451). A.R., M. T., and E.V. are supported by a Royal Society University Research Fellowship through grant URF/R1/201553. 
We thank DESY Theory Group for allowing us to use their computational resources for this project.

\appendix

\section{Origin of the squared logarithms}
\label{app:LogSquared}

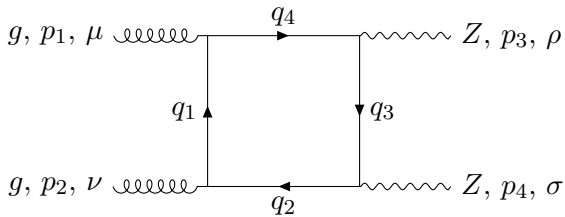
\begin{figure}[h!]
    \centering
\scalebox{1}{
\begin{tikzpicture}
\begin{feynman}
\vertex (i1) {$g$, $p_1$, $\mu$};
\vertex[below=2cm of i1] (i2) {$g$, $p_2$, $\nu$};
\vertex[right=2cm of i1] (l1);
\vertex[right=2cm of l1] (l2);
\vertex[below=2cm of l2] (l3);
\vertex[below=2cm of l1] (l4);
\vertex[right=1.2cm of l3] (f1) {$Z$, $p_4$, $\sigma$};
\vertex[right=1.2cm of l2] (f2) {$Z$, $p_3$, $\rho$};
\diagram* {{[edges={fermion, arrow size=1pt}]
(l1) --[edge label = $q_4$] (l2) -- [edge label = $q_3$] (l3) -- [edge label = $q_2$] (l4) -- [edge label = $q_1$] (l1),},
(i1) -- [gluon] (l1),
(i2) -- [gluon] (l4),
(l3) -- [photon] (f1),
(l2) -- [photon] (f2) };
\end{feynman}
\end{tikzpicture}}
    \caption{Box diagram for $gg\to ZZ$.}
    \label{fig:ZZDiag_log2_example}
\end{figure}

To demonstrate how to extract the leading logarithmic energy behaviour from the loop computation we discuss here as an example the  box diagram for $gg\to ZZ$ as shown in Fig.~\ref{fig:ZZDiag_log2_example}.  For simplicity we only consider SM-like Lorentz structure for all the couplings. We consider the possibility of the $ttZ$ couplings being shifted by dimension-6 operators to make our computation as general as possible. Additionally, we consider the axial $ttZ$ vertex although the procedure can be carried out for vector-like couplings in an identical manner. 

The helicity amplitude for this diagram is,
\begin{equation}
    \mcM^{h_1,h_2,h_3,h_4}_{\Box_1}= -\epsilon_\mu^{h_1}\left(p_1\right) \epsilon_\nu^{h_2}\left(p_2\right)\epsilon_\rho^{h_3}\left(p_3\right)^{*} \epsilon_\sigma^{h_4}\left(p_4\right)^{*}  g_s^2\,\, \tilde{c}_A^2 \,\delta^{ab} \, I_{AA}^{\mu\nu\rho\sigma}\left( p_i \right),  
\end{equation}
where $\epsilon_\alpha^{h}\left(p\right)$ is a polarization tensor, $\tilde{c}_A$ is the general axial $ttZ$ coupling and $a,\,b$ are the gluon color indices.  The integral can be written as,
\begin{align}
    I_{AA}^{\mu\nu\rho\sigma}\left( p_i \right) &= \int\frac{\text{Tr}\left[\gamma^\mu(\slashed{q}_1+m_t)\gamma^\nu(\slashed{q}_2+m_t)\gamma^\sigma\gamma^5(\slashed{q}_3+m_t)\gamma^\rho\gamma^5(\slashed{q}_4+m_t) \right]}{\prod_{i=1}^{4} [q_i^2-m_t^2+i \varepsilon]} \frac{d^4 k}{(2\pi)^4}\\&=\int\frac{\text{Tr}_{1}^{\mu\nu\rho\sigma}(q_i,m_t)}{D_{\Box}(q_i, m_t)} \frac{d^4 k}{(2\pi)^4},
\end{align}
where the internal momenta $q_i$ are linear combinations of the integral variable and the external momenta. Valid choices for these internal momenta are related among them by shifts of the integration variable. 

The part of the integral that generates the $\log^2$ behaviour can be extracted by applying the method of regions \cite{Beneke:1997zp,Becher:2014oda}. One can decompose all the internal and external momenta in the $(\hat{n}_+$, $\hat{n}_-,\hat{n}_\perp)$ basis. We are interested in the soft region, hence we assume that the integration momenta follows the power counting $k\sim(\lambda^2,\lambda^2,\lambda^2)\sqrt{s}$ in the aforementioned basis, with $\lambda\ll 1$, and that all external momenta are $\mcO(1)$. We assign the internal momenta as,
\begin{align}
q_1&=k,\nonumber\\
q_2&=k-p_2,\nonumber\\
q_3&=k+p_1-p_3,\nonumber\\
q_4&=k+p_1.    
\end{align}
We comment later on other possible assignments of the internal momenta.

By expanding each of the factors in $D_{\Box}\left(q_i\right)$ to leading power in $\lambda$, we find that one of them is hard, i.e. $\sim s,\,t$, and hence approximately independent of $k$, $(k+p_1-p_3)^2\simeq (p_1-p_3)^2=t$. Thus, the integral simplifies to,
\begin{equation}
    I_{AA}^{\mu\nu\rho\sigma}\left( p_i \right) \simeq \frac{1}{t}\int\frac{\text{Tr}_{1}^{\mu\nu\rho\sigma}(q_i,m_t)}{[k^2-m_t^2+i\varepsilon][2 k\cdot p_1 - m_t^2 +i\varepsilon][-2 k\cdot p_2 - m_t^2 +i\varepsilon]} \frac{d^4 k}{(2\pi)^4},
    \label{eq:integral_logs}
\end{equation}
where we have used that $k\cdot p_i\gg k^2$ in the soft region, $|t|=|(p_1-p_3)^2|\gg  m_t^2$ in the high-energy limit and the on-shell condition of the external gluons. 

The squared logarithm in the amplitude is associated with a scaleless integral in the region $m_t^2,\,m_{Z}^2\ll k^2\ll s$. By dimensional analysis,  this has to come from the term of the trace that is quadratic in $m_t$. Hence, we keep only that term in Tr$_1\left(q_i,m_t\right)$. We can then drop all the masses from the denominator and reintroduce them later as IR regulators of the integral which would otherwise diverge. The simplified integral is,
\begin{equation}
    I_{AA}^{\mu\nu\rho\sigma}\left( p_i \right) \simeq \frac{m_t^2}{t}\int\frac{\text{tr}_{1}^{\mu\nu\rho\sigma}(q_i)}{[k^2+i\varepsilon][2 k\cdot p_1 +i\varepsilon][-2 k\cdot p_2 +i\varepsilon]} \frac{d^4 k}{(2\pi)^4},
\end{equation}
where we have defined,
\begin{equation}
    \text{tr}_{1}^{\mu\nu\rho\sigma}(q_i) = \frac{1}{2} \frac{\partial^2 \text{Tr}_{1}^{\mu\nu\rho\sigma}}{\partial m_t^2}\Big|_{m_t=0}.
\end{equation}
$\text{tr}_{1}^{\mu\nu\rho\sigma}(q_i)$ is a polynomial in $k$ and only the constant term will yield a non-vanishing integral that can produce a logarithm. Therefore, we define,
\begin{equation}
    \tau_{1}^{\mu\nu\rho\sigma}\left(p_i\right) = \text{tr}_{1}^{\mu\nu\rho\sigma}(q_i)|_{k=0}
\end{equation}
and arrive to,
\begin{equation}
    I_{AA}^{\mu\nu\rho\sigma}\left( p_i \right) \simeq \frac{m_t}{t(2\pi)^4}\tau_{1}^{\mu\nu\rho\sigma}\left(p_i\right)\,\,\mathcal{F}\left(p_1,p_2\right),
\end{equation}
where
\begin{equation}
    \mathcal{F}\left(a_\mu,b_\nu\right) = \int\frac{1}{[k^2+i\varepsilon][2 k\cdot a +i\varepsilon][-2 k\cdot b +i\varepsilon]} d^4 k.
\end{equation}
This integral can be computed by residues with a rapidity regulator and the result is,
\begin{equation}
    \mathcal{F}\left(a_\mu,b_\nu\right) = - \frac{i\pi^2}{8 a\cdot b}\left[ \log^2\left(\frac{s}{m_t^2}\right)-2\pi i \log\left(\frac{s}{m_t^2}\right) \right]+...\,.
\end{equation}
To compute this integral we  employ $\sqrt{s}$ and $m_t$ as the integration limits acting as regulators. Introducing these hard cutoffs is consistent since we only look for the leading squared logarithm piece in the amplitude.  
When $a=p_1$ and $b=p_2$, the prefactor can be simplified to $\frac{i\pi^2}{4 s}$. The behaviour shown by the different helicity configurations is determined upon the contraction of $\tau_{1}^{\mu\nu\rho\sigma}\left(p_i\right)$ with the polarization tensors. 

One then must consider the possible contribution of having chosen a different internal line to be soft, which might also generate a squared logarithm. Each of these possibilities represents a different region of the total external particle phase space, hence they must be summed to obtain the total amplitude. 

This procedure has to be repeated for all the one-loop diagrams either in the SM or when inserting one dimension-6 operator. There are no major differences when it is applied to non-planar diagrams. When one of the vertices is generated by a dipole operator, one needs to take the piece of Tr$_1\left(q_i,m_t\right)$ that is linear on $m_t$ and change accordingly the definition of tr$_{1}^{\mu\nu\rho\sigma}(q_i)$. In the case of triangle diagrams, none of the internal propagators will be hard, thus rendering an integral like Eq.~\eqref{eq:integral_logs} straightforwardly.

For the SM-like example, we found that the only effective contribution is from the soft region of the loop line between $p_1$ and $p_2$ in planar diagrams. Taking as soft other loop lines either gives a vanishing result or are cancelled exactly by the contributions of the non-planar loops. Operators that modify the Lorentz structure of the loop disturb these cancellations and make all contributions relevant in general.

\section{High energy behaviour of helicity amplitudes}
\label{app:helicity_amp}

We present here the high energy behaviour of both growing and decaying helicity amplitudes in $gg \rightarrow HH,\, ZH,\, ZZ$. We follow the same conventions as in the main text with $\lambda_{g_1}, \lambda_{g_2}, \lambda_{H_1/Z_1}, \lambda_{H_2/Z_2}$ representing the polarisation of the two incoming gluons and the two outgoing Higgs or $Z$ bosons respectively. Constant energy behaviours are given by $s^0$ and when a helicity amplitude is equal to $0$, this is denoted by $``-"$. We give the dependence on dimensionful quantities such as the masses and the Higgs vacuum expectation value and we neglect overall numerical quantities. In $gg \rightarrow ZZ$, for the light quark operators, the expressions for amplitudes that decrease with energy were obtained by numerical fits of the amplitudes and for those operators we only give the energy dependence of the helicity amplitudes. In the case of top quark operators, numerical fits were used for amplitudes that decrease faster than $1/\sqrt{s}$. All the other helicity amplitudes were obtained analytically with Mathematica following the procedure described in the main text. Finally, the dependence on the WCs and $\Lambda^2$ is implicit.

\begin{table}[h!]
    \centering
    \resizebox{\textwidth}{!}{
    \begin{tabular}{|c|c|c|c|c|c|}
    \hline
         $\lambda_{g_1}, \lambda_{g_2}, \lambda_{H_1}, \lambda_{H_2}$& $\Oth$ & $\Otg$&$\mathcal{O}_{d \varphi}$&$\mathcal{O}_{\varphi}$&$\Opg$ \\
         \hline
         
         $+, +, 0, 0$&$m_t v\, \logsmtsq $&
         
         $s\,\frac{ m_t}{v} \,\log \Big(\frac{s}{\mu_{EFT}^2}\Big)$&
         $m_t^2\,\logsmtsq$&
         $\frac{m_t^2\, v^2}{s}\, \logsmtsq $& $s$\\
         
         $+, -, 0, 0$&$m_t\,v$&$s \frac{m_t}{v}$&$m_t^2$&$-$&$-$\\
         \hline
    \end{tabular}}
    \caption{High energy behaviour of the $gg \rightarrow HH$ helicity amplitudes modified by SMEFT operators. 
    }
    \label{HHAppendixTable1}
\end{table}

\begin{table}[h]
    \centering
    \resizebox{\textwidth}{!}{
    \begin{tabular}{|c|c|c|c|c|c|}
    \hline
         \small{$\lambda_{g_1}, \lambda_{g_2}, \lambda_{H}, \lambda_{Z}$ }& $\Otg$&$\Opg$&  $\Oht$ & $\OhqTopMinus$ &$\Oth$\\
         \hline
         $+,+, 0, +$&
          $\sqrt{s}\,m_t \,\logsmt$&
          $\frac{m_t^2 \, v }{\sqrt{s}}\,\logsmtsq  $&
          $\frac{m_t^2 \, v}{\sqrt{s}} \,\logsmtsq $&
         $\frac{\,m_t^2 \, v}{\sqrt{s}}\,\logsmtsq   $ 
         &$\frac{m_t \, v^2 }{\sqrt{s}}\,\logsmtsq  $\\
         
         $+, +, 0, -$&
         $\sqrt{s}\,m_t \,\,\logsmtsq$
         &$\frac{m_t^2 m_H^2 \, v }{s^{3/2}}\, \logsmtsq$&
         $\frac{m_t^2\, m_H^2\,v}{s^{3/2}}\,\logsmtsq$
         &$\frac{m_t^2 m_H^2 \, v}{s^{3/2}}\, \logsmtsq  $
         & $\frac{m_t m_H^2 \, v^2}{s^{3/2}}\,\logsmtsq$\\
          
         $+ ,+, 0, 0$&
         $\frac{m_t\,v^2}{m_Z} \, \logsmtsq  $&
         $\frac{m_t^2 \, v }{m_Z}\,\logsmtsq  $&
         $ \frac{m_t^2 \,v}{m_Z}\,\logsmtsq  $
         &$ \frac{m_t^2 \,v }{m_Z}\,\logsmtsq  $
         &$\frac{m_t \, v^2}{m_Z}\,\logsmtsq  $\\

         $+, -, 0, +$&
          $\sqrt{s} \,m_t$&
           $\frac{m_t^2 \, v }{\sqrt{s}}\,\logsmtsq $
           &$\frac{ m_t^2\,v}{\sqrt{s}}\,\logsmtsq$
          &$\frac{ m_t^2\, v }{\sqrt{s}}\,\logsmtsq $
          &$\frac{m_t\, v^2}{\sqrt{s}}\,\logsmtsq  $\\

         $+, -, 0, 0$&
         $s\, \frac{m_t}{m_Z}$
         &$\frac{m_t^2 \, v}{m_Z}\,\logsmtsq $
         &$\frac{m_t^2 \, v}{ m_Z}$
         &$\frac{m_t^2 \, v}{ m_Z}$
         &$\frac{m_t \, v^2 }{m_Z}$\\
         
         \hline
    \end{tabular}}
    \caption{High energy behaviour of the $gg \rightarrow ZH$ helicity amplitudes modified by SMEFT operators. }
    \label{ZHAppendixTable1}
\end{table}

\begin{table}[h]
    \centering
    \resizebox{\textwidth}{!}{
    \begin{tabular}{|c|c|c|c|c|c|}
    \hline
         {\small $\lambda_{g_1}, \lambda_{g_2}, \lambda_{Z_1}, \lambda_{Z_2}$ } & $\Otg$ &  $\mathcal{O}_{tZ}$&  $\Oth$ & $\Oht$ &$\OhqTopMinus$  \\
         \hline
         $+,+, +, +$& 
        $m_t\,v\,
        \log \Big(\frac{\mu_{EFT}^2}{m_t^2}\Big)$&
        $m_t \, v$& 
        $\frac{m_t\, v^3 }{s}\,\logsmtsq  $& $v^2$
        & $v^2$
        \\
        
         $+, +, +, -$&
        $m_t \,v \,\logsmtsq $ & 
        $m_t \, v \,\logsmtsq $ 
        & $-$ & $v^2$&$v^2$
        \\
          
         $+, +, +, 0$&
         $\sqrt{s}\,\frac{m_t\, v}{m_Z}\,\logsmt$& 
         $\frac{m_t\,m_Z\, v}{\sqrt{s}}\,  \logsmtsq  $&$-$
         & $\frac{m_t^2 \, v^2 }{m_Z \sqrt{s}}\, \logsmtsq $&
         $\frac{m_t^2\,  v^2}{m_Z\,\sqrt{s}}\,  \logsmtsq $
         \\
                  
         $+, +, - ,-$& 
         $m_t \, v \,\logsmtsq$&
         $m_t \,v\,\logsmtsq $
         &$\frac{m_t\, v^3 }{s}\,\logsmtsq $& $v^2$ &$v^2$
         \\
         
         $+, +, -, 0$&
         $\sqrt{s}\,\frac{m_t\,v}{m_Z} \, \logsmtsq $&
         $\frac{m_t\,m_Z\,v}{\sqrt{s}}\, \logsmtsq $
         & $-$&$\frac{m_t^2\,m_Z\,v^2}{s^{3/2}}\,  \logsmtsq $&
         $\frac{m_t^2\,m_Z\,v^2}{s^{3/2}}\,  \logsmtsq $
         \\
         
        $+, +, 0, 0$&
        $s \,\frac{m_t\, v}{m_Z^2} \,\log \Big(\frac{\mu_{EFT}^2}{s}\Big) $& 
        $\frac{m_t\,m_Z^2\,v}{s}\,  \logsmtsq $
        & $\frac{m_t\, v^3}{m_Z^2}\,\logsmtsq $&  
        $\frac{m_t^2\, v^2}{m_Z^2} \,\logsmtsq $&
        $\frac{m_t^2\,v^2}{m_Z^2} \,\logsmtsq $
         \\
         
        $+, -, -, +$ &
        $m_t v$& 
        $m_t\, v$&$-$&$v^2$&$v^2$
          \\
         
         $+, -, -, -$&
         $m_t\,v\,\logsmtsq  $&  
         $m_t\,v \,\logsmtsq $& $-$&$v^2$&$v^2$
         \\
         
         $+, -, - ,0$&
         $ \sqrt{s} \,\frac{m_t\, v}{m_Z}$&
         $\frac{m_t\,m_Z\, v}{\sqrt{s}}\,  \logsmtsq $
         &$-$&$\frac{m_t^2 \, v^2}{m_Z \sqrt{s}} \, \logsmtsq $&
         $\frac{m_t^2\,  v^2}{m_Z\,\sqrt{s}}\, \logsmtsq $
       
         \\
         
          $+, -, 0 ,0$&
          $s \,\frac{m_t\, v}{m_Z^2}$& 
          $\frac{m_t\,m_Z^2\,v}{s}\,  \logsmtsq$
          &$-$&$\frac{m_t^2\,v^2}{m_Z^2}$
          &$\frac{m_t^2\,v^2}{m_Z^2}$
         
        \\
         \hline
    \end{tabular}}
    \caption{High energy behaviour of the  $gg \rightarrow ZZ$ helicity amplitudes modified by top operators.}
    \label{ZZAppendixTable1}
\end{table}

\begin{table}[h]
    \centering
    \resizebox{0.65 \textwidth}{!}{
    \begin{tabular}{|c|c|c|c|}
    \hline
        $\lambda_{g_1}, \lambda_{g_2}, \lambda_{Z_1}, \lambda_{Z_2}$ &  $\Opg$ & $\mathcal{O}_{\varphi B}$&$\mathcal{O}_{\varphi W}$ \\
         \hline
         $+,+, +, +$ & $v^2$ &
         
             $m_t^2 \, \logsmtsq  $&
             
             $m_t^2 \, \logsmtsq  $
        \\
         $+, +, +, -$& $-$ & $-$ & $-$
        \\
         $+, +, +, 0$&  $-$ & $-$ & $-$
         \\

         $+, +, - ,-$&$v^2$&
         
         $m_t^2 \, \logsmtsq  $&
         
         $m_t^2 \, \logsmtsq  $

         \\
         
         $+, +, -, 0$& $-$& $-$& $-$
         
         \\
        $+, +, 0, 0$& $s\,\frac{v^2}{m_Z^2}$&
        
       $\frac{m_t^2\, m_Z^2}{s}\,\logsmtsq  $
        &$\frac{m_t^2\, m_Z^2}{s}\,\logsmtsq  $

         \\
         
        $+, -, -, +$&$-$&$-$&$-$
        
          \\
         
         $+, -, -, -$& $-$&$-$&$-$
         
         \\
         
         $+, -, - ,0$&$-$&$-$&$-$
         
         \\
         
          $+, -, 0 ,0$&$-$&$-$&$-$
        \\
         \hline
    \end{tabular}}
    \caption{High energy behaviour of the $gg \rightarrow ZZ$ helicity amplitudes in the presence of purely bosonic operators.}
    \label{ZZAppendixTable2}
\end{table}

\begin{table}[h]
    \centering
    \resizebox{\textwidth}{!}{
    \begin{tabular}{|c|c|c|c|c|c|}
    \hline
       \small{ $\lambda_{g_1}, \lambda_{g_2}, \lambda_{Z_1}, \lambda_{Z_2}$} &  \ensuremath{\mathcal{O}_{\varphi q,i}^{(-)}} &$\Opp{\varphi q_i}{\sss(3)}$& $\Opp{\varphi Q}{\sss(3)}$&  $\Ohu$ &$\Ohd$ \\
         \hline
         $+,+, +, +$& $s^0$& $s^0$&$s^0$& $s^0$& $s^0$
   
        \\
         $+, +, +, -$& 
         $s^0$& $s^0$& $s^0$& $s^0$& $s^0$
   
        \\
          
         $+, +, +, 0$& 
         $\frac{1}{\sqrt{s}}\,  \logsmzsq $&
         $\frac{1}{\sqrt{s}}\,  \logsmzsq $&
         $\frac{1}{\sqrt{s}}\,  \logsmzsq $
         & 
         $\frac{1}{\sqrt{s}}\, \logsmzsq $& 
         $\frac{1}{\sqrt{s}}\, \logsmzsq $
       
         \\
        
         $+, +, - ,-$& $s^0$&$s^0$&$s^0$& $s^0$& $s^0$

         \\
         
         $+, +, -, 0$&
         $\frac{1}{s^{3/2}}\,  \logsmzsq $&
         $\frac{1}{s^{3/2}}\,  \logsmzsq $&
         $\frac{1}{s^{3/2}}\,  \logsmzsq $&
         $\frac{1}{s^{3/2}}\, \logsmzsq $
         &$\frac{1}{s^{3/2}}\, \logsmzsq $

         \\
        $+, +, 0, 0$& 
       $\frac{1}{s}\,  \logsmzsq $&
       $\frac{1}{s}\,  \logsmzsq $&
       $\frac{1}{s}\,  \logsmzsq $& 
       $\frac{1}{s}\,  \logsmzsq $
       & $\frac{1}{s}\,  \logsmzsq $

         \\
         
        $+, -, -, +$& $s^0$& $s^0$&$s^0$& $s^0$& $s^0$
        
          \\
         
         $+, -, -, -$& $s^0$& $s^0$&$s^0$& $s^0$& $s^0$
         
         \\
         
         $+, -, - ,0$&
        $\frac{1}{\sqrt{s}}\,  \logsmzsq $&
        $\frac{1}{\sqrt{s}}\,  \logsmzsq $&
         $\frac{1}{\sqrt{s}}\,\logsmzsq$&
        $\frac{1}{\sqrt{s}}\,  \logsmzsq $&
        $\frac{1}{\sqrt{s}}\,  \logsmzsq $
         
         \\
         
          $+, -, 0 ,0$& 
          $\frac{1}{s}\,  \logsmzsq $&
          $\frac{1}{s}\,  \logsmzsq $&
          $\frac{1}{s}\,\logsmzsq$& 
          $\frac{1}{s}\,  \logsmzsq $& 
          $\frac{1}{s}\,  \logsmzsq $
        \\
         \hline
    \end{tabular}}
    \caption{High energy behaviour of the  $gg \rightarrow ZZ$ helicity amplitudes modified by light quark operators.}
    \label{ZZAppendixTable4}
\end{table}

\section{Simulation details}
\label{app:Sim_details}

In this work, we only consider the 0- and 2-lepton channels of $(W/Z)H$ production, which correspond to the $Z\to\nu\bar\nu$ and $Z\to\ell^{+}\ell^{-}$ decay channels respectively. The signal process in each of the channels is mostly composed of $ZH$ production, with the $Z$ decaying to the appropriate leptons. The 0-lepton channel also receives a significant signal contribution from $WH$ production with $W\to\tau\nu_{\tau}$ and missing products of the tau decay~\cite{ATLAS:2020fcp,ATLAS:2020jwz,Bishara:2022vsc} The production of $Zb\bar b$ is the main background process in both channels and, in fact, is the only relevant one in the 2-lepton channel. The 0-lepton channel background is also composed of $t\bar t$ and $Wb\bar b$ production.

The event generation was performed using \verb|MadGraph5_aMC@NLO v.2.7.3| ~\cite{Alwall:2014hca} and the parton shower was performed with \verb|Pythia 8.2|~\cite{Sjostrand:2014zea}. We used the \verb|SMEFTatNLO v.1.0.3| UFO model for all processes~\cite{Degrande:2020evl}.
The simulation of quark-initiated signal and background processes was performed as described in Ref.~\cite{Bishara:2022vsc}.
This means that, since we limit ourselves to (HL-)LHC, all the relevant processes were simulated at NLO in QCD, except for the $t\bar t$ background process. The latter was simulated at LO with one additional hard jet to account for most of the NLO QCD corrections. Additionally, the quark-initiated part of the signal process includes NLO EW corrections via a k-factor. 
In the following, we detail the generation of the gluon-initiated processes that were added in this work.

\subsection{LO generation}
\label{app:Sim_details_LO}

We generated the signal processes $gg\to \bar{\nu}\nu H$ and $gg \to \ell^{+}\ell^{-} H$ at LO in QCD. The efficiency with which the generated events pass the selection cuts was improved via the application of generation-level cuts and a binning in $p_{T}^{Z}$. These cuts and bins are detailed in Table~\ref{tab:generation_cuts}. 

\begin{table}[htb]
\begin{centering}
\centering\renewcommand*{\arraystretch}{1.5}
\begin{tabular}{|c | m{2.5cm} | m{2.5cm} |}
\hline
 & \centering{$Z\rightarrow\nu\bar{\nu}$}  & \centering{$Z\rightarrow\ell^+ \ell^-$} \tabularnewline
\hline 
$p_{T,\min}^{\ell}$ {[}GeV{]} & \centering{-} &  \centering{$7$}  \tabularnewline
$|\eta_{max}^{\ell}|$ & \centering{-} &  \centering{$2.8$}  \tabularnewline
$p_{T}^{V}$ & \multicolumn{2}{c|}{$\{0,\,200,\,400,\,600,\,800,\,1200,\,\infty\}$} \tabularnewline
\hline
\end{tabular}
\par\end{centering}
\caption{Parton level generation cuts for $gg\to ZH$ at $13\,\mathrm{TeV}$ at LO.}
\label{tab:generation_cuts}
\end{table}

\subsection{$0+1$ jet merged generation}
\label{app:Sim_details_0plus1jet}

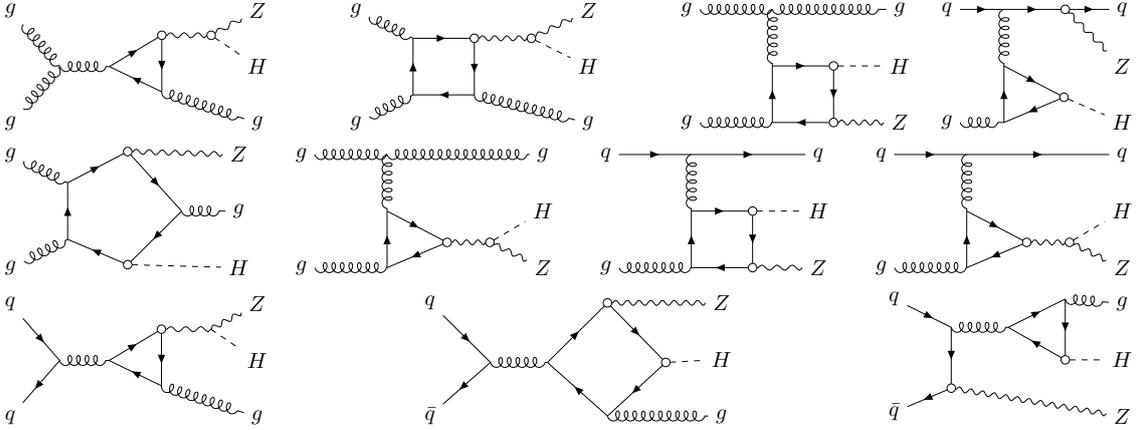
\begin{figure}[h!]
    \centering
\scalebox{0.75}{        
\begin{tikzpicture}
\begin{feynman}
\vertex (a1) {$g$};
\vertex[below right =1cm and 0.866cm of a1] (v1);
\vertex[below=2cm of a1] (a2) {$g$};
\vertex[right=0.866cm of v1] (l1);
\vertex[above right=0.5cm and 0.866cm of l1, empty dot] (l2) {};
\vertex[below=1cm of l2] (l3);
\vertex[right=0.866cm of l2, empty dot] (v2) {};
\vertex[right= 4.332cm of a1] (f1) {$Z$};
\vertex[below= 1cm of f1] (f2) {$H$};
\vertex[below= 1cm of f2] (f3) {$g$};
\diagram*{{[edges={fermion, arrow size=1pt}]
(l1) -- (l2) -- (l3) -- (l1),},
(a1) -- [gluon] (v1),
(a2) -- [gluon] (v1) -- [gluon] (l1),
(l2) -- [photon] (v2) -- [photon] (f1),
(v2) -- [scalar] (f2),
(l3) -- [gluon] (f3) };
\end{feynman}
\end{tikzpicture}}
\hfill
\scalebox{0.75}{
\begin{tikzpicture}
\begin{feynman}
\vertex (a1) {$g$};
\vertex[below=2cm of a1] (a2) {$g$};
\vertex[below right= 0.5cm and 1cm of a1] (l1);
\vertex[right=1cm of l1, empty dot] (l2) {};
\vertex[below=1cm of l2] (l3);
\vertex[below=1cm of l1] (l4);
\vertex[right=1cm of l2, empty dot] (v2) {};
\vertex[above right= 0.5cm and 1cm of v2] (f1) {$Z$};
\vertex[below = 1cm of f1] (f2) {$H$};
\vertex[below = 2cm of f1] (f3) {$g$};
\diagram* {{[edges={fermion, arrow size=1pt}]
(l1) -- (l2) -- (l3) -- (l4) -- (l1)},
(a1) -- [gluon] (l1),
(a2) -- [gluon] (l4),
(l2) -- [photon] (v2) -- [photon] (f1),
(v2) -- [scalar] (f2),
(l3) -- [gluon] (f3)};
\end{feynman}
\end{tikzpicture}}
\hfill
\scalebox{0.75}{
\begin{tikzpicture}
\begin{feynman}
\vertex (i1) {$g$};
\vertex[below=2cm of i1] (i2) {$g$};
\vertex[right=1.5cm of a1] (v1);
\vertex[below=1cm of v1] (l1);
\vertex[right=1cm of l1,empty dot] (l2) {};
\vertex[below=1cm of l2,empty dot] (l3) {};
\vertex[below=1cm of l1] (l4);
\vertex[right=2cm of v1] (f3) {$g$};
\vertex[below=2cm of f3] (f1) {$Z$};
\vertex[below=1cm of f3] (f2) {$H$};
\diagram* {{[edges={fermion, arrow size=1pt}]
(l1) -- (l2) -- (l3) -- (l4) -- (l1),},
(i1) -- [gluon] (v1) -- [gluon] (f3),
(v1) -- [gluon] (l1),
(i2) -- [gluon] (l4),
(l3) -- [photon] (f1),
(l2) -- [scalar] (f2) };
\end{feynman}
\end{tikzpicture}}
\scalebox{0.75}{
\begin{tikzpicture}
\begin{feynman}
\vertex (i1) {$q$};
\vertex[below=2cm of i1] (i2) {$g$};
\vertex[right=1cm of i1] (v1);
\vertex[below=1cm of v1] (l1);
\vertex[below right=0.5cm and 1cm of l1, empty dot] (l2) {};
\vertex[below=1cm of l1] (l3);
\vertex[right=1cm of v1, empty dot] (v2) {};
\vertex[right=1cm of v2] (f3) {$q$};
\vertex[below=1cm of f3] (f1) {$Z$};
\vertex[below=2cm of f3] (f2) {$H$};
\diagram* {{[edges={fermion, arrow size=1pt}]
(l1) -- (l2) -- (l3) -- (l1),},
(i1) -- [fermion, arrow size=1pt] (v1) -- [fermion, arrow size=1pt] (v2)-- [fermion, arrow size=1pt] (f3),
(v1) -- [gluon] (l1),
(i2) -- [gluon] (l3),
(v2) -- [photon] (f1),
(l2) -- [scalar] (f2) };
\end{feynman}
\end{tikzpicture}}
\hfill
\scalebox{0.75}{
\begin{tikzpicture}
\begin{feynman}
\vertex (i1) {$g$};
\vertex[below=2cm of i1] (i2) {$g$};
\vertex[below right=0.5cm and 1cm of a1] (l1);
\vertex[above right=0.5cm and 1cm of l1, empty dot] (l2) {};
\vertex[below right=0.5cm and 2cm of l1] (l3);
\vertex[below=2cm of l2, empty dot] (l4) {};
\vertex[below=1cm of l1] (l5);
\vertex[right=4cm of i1] (f1) {$Z$};
\vertex[below=1cm of f1] (f2) {$g$};
\vertex[below=1cm of f2] (f3) {$H$};
\diagram* {{[edges={fermion, arrow size=1pt}]
(l1) -- (l2) -- (l3) -- (l4) -- (l5) -- (l1),},
(i1) -- [gluon] (l1),
(i2) -- [gluon] (l5),
(l3) -- [gluon] (f2),
(l2) -- [photon] (f1),
(l4) -- [scalar] (f3) };
\end{feynman}
\end{tikzpicture}}
\hfill
\scalebox{0.75}{
\begin{tikzpicture}
\begin{feynman}
\vertex (i1) {$g$};
\vertex[below=2cm of i1] (i2) {$g$};
\vertex[right=1.5cm of i1] (v1);
\vertex[below=1cm of v1] (l1);
\vertex[below right=0.5cm and 1cm of l1, empty dot] (l2) {};
\vertex[below=1cm of l1] (l3);
\vertex[right=0.75cm of l2, empty dot] (v2) {};
\vertex[right=2.5cm of v1](f3){$g$};
\vertex[below=2cm of f3] (f1) {$Z$};
\vertex[below=1cm of f3] (f2) {$H$};
\diagram* {{[edges={fermion, arrow size=1pt}]
(l1) -- (l2) -- (l3) -- (l1),},
(i1) -- [gluon] (v1) -- [gluon] (f3),
(v1) -- [gluon] (l1),
(i2) -- [gluon] (l3),
(l2) -- [photon] (v2),
(v2) -- [photon] (f1),
(v2) -- [scalar] (f2)};
\end{feynman}
\end{tikzpicture}}
\hfill
\scalebox{0.75}{
\begin{tikzpicture}
\begin{feynman}
\vertex (i1) {$q$};
\vertex[below=2cm of i1] (i2) {$g$};
\vertex[right=1.5cm of a1] (v1);
\vertex[below=1cm of v1] (l1);
\vertex[right=1cm of l1, empty dot] (l2) {};
\vertex[below=1cm of l2, empty dot] (l3) {};
\vertex[below=1cm of l1] (l4);
\vertex[right=2cm of v1](f3){$q$};
\vertex[below=2cm of f3] (f1) {$Z$};
\vertex[below=1cm of f3] (f2) {$H$};
\diagram* {{[edges={fermion, arrow size=1pt}]
(l1) -- (l2) -- (l3) -- (l4) -- (l1),},
(i1) -- [fermion, arrow size=1pt] (v1) -- [fermion, arrow size=1pt] (f3),
(v1) -- [gluon] (l1),
(i2) -- [gluon] (l4),
(l3) -- [photon] (f1),
(l2) -- [scalar] (f2) };
\end{feynman}
\end{tikzpicture}}
\hfill
\scalebox{0.75}{
\begin{tikzpicture}
\begin{feynman}
\vertex (i1) {$q$};
\vertex[below=2cm of i1] (i2) {$g$};
\vertex[right=1.5cm of i1] (v1);
\vertex[below=1cm of v1] (l1);
\vertex[below right=0.5cm and 1cm of l1, empty dot] (l2) {};
\vertex[below=1cm of l1] (l3);
\vertex[right=0.75cm of l2, empty dot] (v2) {};
\vertex[right=2.5cm of v1](f3) {$q$};
\vertex[below=2cm of f3] (f1) {$Z$};
\vertex[below=1cm of f3] (f2) {$H$};
\diagram* {{[edges={fermion, arrow size=1pt}]
(l1) -- (l2) -- (l3) -- (l1),},
(i1) -- [fermion, arrow size=1pt] (v1) -- [fermion, arrow size=1pt] (f3),
(v1) -- [gluon] (l1),
(i2) -- [gluon] (l3),
(l2) -- [photon] (v2),
(v2) -- [photon] (f1),
(v2) -- [scalar] (f2) };
\end{feynman}
\end{tikzpicture}}
\hfill
\scalebox{0.75}{        
\begin{tikzpicture}
\begin{feynman}
\vertex (a1) {$q$};
\vertex[below right =1cm and 0.866cm of a1] (v1);
\vertex[below=2cm of a1] (a2) {$q$};
\vertex[right=0.866cm of v1] (l1);
\vertex[above right=0.5cm and 0.866cm of l1, empty dot] (l2) {};
\vertex[below=1cm of l2] (l3);
\vertex[right=0.866cm of l2] (v2);
\vertex[right= 4.332cm of a1] (f1) {$Z$};
\vertex[below= 1cm of f1] (f2) {$H$};
\vertex[below= 1cm of f2] (f3) {$g$};
\diagram*{{[edges={fermion, arrow size=1pt}]
(l1) -- (l2) -- (l3) -- (l1),},
(a1) -- [fermion, arrow size = 1pt] (v1),
(a2) -- [anti fermion, arrow size = 1pt] (v1) -- [gluon] (l1),
(l2) -- [photon] (v2) -- [photon] (f1),
(v2) -- [scalar] (f2),
(l3) -- [gluon] (f3) };
\end{feynman}
\end{tikzpicture}}
\hfill
\scalebox{0.75}{
\begin{tikzpicture}
\begin{feynman}
\vertex (a1) {$q$};
\vertex[below right = of a1] (v1);
\vertex[below=2cm of a1] (a2) {$\bar q$};
\vertex[right=1cm of v1] (l1);
\vertex[right=2cm of l1, empty dot] (l2) {};
\vertex[above right=1cm and 1cm of l1, empty dot] (l3) {};
\vertex[below=2cm of l3] (l4);
\vertex[right= 2cm of l3] (f1) {$Z$};
\vertex[below= 1cm of f1] (f2) {$H$};
\vertex[below= 1cm of f2] (f3) {$g$};
\diagram* {
{[edges={fermion, arrow size=1pt}]
(l1) -- (l3) -- (l2) -- (l4) -- (l1),},
(a1) -- [fermion, arrow size=1pt] (v1),
(a2) -- [anti fermion, arrow size=1pt] (v1) -- [gluon] (l1),
(l3) -- [photon] (f1),
(l2) -- [scalar] (f2),
(l4) -- [gluon] (f3)};
\end{feynman}
\end{tikzpicture}}
\hfill
\scalebox{0.75}{
\begin{tikzpicture}
\begin{feynman}
\vertex (a1) {$q$};
\vertex[below=2cm of a1] (a2) {$\bar q$};
\vertex[below right = 0.5cm and 1cm of a1] (v1);
\vertex[below = 1cm of v1, empty dot] (v2) {};
\vertex[right=1cm of v1] (l1);
\vertex[above right=0.5cm and 1cm of l1] (l2);
\vertex[below=1cm of l2, empty dot] (l3) {};
\vertex[right= 1cm of l3] (f2) {$H$};
\vertex[below= 1cm of f2] (f3) {$Z$};
\vertex[above= 1cm of f2] (f1) {$g$};
\diagram* {
{[edges={fermion, arrow size=1pt}]
(l1) -- (l2) -- (l3) -- (l1),},
(a1) -- [fermion, arrow size=1pt] (v1) --[fermion, arrow size=1pt] (v2) -- [fermion, arrow size=1pt] (a2),
(v1) -- [gluon] (l1),
(v2) -- [photon] (f3),
(l3) -- [scalar] (f2),
(l2) -- [gluon] (f1)};
\end{feynman}
\end{tikzpicture}}
    \caption{Representative diagrams included in the computation of $gg\to ZHj$ in SMEFT with SM topology. The empty dots represent couplings that could be either SM-like or modified by dimension-6 operators. Only one insertion of dimension-6 operators is allowed per diagram.}
    \label{fig:feyn_Zhj_SM}
\end{figure}

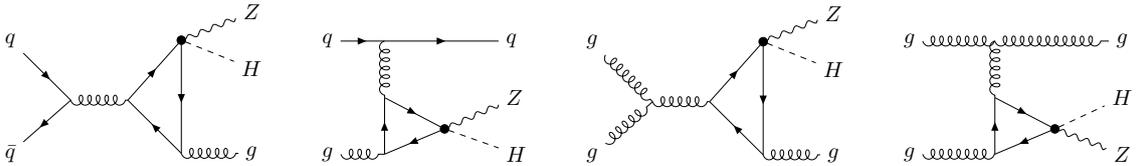
\begin{figure}[h!]
    \centering
\scalebox{0.75}{
\begin{tikzpicture}
\begin{feynman}
\vertex (a1) {$q$};
\vertex[below right = of a1] (v1);
\vertex[below=2cm of a1] (a2) {$\bar q$};
\vertex[right=1cm of v1] (l1);
\vertex[dot, right=3cm of a1] (l2) {};
\vertex[below=2cm of l2] (l3);
\vertex[right= 1cm of l3](f3){$g$};
\vertex[above= 2.5cm of f3] (f1) {$Z$};
\vertex[above= 1.5cm of f3] (f2) {$H$};
\diagram* {
{[edges={fermion, arrow size=1pt}]
(l1) -- (l2) -- (l3) -- (l1),},
(a1) -- [fermion, arrow size=1pt] (v1),
(a2) -- [anti fermion, arrow size=1pt] (v1) -- [gluon] (l1),
(l2) -- [photon] (f1),
(l2) -- [scalar] (f2),
(l3) -- [gluon] (f3) };
\end{feynman}
\end{tikzpicture}}
\hfill
\scalebox{0.75}{
\begin{tikzpicture}
\begin{feynman}
\vertex (i1) {$q$};
\vertex[below=2cm of i1] (i2) {$g$};
\vertex[right=1cm of i1] (v1);
\vertex[below=1cm of v1] (l1);
\vertex[dot, below right=0.5cm and 1cm of l1] (l2) {};
\vertex[below=1cm of l1] (l3);
\vertex[right=2cm of l1] (f1) {$Z$};
\vertex[right=2cm of l3] (f2) {$H$};
\vertex[right=2cm of v1] (f3) {$q$};
\diagram* {{[edges={fermion, arrow size=1pt}]
(l1) -- (l2) -- (l3) -- (l1),},
(i1) -- [fermion, arrow size=1pt] (v1) -- [fermion, arrow size=1pt] (f3),
(v1) -- [gluon] (l1),
(i2) -- [gluon] (l3),
(l2) -- [photon] (f1),
(l2) -- [scalar] (f2) };
\end{feynman}
\end{tikzpicture}}
\hfill
\scalebox{0.75}{        
\begin{tikzpicture}
\begin{feynman}
\vertex (a1) {$g$};
\vertex[below right = of a1] (v1);
\vertex[below=2cm of a1] (a2) {$g$};
\vertex[right=1cm of v1] (l1);
\vertex[dot, right=3cm of a1] (l2) {};
\vertex[below=2cm of l2] (l3);
\vertex[right= 1cm of l3](f3){$g$};
\vertex[above= 2.5cm of f3] (f1) {$Z$};
\vertex[above= 1.5cm of f3] (f2) {$H$};
\diagram* {{[edges={fermion, arrow size=1pt}]
(l1) -- (l2) -- (l3) -- (l1),},
(a1) -- [gluon] (v1),
(a2) -- [gluon] (v1) -- [gluon] (l1),
(l2) -- [photon] (f1),
(l2) -- [scalar] (f2),
(l3) -- [gluon] (f3) };
\end{feynman}
\end{tikzpicture}}
\hfill
\scalebox{0.75}{
\begin{tikzpicture}
\begin{feynman}
\vertex (i1) {$g$};
\vertex[below=2cm of i1] (i2) {$g$};
\vertex[right=1.5cm of i1] (v1);
\vertex[below=1cm of v1] (l1);
\vertex[dot, below right=0.5cm and 1cm of l1] (l2) {};
\vertex[below=1cm of l1] (l3);
\vertex[right=2cm of v1] (f3) {$g$};
\vertex[below=2cm of f3] (f1) {$Z$};
\vertex[below=1cm of f3] (f2) {$H$};
\diagram* {{[edges={fermion, arrow size=1pt}]
(l1) -- (l2) -- (l3) -- (l1),},
(i1) -- [gluon] (v1) -- [gluon] (f3),
(v1) -- [gluon] (l1),
(i2) -- [gluon] (l3),
(l2) -- [photon] (f1),
(l2) -- [scalar] (f2) };
\end{feynman}
\end{tikzpicture}}
    \caption{Feynman diagrams of $gg\to ZH$ with new topologies introduced by the dimension-6 operator $\OhqTopMinus$. The dimension-6 insertion is marked with a black dot.}
    \label{fig:feyn_Zhj_BSM}
\end{figure}

The $0+1$ jet merged samples were generated including all the following processes:
\begin{align}
    gg\to&\, \bar{\nu}\nu H,\\
    gg\to&\, \bar{\nu}\nu H g,\\
    qg\to&\, \bar{\nu}\nu H q, \\
    qq\to&\, g \to \bar{\nu}\nu H g,
\end{align}
and we implemented a loop filter in \textsc{MadGraph} such that only the diagrams in Fig.~\ref{fig:feyn_Zhj_SM} were included in the case of the SM. The dimension-6 operators we study can modify the $Zt\bar t$, $Zq\bar q$, and $t\bar t H$ couplings and introduce the additional topologies shown in Fig.~\ref{fig:feyn_Zhj_BSM}. The latter effect is due only to $\OhqTopMinus$. 

The (0+1) jet process was simulated only in the $Z\to \bar{\nu}\nu$ decay channel since the comparison between QCD orders is independent of the EW decay channel. The generation-level cuts used can be found in Table~\ref{tab:generation_cuts_jet}. We checked that the $k_{\perp}-$ cutoff {[}GeV{]} setting leads to a successful merging by looking at the $p_{T}^{j}$ and $\Delta R_{jj}$ distributions and ensuring they were smooth.

\begin{table}[htb]
\begin{centering}
\centering\renewcommand*{\arraystretch}{1.5}
\begin{tabular}{|c | c  |}
\hline
 & \centering{$Z\rightarrow\nu\bar{\nu}$} \tabularnewline
\hline 
$p_{T,\min}^{j}$ {[}GeV{]} & \centering{$20$}  \tabularnewline
$k_{\perp}-$ cutoff {[}GeV{]} & \centering{$\max\lbrace60,\,\frac{p_{T,\text{min\{bin\}}}^{Z}}{2}\rbrace$}  \tabularnewline
$|\eta_{max}^{j}|$ & \centering{$5$} \tabularnewline
$p_{T}^{Z}$ & \centering{$\{0,\,200,\,400,\,600,\,800,\,\infty\}$} \tabularnewline
\hline
\end{tabular}
\par\end{centering}
\caption{Parton level generation cuts for $gg\to ZH$ at $13\,\mathrm{TeV}$ for $0+1$ jet merged samples.}
\label{tab:generation_cuts_jet}
\end{table}

\section{B-tagging and analysis strategy}
\label{app:Analysis_details}

The classification of collider events as boosted or resolved is done via the scale-invariant tagging procedure~\cite{Gouzevitch:2013qca}, as implemented in Ref.~\cite{Bishara:2016kjn,Bishara:2022vsc}. This procedure looks for boosted Higgs candidate first and then, if the result is negative, looks for two resolved $b$-jets. Boosted jets are reconstructed via the mass-drop-tagging algorithm~\cite{Butterworth:2008iy} and we require them to have two $b$-tags in order to be considered as a Higgs candidate.

The selection cuts used in each of the categories are based on LHC Run 2 analyses performed by the ATLAS collaboration~\cite{ATLAS:2020fcp,ATLAS:2020jwz} but with small differences since they were optimised on the basis of differential distributions obtained from our simulations. Here, we will discuss the most useful cuts for the $0$- and $2$-lepton channels while the full details on them and a cutflow analysis can be found in Ref.~\cite{Bishara:2022vsc}.

The main tool to reject background events is the cut on the invariant mass of the Higgs candidate, $m_{bb}\in [90-120]$~GeV. This reduces the background cross-section by at least one order of magnitude in any of the channels without affecting the signal. In the 0-lepton channel, the application of a jet veto on additional untagged jets is crucial to reduce the very high background generated by $t\bar t$ production, making it relevant only in the lowest energy bins with a resolved Higgs. The same jet veto also helps to control the contributions of $Wb\bar b$ and $Zb\bar b$ to the background, though it reduces the contribution of $WH$ to the signal. The 2-lepton channel has a background composed of only $Zb\bar b$ production, which is greatly reduced by a cut on the lepton $p_T$ imbalance, $\frac{|p_{T}^{\ell_1} - p_{T}^{\ell_2}|}{p_{T}^{Z}}$. Nevertheless, the $Zb\bar b$-dominated background overwhelms the SM signal in most bins for both the $0-$ and $2-$lepton channels.

\section{Projected bounds at LHC Run 3}
\label{app:Bounds_LHCRun3}

In this appendix, we present the projected bounds on dimension-6 SMEFT WCs from $pp\to ZH$ for LHC Run 3 luminosity. These can be found in Table \ref{tab:bounds_summary_NFU_0plus2lep_LHCR3_kF2}.

\begin{table}[htb]
	\begin{centering}
		\begin{tabular}{|c|c|}
			\hline
			WC &  $95\%$ C.L. Bound \tabularnewline
			\hline
			$c_{\varphi q}^{(3)}$ &
			\begin{tabular}{ll}
				\rule{0pt}{1.25em}$[-5.3,\,3.5]\times10^{-2}$ & $1\%$ syst.\\[-0.65em]
				\rule{0pt}{1.25em}$[-6.1,\,3.7]\times10^{-2}$ & $5\%$ syst.\\[-0.65em]
				\rule[-.65em]{0pt}{1.9em}$[-8.2,\,4.2]\times10^{-2}$ & $10\%$ syst.
			\end{tabular}
			\tabularnewline
			
			\hline
			$c_{\varphi q}^{(1)}$ &
			\begin{tabular}{ll}
				\rule{0pt}{1.25em}$[-8.7,\,10.6]\times10^{-2}$ & $1\%$ syst.\\[-0.65em]
				\rule{0pt}{1.25em}$[-9.0,\,10.8]\times10^{-2}$ & $5\%$ syst.\\[-0.65em]
				\rule[-.65em]{0pt}{1.9em}$[-9.5,\,11.4]\times10^{-2}$ & $10\%$ syst.
			\end{tabular}
			\tabularnewline
			
			\hline
			$c_{\varphi u}$ &
			\begin{tabular}{ll}
				\rule{0pt}{1.25em}$[-16.7,\,9.0]\times10^{-2}$ & $1\%$ syst.\\[-0.65em]
				\rule{0pt}{1.25em}$[-16.9,\,9.3]\times10^{-2}$ & $5\%$ syst.\\[-0.65em]
				\rule[-.65em]{0pt}{1.9em}$[-1.8,\,1.0]\times10^{-1}$ & $10\%$ syst.
			\end{tabular}
			\tabularnewline
			
			\hline
			$c_{\varphi d}$ &
			\begin{tabular}{ll}
				\rule{0pt}{1.25em}$[-1.3,\,1.7]\times10^{-1}$ & $1\%$ syst.\\[-0.65em]
				\rule{0pt}{1.25em}$[-1.4,\,1.8]\times10^{-1}$ & $5\%$ syst.\\[-0.65em]
				\rule[-.65em]{0pt}{1.9em}$[-1.5,\,1.9]\times10^{-1}$ & $10\%$ syst.
			\end{tabular} 
			\tabularnewline
                \hline
            \end{tabular}
  		\begin{tabular}{|c|c|}
			\hline
			WC &  $95\%$ C.L. Bound \tabularnewline
			\hline
			$c_{\varphi Q}^{(3)}$ &
			\begin{tabular}{ll}
				\rule{0pt}{1.25em}$[-10.3,\,8.7]\times10^{-1}$ & $1\%$ syst.\\[-0.65em]
				\rule{0pt}{1.25em}$[-10.5,\,9.0]\times10^{-1}$ & $5\%$ syst.\\[-0.65em]
				\rule[-.65em]{0pt}{1.9em}$[-11.2,\,9.7]\times10^{-1}$ & $10\%$ syst.
			\end{tabular}
			\tabularnewline		
			\hline
			$c_{\varphi Q}^{(-)}$ &
			\begin{tabular}{ll}
				\rule{0pt}{1.25em}$[-2.10,\,1.67]$ & $1\%$ syst.\\[-0.65em]
				\rule{0pt}{1.25em}$[-2.14,\,1.74]$ & $5\%$ syst.\\[-0.65em]
				\rule[-.65em]{0pt}{1.9em}$[-2.3,\,\phantom{11}1.9]\phantom{\times10^{-1}}$ & $10\%$ syst.
			\end{tabular}	
			\tabularnewline			
			\hline
			$c_{\varphi t}$ &
			\begin{tabular}{ll}
				\rule{0pt}{1.25em}$[-12.9,\,25.6]$ & $1\%$ syst.\\[-0.65em]
				\rule{0pt}{1.25em}$[-13.8,\,26.1]$ & $5\%$ syst.\\[-0.65em]
				\rule[-.65em]{0pt}{1.9em}$[ -15.4,\,27.4]\phantom{\times10^{-1}}$ & $10\%$ syst.
			\end{tabular}
			\tabularnewline			
			\hline
			$c_{t \varphi }$ &
			\begin{tabular}{ll}
				\rule{0pt}{1.25em}$[-25.4,\,12.8]$ & $1\%$ syst.\\[-0.65em]
				\rule{0pt}{1.25em}$[-25.9,\,13.7]$ & $5\%$ syst.\\[-0.65em]
				\rule[-.65em]{0pt}{1.9em}$[-27.2,\,15.3]\phantom{\times10^{-1}}$ & $10\%$ syst.
			\end{tabular} \\
			\hline
		\end{tabular}
        \end{centering}
	\caption{ Projected bounds at $95\%$ C.L.~from one-dimensional fits on the dimension-6 SMEFT WCs probed by $pp\to ZH$ at LHC Run 3 with integrated luminosity of $300\,\mathrm{fb}^{-1}$. The WCs are in units of TeV$^{-2}$. The gluon-initiated channels were rescaled with a constant k-factor of $2.0$ to account for NLO QCD corrections. {\bf Left: } Light-quark WCs. {\bf Right column:} Heavy-quark WCs.}
	\label{tab:bounds_summary_NFU_0plus2lep_LHCR3_kF2}
\end{table}

\section{Signal and background cross-sections in $ZH$ production}
\label{app:EvtNumbersVH}

In this appendix, we present the number of signal and background events after the selection cuts per bin for each $pp\to ZH$ channel. We present the results for HL-LHC ($14$~TeV and $3$~ab$^{-1}$) only. The corresponding numbers for LHC Run 3 ($300$~fb$^{-1}$) can be obtained by simple rescaling since the change in energy has a negligible impact.
The number of signal events is given as a quadratic function of the studied WCs. The background contributions include the main background processes, as described in Ref.~\cite{Bishara:2022vsc}. 

\subsection{The 0-lepton channel}
\label{app:EvtNumbers_Zh_vv}

\begin{table}[t]
	\centering
		\begin{scriptsize}
			\begin{tabular}{|c|c|c|c@{\hspace{.25em}}|}
				\hline
				\multicolumn{3}{|c|}{0-lepton channel, resolved, HL-LHC} \tabularnewline
				\hline
				\multirow{2}{*}{
				\hspace{-2.5em}
				\begin{tabular}{c}
				$p_{T,\mathrm{min}}$ bin\\
				$[$GeV$]$
				\end{tabular}
				\hspace{-2.5em}} & \multicolumn{2}{c|}{Number of expected events}\tabularnewline
				\cline{2-3} &  \rule{0pt}{1.15em}Signal & Background \tabularnewline
				\hline
				$[0-160]$ & 
				$\begin{aligned} \phSpa 350 \,& + 1130 \,\chqtii + (-32 \pm 23)\,\chqii
                                              + (155 \pm 40 ) \,\chu - (80 \pm 30) \,\chd \\
                                              & + ( 21\pm 4 ) \, \chqTopt + (13 \pm 2) \, \chqTopMinus - (3.7 \pm 0.2) \, \cht + (3.9\pm0.2) \,\cth \\
                                              &+ 1260 \,\left(\chqtii\right)^{2}
                                              + 986 \,\left(\chqii\right)^{2} +(540 \pm 74)\,\left(\chu\right)^{2} \\
                                              &+  (400 \pm 46)\,\left(\chd\right)^{2} + (25 \pm 4)\,\left(\chqTopt \right)^{2} + (8.2 \pm 0.5)\,\left(\chqTopMinus \right)^{2}\\ 
                                              &+  (0.09\pm 0.01) \left( \cht \right)^{2} + (0.104 \pm 0.006)\,\left(\cth \right)^{2}\\ 
                                              \rule[-1.em]{0pt}{1em}& -(200 \pm 240)\,\chqtii\,\chqii + (27 \pm 4)\, \chqTopt\, \chqTopMinus\\ 
                                              &-0.202 \,\chqTopMinus\,\cht +0.200\,\chqTopMinus\, \cth -0.204\,\cht\,\cth \rule[-2ex]{0pt}{0ex}
				\end{aligned}$ & $8000\pm3700$ \tabularnewline \hline
	            $[160-200]$ & $\begin{aligned} \phSpa 800 \,& + 3160 \,\chqtii - (53 \pm 30)\,\chqii
                                              + (596 \pm 50 ) \,\chu - (263 \pm 40) \,\chd \\
                                              & + ( 35\pm 5 ) \, \chqTopt + (31 \pm 3) \, \chqTopMinus - 9.9 \, \cht + 10.1 \,\cth \\
                                              &+ 4340 \,\left(\chqtii\right)^{2} + 3420 \,\left( \chqii \right)^{2} + 1930 \,\left(\chu\right)^{2} \\
                                              & +  1510 \,\left(\chd\right)^{2} + 103 \,\left(\chqTopt \right)^{2} + 23.7 \,\left(\chqTopMinus \right)^{2}\\ 
                                              & +  (0.34\pm0.02) \left( \cht \right)^{2} + 0.35 \,\left(\cth \right)^{2}\\ 
                                              \rule[-1.em]{0pt}{1em}& -(1600 \pm 320)\,\chqtii\,\chqii + 103 \, \chqTopt\, \chqTopMinus\\ 
                                              & - 0.660 \,\chqTopMinus\,\cht + 0.706 \,\chqTopMinus\, \cth - 0.68\,\cht\,\cth \rule[-2ex]{0pt}{0ex}
				\end{aligned}$ & $5350\pm1400$ \tabularnewline
				\hline 
				 $[200-250]$ &  $\begin{aligned} \phSpa 213 \,& + 1150 \,\chqtii - (39 \pm 10)\,\chqii
                                              + (222 \pm 15 ) \,\chu - (90 \pm 13) \,\chd \\
                                              & + ( 10 \pm 1 ) \, \chqTopt + (8.1 \pm 0.7) \, \chqTopMinus - 2.78 \, \cht + 2.9 \,\cth \\
                                              & + 2060 \,\left(\chqtii\right)^{2}  + 1640 \,\left(\chqii\right)^{2} + 950 \,\left(\chu\right)^{2} \\
                                              & +  700 \,\left(\chd\right)^{2} + 39 \,\left(\chqTopt \right)^{2} + 9.4\,\left(\chqTopMinus \right)^{2}\\ 
                                              &+  0.125 \left( \cht \right)^{2} + 0.123 \,\left(\cth \right)^{2}\\ 
                                              \rule[-1.em]{0pt}{1em}& -(460 \pm 140)\,\chqtii\,\chqii + 39\, \chqTopt\, \chqTopMinus\\ 
                                              &- 0.231 \,\chqTopMinus\,\cht +0.286\,\chqTopMinus\, \cth - 0.246 \,\cht\,\cth \rule[-2ex]{0pt}{0ex}
				\end{aligned}$ & $1310\pm90$ \tabularnewline
				\hline 
				 $[250-\infty]$ &  $\begin{aligned}  \phSpa 35.8 \,& + 315 \,\chqtii - (24 \pm 5)\,\chqii + (66 \pm 7) \,\chu - (27 \pm 6) \,\chd \\
                                              & + ( 3.3 \pm 0.6 ) \, \chqTopt + ( 2.0 \pm 0.3 ) \, \chqTopMinus - 0.49 \, \cht + (0.47\pm0.05) \,\cth \\
                                              & + 1025 \,\left(\chqtii\right)^{2} + 883 \,\left(\chqii\right)^{2} + 511 \,\left(\chu\right)^{2} \\
                                              & +  350 \,\left(\chd\right)^{2} + (12.1 \pm 0.7)\,\left(\chqTopt \right)^{2} + 3.2 \,\left(\chqTopMinus \right)^{2}\\ 
                                              & +  0.035 \left( \cht \right)^{2} + (0.034 \pm 0.003)\,\left(\cth \right)^{2}\\ 
                                              \rule[-1.em]{0pt}{1em}& - ( 380 \pm 80 )\,\chqtii\,\chqii + (12.1 \pm 0.7)\, \chqTopt\, \chqTopMinus\\ 
                                              & - 0.066 \,\chqTopMinus\,\cht + 0.085\,\chqTopMinus\, \cth - (0.066 \pm 0.004)\,\cht\,\cth \rule[-2ex]{0pt}{0ex}
				\end{aligned}$ & $265\pm37$ \tabularnewline
				\hline 
\end{tabular}
\end{scriptsize}
	\caption{Number of expected signal events as a function of the WCs (in units of TeV$^{-2}$) and of total background events in the $pp\rightarrow ZH \rightarrow \nu \bar{\nu} b\bar{b}$ channel, resolved category, at HL-LHC.
	The Monte Carlo errors on the fitted coefficients, when not explicitly specified, are $\lesssim 5$ \%.
	}
	\label{tab:App_sigma_full_Zh_neut_HL_LHC_res}
\end{table}
\begin{table}[t]
	\centering
			\begin{scriptsize}
			\begin{tabular}{|c|c|c|c@{\hspace{.25em}}|}
				\hline
				\multicolumn{3}{|c|}{0-lepton channel, boosted, HL-LHC} \tabularnewline
				\hline
				\multirow{2}{*}{
				\hspace{-2.5em}
				\begin{tabular}{c}
				$p_{T,\mathrm{min}}$ bin\\
				$[$GeV$]$
				\end{tabular}
				\hspace{-2.5em}} &
				\multicolumn{2}{c|}{Number of expected events}\tabularnewline
				\cline{2-3} &  \rule{0pt}{1.15em}Signal & Background \tabularnewline
				\hline
				$[0-300]$ & 
				$\begin{aligned} \phSpa 125 \,& + 1177 \,\chqtii - ( 66 \pm 10 ) \,\chqii
                                              + ( 250 \pm 14 ) \,\chu - ( 125 \pm 13) \,\chd \\
                                              & + ( 10 \pm 1 ) \, \chqTopt + ( 5.8 \pm 0.6) \, \chqTopMinus - 1.67 \, \cht + (1.61\pm0.09) \,\cth \\
                                              & + 3490 \,\left(\chqtii\right)^{2} + 2970 \,\left(\chqii\right)^{2} + 1720 \,\left(\chu\right)^{2} \\
                                              & + 1300 \,\left(\chd\right)^{2} + 44 \,\left(\chqTopt \right)^{2} + 11.8 \,\left(\chqTopMinus \right)^{2}\\ 
                                              & +  0.131 \left( \cht \right)^{2} + 0.125 \,\left(\cth \right)^{2}\\ 
                                              \rule[-1.em]{0pt}{1em}& - (1100\pm170) \,\chqtii\,\chqii + 44 \, \chqTopt\, \chqTopMinus\\ 
                                              & - 0.254 \,\chqTopMinus\,\cht + 0.269 \,\chqTopMinus\, \cth - 0.244 \,\cht\,\cth \rule[-2ex]{0pt}{0ex}
				\end{aligned}$ & $492\pm50$\tabularnewline
				\hline 
				 $[300-350]$ &  $\begin{aligned} \phSpa 120 \,& + 1413 \,\chqtii - ( 97 \pm 10 ) \,\chqii
                                              + ( 272 \pm 14 ) \,\chu - (75 \pm 13) \,\chd \\
                                              & + ( 9 \pm 1 ) \, \chqTopt + (6.5 \pm 0.5) \, \chqTopMinus - 1.39 \, \cht + (1.54\pm0.09) \,\cth \\
                                              &+ 5220 \,\left(\chqtii\right)^{2}
                                              + 4580 \,\left(\chqii\right)^{2} + 2670 \,\left(\chu\right)^{2} \\
                                              & +  1810 \,\left(\chd\right)^{2} + 60 \,\left(\chqTopt \right)^{2} + 14.9\,\left(\chqTopMinus \right)^{2}\\ 
                                              &+  0.139 \left( \cht \right)^{2} + 0.156 \,\left(\cth \right)^{2}\\ 
                                              \rule[-1.em]{0pt}{1em}& -(1710 \pm 190)\,\chqtii\,\chqii + 60 \, \chqTopt\, \chqTopMinus\\ 
                                              & - 0.292 \,\chqTopMinus\,\cht + 0.226\,\chqTopMinus\, \cth - 0.298 \,\cht\,\cth \rule[-2ex]{0pt}{0ex}
				\end{aligned}$ & $492\pm43$ \tabularnewline
				\hline
				 $[350-\infty]$ &  $\begin{aligned} \phSpa 113 \,& + 2123 \,\chqtii - 208 \,\chqii + 490 \,\chu - (163 \pm 9) \,\chd \\
                                              & + ( 9.8 \pm 0.6 ) \, \chqTopt + (5.6 \pm 0.3) \, \chqTopMinus - 0.94 \, \cht + 0.99 \,\cth \\
                                              & + 12860 \,\left(\chqtii\right)^{2} + 11640 \,\left(\chqii\right)^{2} + 7050 \,\left(\chu\right)^{2} \\
                                              & +  4650 \,\left(\chd\right)^{2} + 94.9 \,\left(\chqTopt \right)^{2} + 24.1 \,\left(\chqTopMinus \right)^{2}\\ 
                                              & +  0.147 \left( \cht \right)^{2} + 0.153 \,\left(\cth \right)^{2}\\ 
                                              \rule[-1.em]{0pt}{1em}& - 4730 \,\chqtii\,\chqii + 95.0\, \chqTopt\, \chqTopMinus\\ 
                                              & - 0.2985 \,\chqTopMinus\,\cht + 0.317 \,\chqTopMinus\, \cth - 0.304 \,\cht\,\cth \rule[-2ex]{0pt}{0ex}
				\end{aligned}$ & $243\pm16$ \tabularnewline
				\hline
\end{tabular}
\end{scriptsize}
	\caption{Number of expected signal and background events in the $pp\rightarrow ZH \rightarrow \nu \bar{\nu} b\bar{b}$ channel, boosted category, at HL-LHC. The Monte Carlo errors on the fitted coefficients, when not explicitly specified, are $\lesssim 5$ \%.
	}
	\label{tab:App_sigma_full_Zh_neut_HL_LHC_boos}
\end{table}

The number of signal and background events in the 0-lepton channel at HL-LHC is reported in tables~\ref{tab:App_sigma_full_Zh_neut_HL_LHC_res} and \ref{tab:App_sigma_full_Zh_neut_HL_LHC_boos} for the resolved and boosted channels respectively. In this channel, the number of signal events includes the contributions from both $ZH\to\nu\bar\nu b\bar b$ and $WH\to \nu\ell b \bar b$ with a missing lepton. The latter process only contributes to the SM, $\chqtii$ and $(\chqtii)^2$ coefficients.

\clearpage

\subsection{The 2-lepton channel}
\label{app:EvtNumbers_Zh_ll}

In this subsection, we present the number of signal and background events in the 2-lepton channel at HL-LHC. Tables~\ref{tab:App_sigma_full_Zh_lep_HL_LHC_res} and~\ref{tab:App_sigma_full_Zh_lep_HL_LHC_boos} show the results in the resolved and boosted categories respectively. The signal is composed of the processes $q\bar q\to ZH\to \ell^+\ell^- b\bar{b}$ and $g\bar g\to ZH\to \ell^+\ell^- b\bar{b}$.

\begin{table}[t]
	\centering
		\begin{scriptsize}
			\begin{tabular}{|@{\hspace{.35em}}c|c|c@{\hspace{.5em}}|}
				\hline
				\multicolumn{3}{|c|}{2-lepton channel, resolved, HL-LHC} \tabularnewline
				\hline
				\multirow{2}{*}{$p_{T,\mathrm{min}}$ bin [GeV]} & \multicolumn{2}{c|}{Number of expected events}\tabularnewline
				\cline{2-3} &  \rule{0pt}{1.15em}Signal & Background \tabularnewline
				\hline
				 $[175-200]$ &  $\begin{aligned} \phSpa 64 \,& + 276 \,\chqtii - (3 \pm 7) \,\chqii
                                              + (73\pm 5) \,\chu - (19 \pm 4) \,\chd \\
                                              & + ( 4.4 \pm 0.5 ) \, \chqTopt + ( 3.1 \pm 0.3 ) \, \chqTopMinus - 0.9 \, \cht + 0.97 \,\cth \\
                                              &+ 401 \,\left(\chqtii\right)^{2}
                                              + 405 \,\left(\chqii\right)^{2} + 240 \,\left(\chu\right)^{2} \\
                                              & +  172 \,\left(\chd\right)^{2} + (10.0 \pm 0.5)\,\left(\chqTopt \right)^{2} + 2.73 \,\left(\chqTopMinus \right)^{2}\\ 
                                              & +  (0.033\pm0.002) \left( \cht \right)^{2} + (0.036 \pm 0.002)\,\left(\cth \right)^{2}\\ 
                                              \rule[-1.em]{0pt}{1em}& - (140 \pm 50)\,\chqtii\,\chqii + (10.1 \pm 0.6)\, \chqTopt\, \chqTopMinus\\ 
                                              &- 0.0694 \,\chqTopMinus\,\cht + 0.066\,\chqTopMinus\, \cth - 0.069\,\cht\,\cth \rule[-2ex]{0pt}{0ex}
				\end{aligned}$ & $361\pm21$ \tabularnewline
				\hline 
				 $[200-\infty]$ &  $\begin{aligned} \phSpa 52 \,& + 298 \,\chqtii - (5 \pm 6) \,\chqii
                                              + (65\pm 5) \,\chu - (25 \pm 4) \,\chd \\
                                              & + ( 3.6 \pm 0.4 ) \, \chqTopt + (2.6 \pm 0.2) \, \chqTopMinus - 0.76 \, \cht + 0.78 \,\cth \\
                                              & + 580 \,\left(\chqtii\right)^{2}
                                              + 560 \,\left(\chqii\right)^{2} + 330 \,\left(\chu\right)^{2} \\
                                              & +  257 \,\left(\chd\right)^{2} + 12.7\,\left(\chqTopt \right)^{2} + 3.23\,\left(\chqTopMinus \right)^{2}\\ 
                                              & +  0.037 \left( \cht \right)^{2} + 0.039 \,\left(\cth \right)^{2}\\ 
                                              \rule[-1.em]{0pt}{1em}& - ( 110 \pm 50)\,\chqtii\,\chqii + 12.8\, \chqTopt\, \chqTopMinus\\ 
                                              & - 0.0707 \,\chqTopMinus\,\cht + 0.075\,\chqTopMinus\, \cth - 0.077 \,\cht\,\cth \rule[-2ex]{0pt}{0ex}
				\end{aligned}$ & $296 \pm 19$ \tabularnewline
				\hline
		\end{tabular}
		\end{scriptsize}
	\caption{ Number of expected signal and background events in the $pp\rightarrow ZH \rightarrow \ell^+ \ell^- b\bar{b}$ channel, resolved category, at HL-LHC. The Monte Carlo errors on the fitted coefficients, when not explicitly specified, are $\lesssim 5$ \%.}
	\label{tab:App_sigma_full_Zh_lep_HL_LHC_res}
\end{table}

\begin{table}[t]
	\begin{center}
		\setlength{\extrarowheight}{0mm}%
		\begin{scriptsize}
			\begin{tabular}{|@{\hspace{.35em}}c|c|c@{\hspace{.5em}}|}
				\hline
				\multicolumn{3}{|c|}{2-lepton channel, boosted, HL-LHC} \tabularnewline
				\hline
				\multirow{2}{*}{$p_{T,\mathrm{min}}$ bin [GeV]} & \multicolumn{2}{c|}{Number of expected events}\tabularnewline
				\cline{2-3} &  \rule{0pt}{1.15em}Signal & Background \tabularnewline
				\hline
				 $[250-\infty]$ &  $\begin{aligned} \phSpa 119 \,& + 977 \,\chqtii - (53 \pm 11) \,\chqii
                                              + 231 \,\chu - (80 \pm 7) \,\chd \\
                                              & +  14.9 \, \chqTopt + 11.6 \, \chqTopMinus - 3.89 \, \cht + 3.86 \,\cth \\
                                              &+ 2800 \,\left(\chqtii\right)^{2}
                                              + 2840 \,\left(\chqii\right)^{2} + 1660 \,\left(\chu\right)^{2} \\
                                              & +  1150 \,\left(\chd\right)^{2} + 69.2 \,\left(\chqTopt \right)^{2} + 17.6 \,\left(\chqTopMinus \right)^{2}\\ 
                                              & +  0.283 \left( \cht \right)^{2} + 0.282 \,\left(\cth \right)^{2}\\ 
                                              \rule[-1.em]{0pt}{1em}& -(1070 \pm 95)\,\chqtii\,\chqii + 69.5 \, \chqTopt\, \chqTopMinus\\ 
                                              &- 0.573 \,\chqTopMinus\,\cht + 0.570 \,\chqTopMinus\, \cth - 0.565 \,\cht\,\cth \rule[-2ex]{0pt}{0ex}
				\end{aligned}$ & $370\pm21$ \tabularnewline
				\hline
		\end{tabular}
		\end{scriptsize}
	\end{center}
	\caption{ Number of expected signal and background events in the $pp\rightarrow ZH \rightarrow \ell^+ \ell^- b\bar{b}$ channel, boosted category, at HL-LHC. The Monte Carlo errors on the fitted coefficients, when not explicitly specified, are $\lesssim 5$ \%.}
	\label{tab:App_sigma_full_Zh_lep_HL_LHC_boos}
\end{table}

\bibliographystyle{JHEP.bst}
\bibliography{bibGGZH.bib}
\end{document}